\documentclass[a4paper,11pt]{article}
\usepackage{jheppub} 
\usepackage{amsmath}
\usepackage{amssymb}
\usepackage{wasysym}
\usepackage{etoolbox}
\usepackage{soul}
\usepackage{tabularray}

\newcommand{\beq}{\begin{equation}}
\newcommand{\eeq}{\end{equation}}
\newcommand{\z}[1]{\zeta_{#1}}
\newcommand{\s}[1]{\text{S}_{#1}}
\newcommand{\sC}[1]{\tilde{\text{S}}_{#1}}
\newcommand{\bI}[1]{\text{I}_{#1}}

\def\A{\mathcal{A}}
\def\B{\mathcal{B}}

\def\D{\mathcal{D}}
\def\E{\mathcal{E}}

\def\G{\mathcal{G}}
\def\I{\mathcal{I}}
\def\J{\mathcal{J}}

\def\L{\mathcal{L}}
\def\N{\mathcal{N}}

\def\Op{\mathcal{O}}
\def\S{\mathcal{S}}
\def\Z{Z}
\def\bt{\bold{t}}
\def\bT{\bold{T}}
\def\bP{\bold{P}}

\def\R{R}
\def\cl{\text{cl}}

\def\Area{\textrm{Area}}
\def\Em{\tilde{E}}
\def\mum{\tilde{\mu}}

\def\Xint#1{\mathchoice
   {\XXint\displaystyle\textstyle{#1}}%
   {\XXint\textstyle\scriptstyle{#1}}%
   {\XXint\scriptstyle\scriptscriptstyle{#1}}%
   {\XXint\scriptscriptstyle\scriptscriptstyle{#1}}%
   \!\int}
\def\XXint#1#2#3{{\setbox0=\hbox{$#1{#2#3}{\int}$}
     \vcenter{\hbox{$#2#3$}}\kern-.5\wd0}}

\def\dashint{\Xint-}

\title{\boldmath Exploring Structure Constants in Planar $\mathcal{N} = 4$ SYM: From Small Spin to Strong Coupling}

\author[a]{Benjamin Basso}
\author[b]{and Alessandro Georgoudis}

\affiliation[a]{Laboratoire de Physique de l’Ecole Normale Sup\'erieure, ENS, Universit\'e PSL, CNRS, Sorbonne Universit\'e, Universit\'e de Paris, 24 rue Lhomond, F-75005 Paris, France}
\affiliation[b]{Centre for Theoretical Physics, Department of Physics and Astronomy, Queen Mary University of London, Mile End Road, London E1 4NS, United Kingdom}
\emailAdd{benjamin.basso@phys.ens.fr} 
\emailAdd{a.georgoudis@qmul.ac.uk}
\abstract{We study the structure constants of two conformal primary operators and one spinning operator in planar $\mathcal{N} = 4$ Super-Yang-Mills theory using the hexagon formalism. By analytically continuing in the spin, we derive a formula for computing these structure constants at any coupling in the small-spin limit, up to a normalization factor. This formula allows us to explore their analytical properties at strong coupling. In this regime, using classical string calculations and a suitable ansatz, we extend our analysis to finite-spin operators, verifying recent two-loop results for structure constants in string theory and generalizing them to operators with arbitrary R-charges.
}

\preprint{QMUL-PH-25-05}

\begin{document}
           \maketitle
\flushbottom

\section{Introduction}

Since its discovery more than two decades ago, integrability in planar $\mathcal{N} = 4$ Super Yang-Mills (SYM) has been pivotal in advancing our understanding of correlation functions at finite ’t Hooft coupling~\cite{Beisert:2010jr}. A central example is the spectrum of scaling dimensions of single-trace operators, which, after extensive developments, was captured exactly by the Quantum Spectral Curve (QSC) formalism~\cite{Gromov:2013pga,Gromov:2014caa}. This framework, along with its Bethe-ansatz predecessors~\cite{Beisert:2010jr}, enabled some of the most striking tests of the AdS/CFT correspondence, including the computation of the Konishi operator’s dimension~\cite{Gubser:1998bc,Gromov:2009bc,Marboe:2014gma,Roiban:2009aa,Roiban:2011fe,Vallilo:2011fj,Gromov:2011de,Basso:2011rs,Gromov:2011bz,Frolov:2010wt,Gromov:2014bva,Ekhammar:2024rfj} and the cusp anomalous dimension~\cite{Gubser:2002tv,Frolov:2002av,Beisert:2006ez,Belitsky:2006en,Bern:2006ew,Benna:2006nd,Basso:2007wd,Kruczenski:2007cy,Roiban:2007dq} at finite coupling.

Beyond the spectrum, integrability has been instrumental in establishing sharp numerical bounds on three-point functions at finite coupling via the conformal bootstrap~\cite{Caron-Huot:2022sdy,Cavaglia:2022qpg,Chester:2023ehi,Caron-Huot:2024tzr,Cavaglia:2024dkk} and in extracting detailed results at strong coupling through string theory methods~\cite{Kazama:2016cfl,Alday:2022uxp,Alday:2022xwz,Gromov:2023hzc,Julius:2023hre,Julius:2024ewf}. It has also inspired direct analytic approaches, such as the hexagon formalism~\cite{Basso:2015zoa,Fleury:2016ykk,Eden:2016xvg,Bargheer:2017nne,Eden:2017ozn}, which offers a systematic approach to studying correlation functions in the planar limit. In particular, its application to four-point functions in the large R-charge regime has led to remarkable finite-coupling results and uncovered unexpected mathematical structures~\cite{Basso:2017khq,Coronado:2018cxj,Kostov:2019stn,Belitsky:2020qrm,Caron-Huot:2023wdh}, pointing to deeper underlying symmetries. Nonetheless, computations involving ``short operators'' with finite quantum numbers remain challenging, even for three-point functions, due primarily to the presence of wrapping corrections~\cite{Basso:2017muf,Basso:2018cvy,Ferrando:2025qkr}. 

While a complete integrability-based solution for three-point functions of generic operators, comparable to the QSC formalism for the spectrum, is still out of reach, significant progress has been achieved for structure constants involving two protected chiral primary operators (CPOs) $\textrm{Tr}\, Z_{i}^{J_{i}}$ and one spin-$S$ operator on the leading Regge trajectory.
Stripping away the kinematic dependence fixed by superconformal symmetry, these structure constants take the form
\beq\label{eq:intro}
C_{123} \sim \langle \textrm{Tr}\, Z_{1}^{J_{1}}\, \textrm{Tr}\, Z_{2}^{J_{2}} \, \textrm{Tr}\, D^{S}Z^{J}\rangle\, ,
\eeq
where $Z_{1},Z_{2}$ and $Z$ are generic complex scalar fields, and $D$ is a light-cone derivative. These quantities depend on the spin $S$, the R-charges $J,J_{1},J_{2}$, and the coupling constant $g^{2} = \lambda/(4\pi)^2$, where $\lambda$ is the 't Hooft coupling. Explicit results are known up to five loops at weak coupling in gauge theory~\cite{Eden:2012rr,Goncalves:2016vir,Eden:2016aqo,Georgoudis:2017meq}, and up to two loops at strong coupling from string theory~\cite{Costa:2012cb,Bargheer:2013faa,Minahan:2014usa,Alday:2022uxp,Alday:2022xwz,Alday:2023flc,Alday:2023jdk,Fardelli:2023fyq,Alday:2023mvu,Caron-Huot:2024tzr,Wang:2025pjo}, for various values of spin and R-charges. The hexagon formula recently introduced in~\cite{Basso:2022nny} describes these structure constants in principle at any coupling in the planar limit, including wrapping effects.

Despite the power of integrability, analytic progress often relies on simplifying limits. As noted earlier, one effective approach is to consider the large R-charge limit, where $J\rightarrow \infty$. In this regime, wrapping corrections are suppressed, allowing the asymptotic hexagon construction to be applied~\cite{Basso:2015zoa}. Another useful, though less conventional, approach involves analytic continuation in the spin. At spin $S=0$, both operators and structure constants are protected from quantum corrections. Expanding around this point introduces notable simplifications, making computations at finite coupling and R-charge more tractable.

In this work, we explore the hexagon formula in this small-spin limit, $S\rightarrow 0$. Similar to scaling dimensions~\cite{Basso:2011rs,Gromov:2011bz,Basso:2012ex,Gromov:2014bva}, we will find that structure constants admit concise representations in this regime. Specifically, we will derive all-loop expressions for structure constants at small spin, up to a normalization factor that depends nontrivially on the R-charge of the spinning operator but not on that of the chiral primary operators.

At strong coupling, $\lambda\rightarrow \infty$, we will examine the connection between these expressions and results for physical operators with integer spins. Drawing on insights from string theory, we will argue that, with a simple refinement, the structure constants admit a strong coupling expansion whose coefficients follow a polynomial pattern in the quantum numbers of the operators, with a structure that smoothly interpolates between the small-spin regime and the classical limit.

We will perform extensive tests of this behavior in the classical regime, where structure constants are obtained from the area of a minimal surface spanned by strings in Anti-de Sitter (AdS) space~\cite{Kazama:2016cfl,Kazama:2011cp,Janik:2011bd}. To this end, we will introduce a method for systematically analyzing the classical string formula in the short-string limit and verify consistency with the expected polynomial structure at higher orders.

Finally, by combining the classical and small-spin approaches, we will perform checks of recent two-loop data from string theory~\cite{Alday:2023mvu} and extract new two-loop predictions for structure constants at strong coupling, valid for operators of arbitrary spin and length.

The paper is organized as follows. In Section~\ref{sec:hexagons}, we describe how to evaluate structure constants at small spin using the hexagon formula, up to a normalization factor. We then analyze the term linear in $S$ at weak and strong coupling, gaining insight into its analytic properties in these regimes. In Section~\ref{sec:ansatz}, we introduce a polynomial ansatz for structure constants at strong coupling, motivated by string theory data, and test it in the classical limit in Section~\ref{sec:classical}. In Section~\ref{sec:two-loops}, we determine the missing information about the normalization factor in the hexagon framework at strong coupling and derive a two-loop formula for structure constants of operators of any length. Section~\ref{sec:conclusion} contains our concluding remarks. Additional details are provided in the appendices.

\section{Small spin limit of hexagons}\label{sec:hexagons}
In this paper, we study the structure constants of single-trace operators in the $sl(2)$ sector, see eq.~\eqref{eq:intro}. In this set-up, two of the operators are protected, with dimension $\Delta_i = J_i$ for all values of $g$. The third operator takes the schematic form
\beq\label{eq:excited}
\textrm{Tr}\, D^{S} Z^{J} +\ldots\, ,
\eeq
where $Z$ is a complex scalar field and $D$ a light-cone derivative. The ellipsis represents mixing with other operators of the same spin $S$ and R-charge $J$ but with different derivative distributions among the scalar fields. The precise form of this operator is not needed here, as we use integrability to characterize the corresponding eigenstate. What matters is that, for any value of $J$, we focus on the operator with the lowest scaling dimension $\Delta$ at given spin $S$. In string theory, these operators map to states on the leading Regge trajectory. They are non-degenerate, exist for all even spin $S$, and have been extensively studied in the past. In particular, their scaling dimensions $\Delta = \Delta_{J}(S)$ have been analyzed in detail at both weak and strong coupling.

\begin{figure}[h]
\centering
\includegraphics[width=6cm]{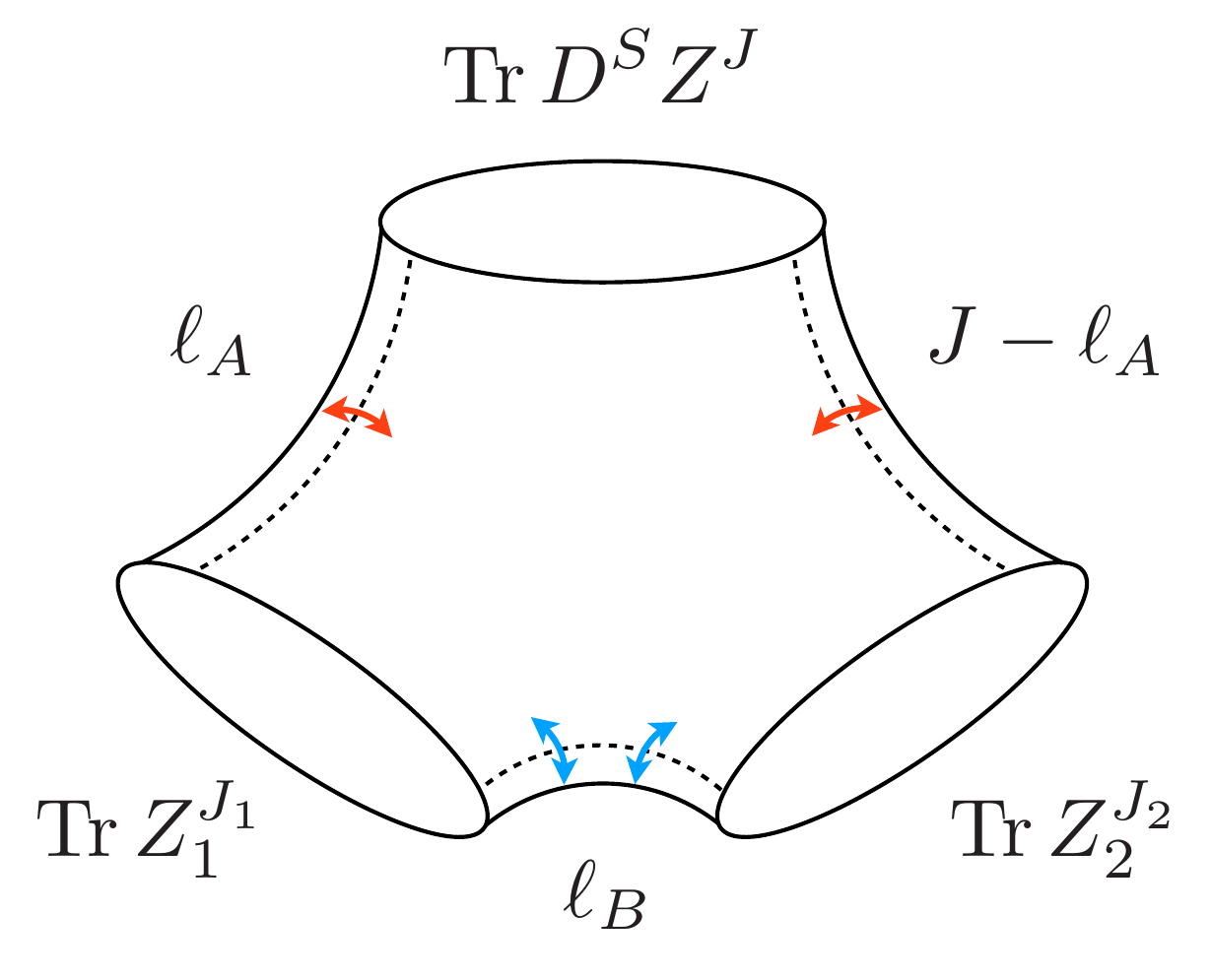}
\caption{Pair-of-pants diagram representing the structure constant in the planar limit. The diagram is decomposed into two hexagons stitched together across three mirror bridges (dashed lines) of lengths $\ell_{A}, \ell_{B}$ and $J-\ell_{A}$, respectively.
The $\B$ factor in eq.~\eqref{eq:NAB} arises from the sum over mirror magnons (blue arrows) in the bottom bridge, while the $\A$ factor comes from mirror magnons (red arrows) on the two side channels of the spinning operator.}\label{fig:Pair-of-Pants}
\end{figure}

In the integrability framework, the fusion of operators in a three-point function is described by gluing two hexagons along the seams of a pair-of-pants diagram~\cite{Basso:2015zoa}, as illustrated in figure~\ref{fig:Pair-of-Pants}. In this representation, each operator corresponds to a string whose length is determined by its R-charge, while the derivatives on the third operator correspond to string excitations, or magnons. The hexagons are stitched together by summing over a complete basis of mirror magnons along each seam, known as mirror bridges.

Structure constants derived using this approach were first formulated in~\cite{Basso:2015zoa,Basso:2015eqa}. However, this initial formulation did not account for wrapping corrections. A method to incorporate them was later introduced in ref.~\cite{Basso:2022nny}, building on the analysis of leading wrapping corrections~\cite{Basso:2017muf,Basso:2018cvy}. Below, we examine this formula around $S = 0$ after analytically continuing in the spin.

\subsection{Hexagon formula}

The proposal in~\cite{Basso:2022nny} expresses the structure constant as a product of three factors, each capturing distinct magnonic processes on the pair-of-pants. It takes the form
\beq\label{eq:NAB}
\frac{C_{123}}{C_{123}^{(0)}} = \N \A\, \B\, ,
\eeq
where the structure constant is normalized by its value at spin $S=0$, i.e., for three half-BPS operators, $C_{123}^{(0)}= \sqrt{J_{1}J_{2}J}/N$, with $N\gg 1$ representing the number of colors. Under this normalization, the left-hand side approaches 1 as $S\rightarrow 0$, as does each factor in the equation.

The first factor, $\N$, accounts for the normalization of the excited state in the integrability framework and is expressed in terms of determinants associated with the (Thermodynamic) Bethe Ansatz equations for this operator. Its analysis at finite coupling is beyond the scope of this paper; however, we will examine it at strong coupling in Section~\ref{sec:two-loops}. For now, we note that this factor depends only on the quantum numbers of the excited operator, $\N = \N(J, S)$, and not on those of the two chiral primary operators.

The remaining two factors in~\eqref{eq:NAB} depend on the three operators on the pair-of-pants diagram. They are referred to as the adjacent $(\A)$ and bottom ($\B$) contributions, and they depend not only on $S$ and $J$ but also on the bridge lengths
\beq
\ell_{A} = \frac{|J_1 -J_2| +J}{2}\, , \qquad \ell_{B} = \frac{J_1 +J_2 - J}{2}\, ,
\eeq
as illustrated in figure~\ref{fig:Pair-of-Pants}. Put differently, $\A$ governs the dependence of the structure constant on the difference $|J_1-J_2|$, while $\B$ controls its dependence on the sum $J_1+J_2$. A remarkable prediction of the hexagon proposal is that these two dependencies factorize~\cite{Basso:2015eqa}. Without loss of generality, we may assume $J_1 \geqslant J_2$, allowing us to remove the absolute value in $\ell_{A}$; we then have $\A(\ell_{A}) = \A(J-\ell_{A})$ due to the $J_{1}\leftrightarrow J_2$ symmetry.

In the hexagon construction, both $\A$ and $\B$ are expressed as infinite sums over exchanged mirror magnons, where each mirror magnon carries a weight determined by a transfer matrix that characterizes the excited operator. Importantly, for states on the leading trajectory, these transfer matrices can be easily continued beyond integer spins using the QSC formalism. Moreover, they vanish at small spin, meaning that the exchange of $M$ mirror magnons contributes a factor of $S^M$ as $S\rightarrow 0$. Consequently, for the term linear in $S$, we can truncate the infinite sums to their leading nontrivial contributions, which correspond to the one-magnon exchange.

To illustrate this, let us consider the $\B$ factor in more detail. It takes the form~\cite{Basso:2012ex}
\beq\label{eq:Bsum}
\B = 1 + \sum_{a\, =\, 1}^{\infty} \int \frac{du}{2\pi} e^{-\frac{1}{2}(J_{1}+J_{2}) \Em_{a}(u)} \mum_{a}(u) \, \bt_{a,1}(u) + \ldots\, ,
\eeq
where the dots represent integrals involving $M\geqslant 2$ mirror magnons. The `$1$' corresponds to the vacuum contribution, while the next term sums over a complete basis of single mirror-magnon states, labeled by spin $a$ and rapidity $u$. The integration measure $\mum$ and the mirror energy $\Em$ are given by
\beq\label{eq:mum}
\mum_{a}(u) = \frac{a}{g^2 \left(x^{[+a]}x^{[-a]}\right)^2}\prod_{\sigma_{1}, \sigma_{2}\, =\, \pm} \left(1-\frac{1}{x^{[\sigma_1 a]}x^{[\sigma_2 a]}}\right)^{-1}\, ,
\eeq
and
\beq\label{eq:Em}
\Em_{a}(u) = \log{\left(x^{[+a]} x^{[-a]}\right)}\, ,
\eeq
respectively, where $x^{[\pm a]} = x(u\pm i a/2)$, with $x(u)$ the Zhukovsky variable,
\beq\label{eq:x-u}
x(u) = \frac{u+\sqrt{u^2-4g^2}}{2g}\, .
\eeq
The key objects are the functions $\bt_{a,1}(u)$, which capture the full spectral data of the excited state. They are (the eigenvalues of) the L-hook transfer matrices, whose general expressions in terms of QSC functions were derived in ref.~\cite{Kazakov:2015efa}. Their continuation at small spin follows directly from the small spin expansion of the QSC solution constructed in ref.~\cite{Gromov:2014bva}; see also~\cite{Basso:2012ex,Gromov:2012eg} for earlier results in the Bethe ansatz framework.

Using the general formulas in~\cite{Kazakov:2015efa} and expanding at small $S$ with the results of~\cite{Gromov:2014bva}, one finds%
\footnote{For $a=1$ the formula is exact for any spin $S$, without any $\Op\left(S^2\right)$ corrections. For higher $a$, the corresponding corrections involve terms that are quadratic or higher in the $\bP$-functions~\eqref{eq:Ps}.}
\beq\label{eq:b-toP}
\bt^{\textrm{phys}}_{a,1}(u) = - \sum_{b\, =\, 1}^{4} \bP_{b}^{[+a]}(u) \bP^{b\, [-a]}(u) + \Op\left(S^2\right)\, ,
\eeq
where $\bP_{b}$ and $\bP^{b}$ are the $\bP$-functions of the QSC formalism, and $\bP^{[\pm a]}(u) = \bP(u\pm i a/2)$. The superscript `phys' indicates that the expression holds on the physical sheet (i.e., spin-chain kinematics). At small spin, the $\bP$-functions are suppressed and exactly known. Their form depends on whether $J$ is even or odd. For simplicity, we assume $J$ is even in what follows. The leading-order solution is then given by~\cite{Gromov:2014bva}
\beq\label{eq:Ps}
\begin{aligned}
&\bP_{1} = \bP^{4} = \epsilon x^{-J/2}\, , \qquad \bP_{2} = -\bP^{3} = -\epsilon x^{J/2} \sum_{n\, =\, J/2+1}^{\infty} I_{2n-1} x^{1-2n}\, , \\
&\bP_{3} = \bP^{2} = \epsilon\left(x^{-J/2}-x^{J/2}\right)\, ,\\
&\bP_{4} = -\bP^{1} = \epsilon x^{J/2}\sum_{n\, =\, J/2+1}^{\infty} I_{2n-1} x^{1-2n} -\epsilon x^{-J/2}\sum_{n\, = \, 1-J/2}^{\infty} I_{2n-1} x^{2n-1}\, ,
\end{aligned}
\eeq
with
\beq
\epsilon^2 =  \frac{2\pi i S}{J I_{J}(4\pi g)}\, .
\eeq
Here, $I_{k} = I_{k}(4\pi g)$ is the modified Bessel function of the 1st kind.
Up to a proportionality factor, the coefficients in the infinite sums above match the conserved charges of the excited state at small spin, $\epsilon^2\rightarrow 0$. In particular, the leading coefficient in $\bP_{2}$ is related to the operator's anomalous dimension~\cite{Basso:2011rs,Gromov:2012eg,Basso:2012ex}
\beq\label{eq:slope}
\gamma = \Delta-S-J = \gamma^{(1)}_{J} \, S + \Op\left(S^2\right)\, , \qquad \gamma^{(1)}_{J} = \frac{4\pi gI_{J+1}(4\pi g)}{J I_{J}(4\pi g)}\, .
\eeq
The next term in the small-spin expansion of the anomalous dimension, known as the curvature function, was studied in ref.~\cite{Gromov:2014bva} using the second-order solution to the QSC equations and can also be determined exactly.
 
To apply formula~\eqref{eq:b-toP} for $\bt_{a,1}$ in the integral~\eqref{eq:Bsum}, we must analytically continue the $\bP$-functions to the mirror kinematics by crossing the Zhukovsky cuts. Either of the two cuts at $\textrm{Im}\, (u) = \pm a/2$ in eqs.~\eqref{eq:b-toP} and~\eqref{eq:Ps} can be crossed, as both paths yield the same result.%
\footnote{This is a general property for $\bt$ transfer matrices, valid at any spin for any left-right symmetric state, $\bt_{a,1}(1/x^{[+a]}, 1/x^{[-a]}) = \bt_{a,1}(x^{[+a]}, x^{[-a]})$.} Choosing, for instance, $x^{[-a]} \rightarrow 1/x^{[-a]}$, we obtain
\beq\label{eq:t-mirror}
\bt_{a,1}(u) = - \sum_{b\, =\, 1}^{4} \bP_{b}^{[+a]}(u) \widetilde{\bP}^{b\, [-a]}(u) + \Op\left(S^{2}\right)\, ,
\eeq
where the transformed $\widetilde{\bP}$-functions are obtained by applying the mapping $x\rightarrow 1/x$ to eq.~\eqref{eq:Ps}.

Generally, under such an analytic continuation, we would expect the $\bt$-functions to develop infinitely many cuts on the mirror sheet, as typically occurs at finite spin (see ref.~\cite{Gromov:2013pga}). However, at small spin, this does not happen; the two-cut structure remains unchanged on every sheet.

A significant change does occur at large rapidity, though, where the transfer matrix exhibits an exponentially large behavior as $u \rightarrow \infty$,
\beq
\bt_{a,1}(u) \sim e^{2\pi u}\, .
\eeq
This behavior poses a problem, as $\bt_{a,1}(u)$ is integrated over the entire rapidity line. Similar integration difficulties at large $u$ have appeared in previous studies of the $\B$ contribution at finite spin~\cite{Basso:2015zoa,Basso:2015eqa}. In those cases, however, the large-rapidity behavior was power-like, making the analysis more tractable.%
\footnote{The modification of the large-$u$ asymptotics and the emergence of exponential behavior are linked to the continuation to non-integer spins~\cite{Basso:2012ex,Gromov:2014bva,Alfimov:2014bwa,Brizio:2024nso}.} A general solution for any operator remains unknown.

In our case, the most natural approach is to assume that the integration contour in~\eqref{eq:Bsum} can be deformed to encircle a cut of the transfer matrix. This prescription ensures the integral remains finite. Moreover, while our problem involves two distinct cuts, both yield the same result. For concreteness, we will work with a contour that runs clockwise around the cut at $\textrm{Im}\, (u) = a/2$ in the following.

The analysis for $\A$ follows a similar approach but is more technically involved. In this case, multiple integrals contribute to the single-magnon exchange. They are given by~\cite{Basso:2022nny}
\beq\label{eq:A-hex-int}
\begin{aligned}
\A = 1 + \sum_{a\, \geqslant\, 1} &\left(\oint_{\textrm{roots}} + \int\right) \frac{du}{2\pi} e^{-\frac{1}{2}(J_{1}-J_{2}) \Em_{a}(u)}\mum_{a}(u) \frac{\bT_{a, 1}(u)}{\bT_{a, 0}^{+}(u)} \\
& \qquad \,\,\,\,\,\,\, + \sum_{a\,\geqslant\, 1}\int \frac{du}{2\pi} e^{-\frac{1}{2}(J_{2}-J_{1}) \Em_{a}(u)}\mum_{a}(u)  \frac{\bT_{a, 1}(u)}{\bT_{a, 0}^{-}(u)} + \ldots\, .
\end{aligned}
\eeq
The dots represent contributions from multi-magnon processes, which are further suppressed as $S\rightarrow 0$. The functions $\bT_{a, 1}$ and $\bT_{a, 0}$ are the T-hook transfer matrices constructed in~\cite{Gromov:2013pga}, with $\bT_{a, 0}^{\pm}(u) = \bT_{a, 0}(u\pm i/2)$. We obtain two integrals over $u\in \mathbb{R}$ because a mirror magnon can propagate either to the left or to the right of the excited operator on the pair-of-pants diagram (see fig.~\ref{fig:Pair-of-Pants}).

An additional contour integral arises from the so-called asymptotic contribution to the structure constant~\cite{Basso:2015zoa,Basso:2022nny}. This term receives contributions from the zeros of $\bT^{+\, \textrm{phys}}_{a, 0}$ on the physical sheet for $a=1$, corresponding to the Bethe roots of the excited operator. In other words, to evaluate this term, the integrand must be analytically continued to the physical sheet by crossing the cut at $\textrm{Im}\, (u) = -a/2$ when $a=1$. (For $a\neq 1$, the contour integral vanishes.)

Remarkably, the three contributions in eq.~\eqref{eq:A-hex-int} can be combined by deforming the integration contours, allowing the expression for $\A$ to be brought to the same form as that for $\B$.

\begin{figure}[h]
\centering
\includegraphics[width=12cm]{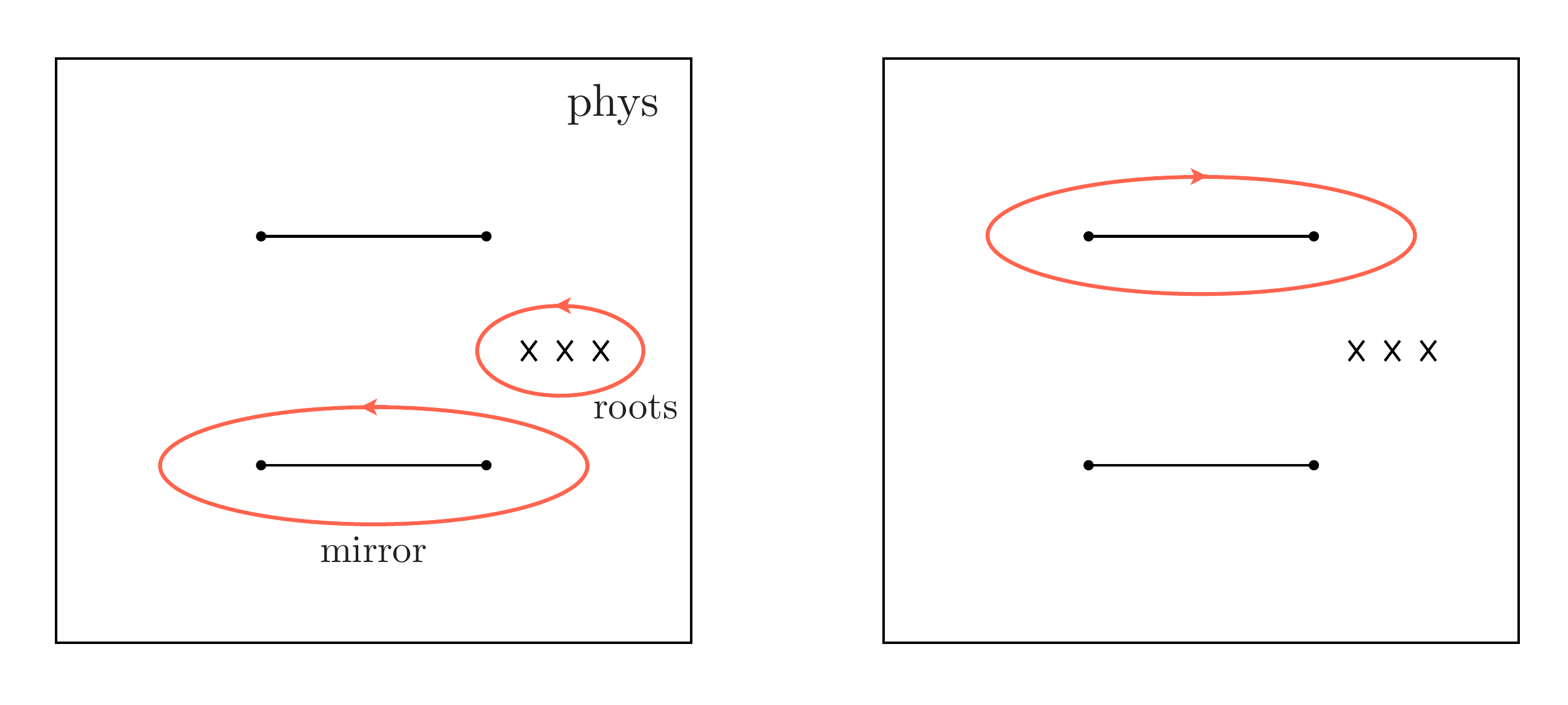}
\caption{Physical sheet of the first integrand in eq.~\eqref{eq:A-hex-int}, showing the two Zhukovsky cuts. The Bethe roots are represented by crosses. Left panel: The integration contour consists of two components, one encircling the cut in the lower half-plane and another surrounding the Bethe roots. Right panel: By flipping the contour, we obtain an equivalent integral around the cut in the upper half-plane.}\label{fig:sheets}
\end{figure}

The starting point is that, at small spin, the $\bT$-functions, like the $\bt$-functions, have only two branch cuts, corresponding to the Zhukovsky variables $x^{[\pm a]}$. By deforming the contour in the second term in~\eqref{eq:A-hex-int}, we convert it into an integral around the cut at $\textrm{Im}\, (u) = -a/2$. Continuing this contour through the cut brings the integral onto the physical sheet, albeit with reversed orientation. As a result, the first two integrals in~\eqref{eq:A-hex-int} merge into a single contour encircling both the cut and the Bethe roots, as illustrated in fig.~\ref{fig:sheets}. Finally, flipping this contour across infinity maps it to an integral around the cut at $\textrm{Im}\, (u) = a/2$, which can then be crossed to the mirror sheet on the other side of the cut. This operation removes the contribution from the Bethe roots, leaving a simplified mirror-like integral with an analytically continued integrand.%
\footnote{The analysis is heuristic for $a>1$ since the transfer matrices may have additional singularities on the physical sheet, which are not captured by the original contour integral. However, we expect such spurious contributions to follow from the fusion of elementary transfer matrices with $a=1$, hence producing further suppressed corrections at small spin.}

By following these lines, the integrals in eq.~\eqref{eq:A-hex-int} are replaced by
\beq\label{eq:A-hex-int-bis}
\begin{aligned}
\sum_{a\, \geqslant\, 1} &\oint \frac{du}{2\pi} \left[e^{-\frac{1}{2}(J_{1}-J_{2}) \Em_{a}(u)}\mum_{a}(u) \, \frac{\bT_{a, 1}(u)}{\bT_{a, 0}^{+}(u)}\right]^{\circlearrowleft} \\
& \qquad \,\,\,\,\,\, + \sum_{a\,\geqslant\, 1}\oint \frac{du}{2\pi} e^{-\frac{1}{2}(J_{2}-J_{1}) \Em_{a}(u)}\mum_{a}(u) \, \frac{\bT_{a, 1}(u)}{\bT_{a, 0}^{-}(u)}\, ,
\end{aligned}
\eeq
where the contour encircles the upper cut at $\textrm{Im}\, (u) = a/2$. The superscript $\circlearrowleft$ indicates analytic continuation through the two Zhukovsky cuts,
\beq
\left(x^{[\pm a]}\right)^{\circlearrowleft} = \frac{1}{x^{[\pm a]}}\, .
\eeq
Under this mapping, the mirror energy flips sign, $\Em \rightarrow -\Em$, while the measure $\mum$ remains unchanged. Moreover, at small spin, the sum of $\bT$'s in eq.~\eqref{eq:A-hex-int-bis} can be rewritten directly in terms of $\bt$'s, using%
\footnote{The equality is exact for any spin when $a=1$.}
\beq
\bt_{a, 1}(u) = \left[\frac{\bT_{a, 1}(u)}{\bT_{a, 0}^{+}(u)}\right]^{\circlearrowleft} + \frac{\bT_{a, 1}(u)}{\bT_{a, 0}^{-}(u)} + \Op\left(S^2\right) \, .
\eeq
This identity follows directly from the crossing properties of $\bP$-functions and the leading expressions for $\bT$'s at small spin,
\beq
\bT_{a, 1}(u) = -\left(\bP_{1}^{[+a]}\bP_{2}^{[-a]} - \bP_{2}^{[+a]}\bP_{1}^{[-a]}\right)\, , \qquad  \bT_{a, 0}(u) = 1\, ,
\eeq
both of which hold up to $\Op\left(S^2\right)$ corrections.

Finally, we reach the remarkable conclusion that both $\A$ and $\B$ can be expressed in terms of the same single-magnon integral at leading order in the small-spin expansion,
\beq\label{eq:A-to-F}
\A = 1 + S \, F_{J}(-\ell_{A}) + \Op\left(S^2\right)\, , \qquad \B = 1 + S\, F_{J}(\ell_{B}) + \Op\left(S^2\right)\, ,
\eeq
where $F_J(\ell)$ is given by
\beq\label{eq:F-l}
F_{J}(\ell) = \sum_{a\, =\, 1}^{\infty} \oint \frac{du}{2\pi} e^{-(\ell+\frac{1}{2}J)\Em_{a}(u)} \mum_{a}(u)\,  t_{a}(u)\, ,
\eeq
with the contour running clockwise around the Zhukovsky cut in the upper half-plane. The measure and energy are defined in eqs.~\eqref{eq:mum} and~\eqref{eq:Em}, and $t_{a}$ denotes the leading small-spin contribution to the transfer matrix, given by $\bt_{a,1} = S\, t_{a} + \Op\left(S^2\right)$. The only distinction between $\A$ and $\B$ lies in the argument of the function $F_{J}(\ell)$: for $\B$, it corresponds to the bottom bridge length $\ell_{B}$, while for $\A$, it is given by the negative of the adjacent bridge length, $-\ell_{A}$. 

A few remarks are in order.

First, the minus sign in front of $\ell_A$ in~\eqref{eq:A-to-F} is not entirely trivial, as the sum in~\eqref{eq:F-l} diverges logarithmically for $\ell < 0$, with a divergence controlled by the anomalous dimension $\sim \sum_{a} (1+\gamma^{(1)}_{J})/a$. More precisely, the sum diverges for $\ell\in \{1-J, \ldots, -1\}$, where the lower bound follows from the symmetry of $F$, as discussed below (see eq.~\eqref{eq:SymmAd}).%
\footnote{A divergence also arises for $\ell = 0$ and $\ell = -J$, which correspond to extremal structure constants. These cases will not be considered here.}

This divergence is not present in the original expression for $\A$ and arises as an artifact of the contour deformation. Specifically, it appears when shifting the integration contour from the left to the right panel in fig.~\ref{fig:sheets}, effectively passing through $u=\infty$ on the physical sheet. Fortunately, the issue is easily addressed by subtracting the spurious pole $\sim 1/a$ from the summand. To confirm that this procedure preserves the correct finite part, we compared our result with a direct analytic continuation at small spin of known finite-spin expressions for $\A$. This comparison can be carried out at weak coupling for $J = 2\ell_{A} = 2$, using closed-form results expressed in terms of nested harmonic sums (see e.g.~\cite{Eden:2012rr}), as well as in the large $J$-limit.
A detailed discussion is provided in Appendices~\ref{App:PerturbTest} and~\ref{app:largeJ}.

Second, the function $F_{J}(-\ell_{A})$ is symmetric under the exchange of $J_{1}$ and $J_{2}$, i.e., under $\ell_{A} \leftrightarrow J-\ell_{A}$. This symmetry was manifest in the original hexagon formula and implies that $F_{J}(\ell)$ satisfies
\beq\label{eq:SymmAd}
F_{J}(\ell) = F_{J}(-\ell-J)\, .
\eeq
This identity can be derived by contour manipulations and using the crossing properties of the transfer matrices. It can also be verified through direct computation, as shown in eq.~\eqref{eq:F-to-f} below.

\subsection{Weak coupling series}

As we have shown, the integrand for the adjacent and bottom bridge can be constructed using the same QSC transfer matrix, encoded in the function $F_{J}(\ell)$ in eq.~\eqref{eq:F-l}. To perform the summation and integration more efficiently in this formula, it is convenient to change variables from $(u, a)$ to $(x, y)$, using
\beq
x =  x^{[+a]}\, , \qquad y = x^{[-a]}\, .
\eeq
It is then possible to treat these two variables as independent continuous variables by following the approach in ref.~\cite{Basso:2021omx}. This method allows us to express the sum over $a$ as an integral over an auxiliary variable $t$. Following this procedure, we find that $F_J$ can be written as
\beq \label{eq:integrandB}
F_{J}(\ell)= \int_0^\infty \frac{dt~e^t}{(1-e^t)^2} \oint \frac{dx dy}{(2\pi)^2}  \frac{e^{i(u-v)t}\, t_J(x,y)}{(x  y)^{\ell+J/2}(x y-1)^2}\, ,
\eeq
where $u = g(x+1/x)$ and $v = g(y+1/y)$. The contour in $x$ runs counterclockwise around the unit circle $|x|=1$, and similarly for $y$.

To avoid cluttering the formula, we have omitted the prescription required to handle the aforementioned logarithmic divergences. In the new variables, this behavior manifests as a $1/t$ divergence of the integrand at small $t$. It can be regulated by replacing the exponential in~\eqref{eq:integrandB} with
\beq
e^{i(u-v)t} \rightarrow e^{i(u-v)t}-1-i(u-v)t\, .
\eeq
This modification is necessary to ensure the integral is well-defined when $1-J\leqslant \ell \leqslant -1$. For the bottom channel, where $\ell$ falls outside this range, the subtraction is not strictly required.

Finally, $t_{J}(x, y)$ follows directly from the expression for the transfer matrix $\bt_{a, 1}$, after factoring out $S$, see eqs.~\eqref{eq:t-mirror} and~\eqref{eq:b-toP}. For a state of even twist $J$, it is given by
\begin{align}\label{eq:btJ}
t_{J}(x,y)= \frac{i}{(xy)^{J/2}} & \bigg[ \sum_{n\, =\, 1}^{J/2} \bI{2n-1} \left( x^{J}\left(x^{1-2n}-y^{2n-1} \right)+y^{J}\left(x^{2n-1}-y^{1-2n} \right)\right) \nonumber \\ &  + \sum_{n\, =\, 1}^{\infty}\bI{2n-1} (x^{1-2n}-y^{2n-1}) (y^J-x^J)\bigg]\, ,
\end{align}
where the coefficient $\bI{n}$ denotes the ratio of Bessel $I$ functions
\beq
\bI{n} = \frac{2\pi I_{n}(4\pi g)}{J I_{J}(4\pi g)}\, .
\eeq

The integrals in~\eqref{eq:integrandB} can be readily evaluated at weak coupling, $g^2\rightarrow 0$, for any values of $\ell$ or $J$. The first step is to expand the integrand to the desired loop order, truncating the infinite sum in~\eqref{eq:btJ} according to $\bI{n} = \Op(g^{n-J})$. The two contour integrals can then be performed straightforwardly, followed by the integration over $t$, using
\beq\label{eq:ttozeta}
\int_0^\infty \frac{dt~e^t}{(1-e^t)^2}~ t^n = \Gamma(n+1)\z{n}\, , \quad \text{for} \quad n >1 \, ,
\eeq
where $\zeta_n = \zeta(n) = \sum_{k\geqslant 1}1/k^n$ is the Riemann zeta function.

As an illustration, let us compute $F_{J}(\ell)$ for small values of $J$ and $\ell$. When $J=2$, the only possible bridge length for $\A$ is $\ell_A=1$. In this case, we find
\begin{equation}\label{eq:Adjacent-weak-coupling}
\begin{aligned}
     F_{J\, =\, 2}(-1) =&-8g^2 \z{3} + g^4(-32 \z{2} \z{3} + 90 \z{5}) + g^6(160 \z{3} \z{4} + 288 \z{2} \z{5} - 1120 \z{7}) \\ & +g^8(-1440 \z{4} \z{5} - 896 \z{3} \z{6} - 3360 \z{2} \z{7} + 14700 \z{9})+\Op\left (g^{10}\right)\, .
\end{aligned}
\end{equation}
For the $\B$ factor, $\ell_B$ can take any positive integer value. The first three cases, corresponding to $\ell_{B} = 1,2,3$, are detailed in Table~\ref{tab:C123BJ2} up to four loops. In Appendix \ref{App:PerturbTest} we verify that the expressions in eq.~\eqref{eq:Adjacent-weak-coupling} and in Table~\ref{tab:C123BJ2} agree with the results obtained via direct analytical continuation of the finite-spin formulae.

We can also derive expressions for operators of lengths $J>2$, as shown in Table~\ref{tab:C123BJ4} for $J=4$. In this case, inverse zeta values appear, similar to what was found for the curvature function~\cite{Gromov:2014bva}. Testing these higher-length results is more challenging, as no known method currently exists to interpolate the finite-spin data.

\renewcommand{\arraystretch}{1.25}
\begin{table}[!t]
\centering
\begin{tabular}[t]{c|l}
$\ell_{B}$ & $F_{J}(\ell_{B})$ for $J=2$ \\
\hline\hline
     1 & $3g^4(4 \z{2} \z{3} + 5 \z{5}) -48 g^6(\z{3} \z{4} + \z{2} \z{5}+7 \z{7}) + 4g^8(15 \z{4} \z{5} + 63 \z{3} \z{6} - 56 \z{2} \z{7}+ 1470 \z{9})$ \\
     2 & $4g^6(-3 \z{3} \z{4} + 9 \z{2} \z{5} + 14 \z{7}) + 84 g^8(\z{3} \z{6}- 4 \z{2} \z{7} - 20 \z{9})$ \\
     3 & $2g^8(-30 \z{4} \z{5} + 56 \z{2} \z{7} + 105 \z{9})$ \\
\end{tabular}
\caption{Four-loop results for $F_{J}(\ell_{B})$ for $J=2$ and bridge lengths $\ell_{B}=1,2,3$.}
\label{tab:C123BJ2}
\end{table}
\renewcommand{\arraystretch}{1.25}
\begin{table}[!t]
\centering
\begin{tabular}[t]{c|l}
     $\ell_{B}$ & $F_{J}(\ell_{B})$ for $J=4$ \\
     \hline \hline
     $1$ & $\frac{g^4}{\z{2}}(9 \z{3} \z{4} + 10 \z{2}\z{5}-7 \z{7})+ \frac{g^6}{\z{2}}(-60 \z{4} \z{5} - \frac{48}{5}  \z{3} \z{6}- \frac{1152}{5} \z{2} \z{7}+240 \z{9})$ \\
     $2$ & $\frac{g^6}{\z{2}}(30 \z{4} \z{5} - 3  \z{3} \z{6}+ 36 \z{2} \z{7}-30 \z{9})$ \\
\end{tabular}
\caption{Three-loop results for $F_{J}(\ell_{B})$ with $J=4$ and bridge lengths $\ell_{B} =1,2$.}
\label{tab:C123BJ4}
\end{table}
Finally, we can obtain closed-form expressions at generic lengths. To achieve this, we first integrate by parts in $t$ in eq.~\eqref{eq:integrandB} and use the following integral
\beq
\int_{0}^{\infty} dt\, \frac{e^{i(u-v)t}-1}{e^{t}-1} = \psi(1)-\psi(1-iu+iv)\, ,
\eeq
where $\psi(z) = \partial_{z}\log{\Gamma(z)}$. After symmetrizing in $u-v$, we obtain
\beq
F_{J}(\ell) = -\frac{i g}{2}\oint \frac{dx dy}{(2\pi)^2 xy} \frac{x-y}{x y-1} t_{J}(x, y) (\psi(1+iu-iv)-\psi(1) + (u\leftrightarrow v))\, .
\eeq
In this form, the result bears a striking resemblance to the generic formula for the curvature function~\cite{Gromov:2014bva}. The appearance of odd zeta values in the weak coupling expressions follows directly from the power series expansion of the $\psi$-function,
\beq
\frac{1}{2}(\psi(1+iu-iv)-\psi(1) + (u\leftrightarrow v)) = \sum_{k\, =\, 1}^{\infty} (-1)^{k+1} g^{2k}\zeta_{2k+1} (x-y)^{2k} (1-1/xy)^{2k}\, .
\eeq
The contour integrals in $x$ and $y$ can be evaluated exactly for any given $k$. The final result is manifestly symmetric under $\ell \rightarrow -J-\ell$ and takes the form
\beq\label{eq:F-to-f}
F_{J}(\ell) = f_{J}(\ell) + f_{J}(-J-\ell)\, ,
\eeq
where the function $f_{J}(\ell)$ is given by the convergent series
\beq\label{eq:fofell}
f_{J}(\ell) = \sum_{k\, =\, 1}^{\infty}\sum_{n\, \in \, \mathbb{Z}} \frac{(-1)^{k+\ell+1} g^{2k+1}\Gamma(2k)\Gamma(2k+2)\, \zeta_{2k+1}\, \varepsilon(n)\, \bI{2n-J-1}}{\Gamma(1+k+n)\Gamma(2+k-n)\Gamma(k+\ell+n)\Gamma(1+k-\ell-n)}\, .
\eeq
The symbol $\varepsilon(n)$ is a sign function, equal to $-1$ if $n \leqslant 0$ and $1$ otherwise. Specializing~\eqref{eq:F-to-f} and~\eqref{eq:fofell} to low values of $J$ and $\ell$, one can easily verify their agreement with the explicit expressions reported earlier.

\subsection{Strong coupling analysis}\label{sec:hexagon-sc}
Let us now examine the strong coupling limit, $g\rightarrow \infty$. To analyze this regime, it is convenient to reverse some of our previous steps and reintroduce the integral over the variable $t$ using eq.~\eqref{eq:ttozeta}. The sum over $k$ in \eqref{eq:fofell} can then be expressed in terms of Mellin-Barnes integrals for the product of Bessel $J$ functions,
\beq\label{eq:mbBessel}
J_{\mu}(2gt)J_{\nu}(2gt)=\int\limits \frac{dz}{2 \pi i} \frac{\Gamma(-z)\Gamma(2 z+\mu+\nu+1)(gt)^{2z+\mu+\nu}}{\Gamma(z+\mu+1)\Gamma(z+\nu+1)\Gamma(z+\mu+\nu+1)}\, ,
\eeq
with the contour running from $-i\infty$ to $i\infty$, with a small negative real part.
After simple algebraic manipulations, it allows us to rewrite eq.~\eqref{eq:ttozeta} in the form
\begin{align}\label{eq:fBesselF}
f_{J}(\ell)= g\sum_{n\, \in \, \mathbb{Z}} (-1)^{n+1} \varepsilon(n)\, \bI{2n-J-1} \sum_{m\, =\, 0}^{\infty} \left( c_{\ell+m,\, \ell+2n+m} + c_{\ell+m+1,\, \ell+2n+m-1}\right) \, ,
\end{align}
where the coefficient $c_{i, j}$ is given by%
\footnote{The structure of the coefficients $c_{i,j}$ is reminiscent of that found in the Tracy-Widom distribution~\cite{Bajnok:2024epf}.}
\beq\label{eq:cij}
c_{i, j} = \int_0^\infty \frac{dt}{e^t-1} \left(J_{i}(2 g t)J_{j}(2 g t)-\delta_{i,0}\,\delta_{j,0}\right)\, .
\eeq
The sum over $m$ can be further simplified using the identity 
\beq
\begin{aligned}
&\sum_{m\, =\, 0}^\infty J_{\ell+m+\delta}(2gt) J_{\ell+2n+m-\delta}(2gt) \\
&\qquad \qquad = \frac{gt}{2(n-\delta)}(J_{\ell+\delta}(2gt) J_{\ell+2n-\delta-1}(2gt)-J_{\ell+\delta-1}(2gt) J_{\ell+2n-\delta}(2gt)) \, ,
\end{aligned}
\eeq
where $\delta = 0,1$ in our case, except when $n=\delta$, where one can use
\beq
\sum_{m\, = \, 0}^{\infty} J_{\ell+m+\delta}(2gt) J_{\ell+m+\delta}(2gt) = \frac{1}{2}\left(1-\varepsilon\,(\ell+\delta)\sum_{m \, =\, \frac{1}{2}-|\ell+\delta-\frac{1}{2}|}^{|\ell+\delta-\frac{1}{2}|-\frac{1}{2}}J_{m}(2gt) J_{m}(2gt)\right)\, .
\eeq
At leading order in the strong coupling limit, the integrals over the Bessel functions and the sum over $n$ can be evaluated straightforwardly. It yields
\beq
f_{J}(\ell) = \frac{2\pi g}{J}\, \left[ \delta_{\ell\, >\, 0} \left(\log{g}-\psi(\ell)\right) - \delta_{\ell\, <\, 0} \left(\log{g}-\psi(-\ell)\right) \right] +\Op(1)\, ,
\eeq
for $\ell\, \neq\, 0$, with $\delta_{\ell \, >\, 0} = 1$ if $\ell>0$ and $0$ otherwise, and similarly for $\delta_{\ell\, < \, 0}$.

The analysis becomes more involved for the $1/g$-suppressed corrections. This difficulty arises because the sum over in $n$ in~\eqref{eq:fBesselF} fails to converge at higher orders; instead, the summand grows as $(n^2/g)^{k}$ at the $k$-th order. Therefore, to compute these corrections, we must first evaluate the sum over $n$ and then expand the resulting function at large $g$.

To address this issue, we implemented a numerical routine to interpolate the values of $f_J(\ell)$ for sufficiently large $g$, ranging from $2$ to $20$.  As a first step, we truncated the sum over $n$ in eq.~\eqref{eq:fBesselF},
\beq
f_J(\ell)\approx  \sum_{n\, =\, n_\text{max}^-(g)}^{n_\text{max}^+(g)}\tilde{f}_J(\ell,n) \, ,
\eeq
where the cut-offs $n_{\text{max}}^\pm(g)$ are chosen such that
\beq
\tilde{f}_J(\ell, \, n_\text{max}^\pm(g)) \leq \text{precision} \, .
\eeq
Next, we numerically evaluated the integrals over the Bessel functions for a range of $\ell$ and $J$. Finally, we fitted the resulting data to the series expansion
\beq
f_J(\ell) \approx \sum_{i\, =\, -1}^{i_\text{max}} g^{-i}(c_i+d_i \log{g}) \,.
\eeq
Here, $\{c_{i},d_{i}\}$ are numerical constants that can be expressed as linear combinations of transcendental numbers. A natural basis for these constants is suggested by the strong coupling expansion of $\tilde{f}_J(\ell,n)$ at a fixed $n$. In addition to the expected zeta values, the basis includes $\log{2}$ and $\log{\pi}$.

By following this approach, we find
\begin{equation}
\begin{aligned}
f_J(\ell)& = \frac{\gamma^{(1)}_{J}}{4}\left(\log{\lambda}-2\log{(4\pi)}-2\psi(\ell)\right)-\frac{1}{2 J}+\frac{1}{2}(\gamma_\text{E}-\log(8 \pi))+\frac{3+4 J}{8\sqrt{\lambda}} \\
&\,\,\,\,\, + \frac{J+2\ell}{4 J} \left(\log(8\sqrt{\lambda})+\gamma_\text{E} \right)+ \frac{J+2 \ell}{\sqrt{\lambda} J }\left(\frac{2 J^2-1}{8}+\frac{1-J^2-\ell(J+\ell)}{6}\z{2} \right) \\
&\,\,\,\,\, +\Op\left(\frac{1}{\lambda}\right)\, ,
\end{aligned}
\end{equation}
for $\ell>0$, and $f_J(\ell) = f_{-J}(-\ell)$ for $\ell<0$, where $\gamma^{(1)}_{J}$ is defined in eq.~\eqref{eq:slope} and $\gamma_{\textrm{E}}$ is the Euler-Mascheroni constant. Note that we switched to $\sqrt{\lambda} = 4\pi g$ to absorb powers of $\pi$ in the denominators. Note also that the second line on the right-hand side is odd under $\ell \rightarrow -J-\ell$ and thus vanishes for physically relevant quantities such as $F_J(\ell)$ in \eqref{eq:F-to-f}. At higher orders, this zero-mode contribution transforms into a higher-degree rational function, making interpolation increasingly difficult. Therefore, for higher-order expansions, it is more practical to work directly with $F_J(\ell)$.

By applying the numerical routine to $F_J(\ell)$ directly, we could extract two additional coefficients in the strong coupling expansion. The final result can be written as
\beq\label{eq:FpsiP}
F_J(\ell)=\Psi_{J}(\ell)+P_{J}(\ell)\, ,
\eeq
where
\begin{align}
\Psi_{J}(\ell) = &\, \frac{\gamma^{(1)}_J}{2}\left( \psi(J+\ell)-\psi(\ell)\right)+\psi(J+\ell)+C\, , \\
\Psi_{J}(-\ell)= &\,  \frac{\gamma^{(1)}_J}{2}\left( \psi(J-\ell)+\psi(\ell)-\log{\lambda}+2\log{(4\pi)}\right)+\psi(J-\ell)+\psi(\ell)+C\, , 
\end{align}
for $\ell>0$ and $J>\ell>0$, respectively, and $C = \gamma_{\text{E}}-\log{(2\sqrt{\lambda})}$. Here,
\beq\label{eq:PJell}
\begin{aligned}
P_{J}(\ell) &= \frac{3}{4\sqrt{\lambda}}+\frac{1}{\lambda}\left(\frac{17-8 J^2}{32}+ \frac{11 - 8 J^2 - 24 \ell(J+\ell)}{48}\z{3}\right)\\
&\,\,\,\, + \frac{1}{\lambda^{3/2}}\left(\frac{11-9 J^2}{16}+\frac{11 - 8 J^2 - 18 \ell(J+\ell)}{48}\z{3}\right)\\
&\,\,\,\, + \delta_{-J\, <\, \ell\, <\, 0}\left( \frac{1}{J} - \frac{J}{\sqrt{\lambda}}-\frac{J}{2\lambda}-\frac{13 J-4 J^3}{24 \lambda^{3/2}}\right)\\
&\,\,\, + \Op\left(\frac{1}{\lambda^2}\right)\, ,
\end{aligned}
\eeq
for $\ell\neq 0, -J$, where $\delta_{-J\, <\, \ell\, <\, 0}$ is 1 for $-J<\ell<0$ and $0$ otherwise.

In this result, the most intricate dependence on the lengths is encoded in $\Psi_{J}(\ell)$, and is essentially governed by the anomalous dimension of the operator. The remaining term, $P_{J}(\ell)$, is polynomial in $\ell$ and $J$, except for a simple pole $\sim 1/J$ when $0>\ell>-J$. This structure is consistent with earlier results for structure constants at strong coupling~\cite{Basso:2019diw}, and aligns with a natural factorization formula in string theory, which we will discuss shortly.

These observations motivate a redefinition of the components $\A$ and $\B$ in the hexagon formula, valid for arbitrary spin,
\begin{equation}
\begin{aligned}\label{eq:DABfirst}
\B= &\, \left(\frac{e^{\gamma_{\textrm{E}}S}}{2^{S}\lambda^{S/2}}\right)\frac{\Gamma(\ell_B-\gamma/2)\Gamma(J+\ell_B+S+\gamma/2)}{\Gamma(\ell_B)\Gamma(J+\ell_B)}\times \mathcal{D}_{\B} \, , \\
\A=&\, \left(\frac{e^{\gamma_{\textrm{E}}S}(4\pi)^{S+\gamma}}{2^{S}\lambda^{S+\gamma/2}}\right)\frac{\Gamma(\ell_A+\gamma/2+S)\Gamma(J-\ell_A+S+\gamma/2)}{\Gamma(\ell_A)\Gamma(J-\ell_A)}\times \mathcal{D}_{\A} \, ,
\end{aligned}
\end{equation}
where the ratios of Gamma functions and overall power factors are introduced to cancel the $\psi$-functions and logarithmic terms associated with the spin or anomalous dimension in eq.~\eqref{eq:FpsiP} in the small-spin limit. The remaining factors, $\D_{\A}$ and $\D_{\B}$, then admit a small-spin expansion with simpler coefficients,
\begin{equation}\label{eq:DABsecond}
\log{\D_{\B}} = S\, P_{J}(\ell_{B}) + \Op\left(S^2\right)\, , \qquad \log{\D_{\A}} = S\, P_{J}(-\ell_{A}) + \Op\left(S^2\right)\, ,
\end{equation}
where $P_{J}(\ell)$ is given in eq.~\eqref{eq:PJell}.

\section{Structure constants at strong coupling}\label{sec:ansatz}

In this section, we introduce a factorization formula for structure constants of operators with arbitrary spin at strong coupling, connecting the small-spin analysis with recent findings on structure constants of short operators in string theory.

\subsection{Factorization and regularity assumptions}

To explore the simplifications that arise at strong coupling and finite spin, we adopt the following ansatz
\beq\label{eq:main-ansatz}
\frac{C_{123}}{C^{(0)}_{123}} = \frac{\Gamma\left[AdS\right]}{\Gamma[Sphere]} \times \frac{\D_{J_{1}J_{2}J}(S)}{\lambda^{S/4}\Gamma\left(1+\frac{S}{2}\right)} \, ,
\eeq
where the prefactor is expressed in terms of Gamma functions, with arguments given as linear combinations of the global charges of the operators. Specifically, the AdS contribution is given by
\beq\label{eq:G-factor}
\Gamma[AdS] = \frac{\Gamma\left(\tfrac{\Delta_{1}-\Delta_{2}+\Delta+S}{2}\right)\Gamma\left(\tfrac{\Delta_{2}-\Delta_{1}+\Delta+S}{2}\right)\Gamma\left(\tfrac{\Delta_{1}+\Delta_{2}-\Delta+S}{2}\right)\Gamma\left(\tfrac{\Delta_{1}+\Delta_{2}+\Delta+S}{2}\right)}{\sqrt{\Gamma(\Delta+S)\Gamma(\Delta+S-1)}}\, ,
\eeq
while the sphere contribution, $\Gamma[Sphere]$, is obtained by replacing the scaling dimension $\Delta$ by the R-charge $J$ and setting the spin $S$ to zero,
\beq
\Gamma[Sphere] = \Gamma[AdS]_{\Delta \rightarrow J, S\rightarrow 0}\, .
\eeq
It follows that the prefactor is normalized to $1$ when $S=0$, since $\Delta(S=0) = J$ and $\Delta_{1,2} = J_{1,2}$ for the half-BPS operators.

Formula~\eqref{eq:G-factor} parallels the structure of the ansatz proposed in refs.~\cite{Bargheer:2013faa,Minahan:2014usa} to relate structure constants at strong coupling to flat-space string amplitudes. In this framework, the prefactor in~\eqref{eq:G-factor} captures the global contribution to the three-point function, arising from the cubic Witten diagram in $AdS_5\times S^5$, while the remaining factor encodes local stringy corrections. The formula is constructed so that the Gamma functions in the prefactor align with those observed at small spin in the previous section (see eq.~\eqref{eq:DABfirst}). Furthermore, at leading order in the strong coupling limit, $\D\rightarrow 1$, and the second factor in eq.~\eqref{eq:G-factor} is designed to reproduce the three-point coupling in flat-space string theory, see ref.~\cite{Alday:2023flc} for a recent discussion.

The central quantity in eq.~\eqref{eq:main-ansatz} is the function $\D = \D_{J_{1}J_{2}J}(S)$, which we refer to as the reduced structure constant. Our key working assumption is that $\log{\D}$ admits a regular expansion at small spin, valid up to the semiclassical regime $S\leqslant \sqrt{\lambda}$. More precisely, we assume
\beq\label{eq:logDS}
\log{\D} = \D_{1}\, S + \frac{\D_{2}}{\sqrt{\lambda}}\, S^2 + \frac{\D_{3}}{\lambda}\, S^3 + \Op\left(\frac{S^{4}}{\lambda^{3/2}}\right)\, ,
\eeq
where each coefficient $\D_{n} = \D_{n}(\lambda)$ admits a strong coupling expansion of the form
\beq\label{eq:Dexpansion}
\D_{n} = \D_{n}^{(0)} + \frac{\D_{n}^{(1)}}{\sqrt{\lambda}} + \frac{\D_{n}^{(2)}}{\lambda} + \ldots\, ,
\eeq
with $\D^{(k)}_n$ independent of the coupling.

The scaling in eq.~\eqref{eq:logDS} is motivated by the requirement to recover the correct behavior in the classical limit, $S\sim \sqrt{\lambda}$. In this regime, the structure constant is expected to scale as
\beq\label{eq:Dcl}
\log{\D} = \sqrt{\lambda} \, \D^{\textrm{cl}} + \Op(1)\, ,
\eeq
in agreement with the minimal area prediction. Here, $\D^{\cl} = \sum_{n\, = \, 1}^{\infty}\D_{n}^{(0)}\S^{n}$ is a function of the classical spin $\S = S/\sqrt{\lambda}$. Our ansatz not only satisfies this requirement, but also enforces that both $\D^{\cl}$ and its semiclassical corrections, obtained by resumming the subleading terms, are smooth functions of $\S$ around $\S= 0$.

An equivalent formulation of our assumption is to express $\log{\D}$ as a strong coupling expansion in $1/\sqrt{\lambda}$, where the $(k-1)$-loop coefficient is a polynomial of degree $k$ in $S$,
\beq\label{eq:polynomial-ansatz}
\log{\D} = \sum_{k\, =\, 1}^{\infty} \frac{1}{(\sqrt{\lambda})^{k-1}} \sum_{n\, =\, 1}^{k} \D_{n}^{(k-n)}\, S^{n}\, ,
\eeq
with each term vanishing at $S=0$ for all $k$.

Equation~\eqref{eq:logDS}, or alternatively~\eqref{eq:polynomial-ansatz}, provides a basis for analyzing structure constants across a wide range of strong-coupling regimes. In particular, it extends the small-spin expansion beyond its naive domain of validity, $S \ll 1/\sqrt{\lambda}$, where the anomalous dimension $\gamma$ is small. Likewise, it generalizes the semiclassical regime, originally limited to $ 1 \ll S \ll \sqrt{\lambda}$, where $\gamma = \mathcal{O}(\lambda)$. The polynomial structure in~\eqref{eq:polynomial-ansatz} bridges these two distinct limits and grants access to intermediate regimes, including the short-string domain, $S = \Op(1)$, discussed below, and the Regge regime, $S \sim (\Delta^2-J^2)/2\sqrt{\lambda}$ with $\Delta = \Op(1)$, which is analyzed in Appendix~\ref{app:Regge}.

A similar assumption was made in the computation of scaling dimensions for short operators on the leading Regge trajectory, where it was proposed that $\Delta^2$ admits a regular expansion at small spin~\cite{Basso:2011rs,Gromov:2011bz,Beccaria:2012xm,Gromov:2014bva,Ekhammar:2024rfj}
\beq
\Delta^2 = J^2 + \sqrt{\lambda} \left(A_{1} S + \frac{A_{2}}{\sqrt{\lambda}} S^2 + \frac{A_{3}}{\lambda} S^{3} + \Op\left(\frac{S^4}{\lambda^{3/2}}\right)\right)\, ,
\eeq
with the coefficients $A_n$ matching the classical string result at strong coupling. While adopting a similar structure for the structure constants may seem a strong assumption, we will show that it is fully consistent with existing data, both from short-string computations and from classical string theory. Furthermore, the same semiclassical scaling appears to govern the dependence on the operator lengths: the coefficients $\D_{n}^{(k)}$ in eq.~\eqref{eq:Dexpansion} are observed to be polynomials of degree $k$ in $J_{1}, J_{2}$ and $J$, as already suggested by the small-spin coefficient in~\eqref{eq:PJell} for $J_{1}$ and $J_{2}$.

\subsection{Testing with short-string data}

As an initial test, we can check whether the $\D$-coefficients satisfy our assumptions at strong coupling for quantum numbers of order $\Op(1)$. Numerous results in this regime have been obtained recently using string-inspired techniques and the conformal bootstrap~\cite{Alday:2022uxp,Alday:2022xwz,Alday:2023flc,Alday:2023jdk,Alday:2023pzu,Fardelli:2023fyq,Alday:2023mvu,Alday:2024yax,Alday:2024ksp}. In particular, structure constants of two length-2 chiral primary operators and one length-2 spinning operator, corresponding to $J_1 = J_2 = J = 2$ in our notation, have been constructed up to two loops at strong coupling in~\cite{Alday:2023mvu}.

To facilitate the comparison in this case, we will use the expressions provided in ref.~\cite{Caron-Huot:2024tzr}. There, the two-loop data was simplified and reformulated in a way similar to our decomposition. The resulting quantity, which exhibits smoother properties, was denoted $\tilde{\lambda}_{\Delta, S}$ in ref.~\cite{Caron-Huot:2024tzr}, see eq.~(C.5) therein.%
\footnote{To be precise, the rescaled structure constant in ref.~\cite{Caron-Huot:2024tzr} was denoted as $\tilde{\lambda}_{\Delta, J}$, where $\Delta$ and $J$ represent the scaling dimension and spin of the length-2 superconformal primary. Our labels, however, refer to the dimension and spin of the $sl(2)$ primary, with $\Delta_{\textrm{\cite{Caron-Huot:2024tzr}}} = \Delta-2$ and $J_{\textrm{\cite{Caron-Huot:2024tzr}}} = S-2$.} It reads
\beq\label{eq:tilde-lambda}
\begin{aligned}
\log{\tilde{\lambda}^2_{\Delta, S}} &= \frac{S \left(\frac{17}{6} + S +\left(-\frac{7}{12} + \zeta_3\right) S^2\right)}{\Delta^2} \\
&\,\, + \frac{S \left(\frac{511}{60} + 6 S + \left(\frac{1}{12} - 2 \zeta_3\right) S^2 - \left(\frac{13}{8} + 6 \zeta_3\right) S^3 + \left(\frac{31}{40} - \frac{3}{2} \zeta_5\right) S^4 \right)}{\Delta^4} \\
&\,\, + \Op\left(\frac{1}{\Delta^6}\right)\, .
\end{aligned}
\eeq
It takes the form of an expansion in inverse powers of $\sqrt{\lambda}$ after substituting the expression for the scaling dimension $\Delta$ for $J=2$~\cite{Gubser:1998bc,Gromov:2009bc,Gromov:2011de,Basso:2011rs,Gromov:2011bz}
\beq\label{eq:Delta2}
\Delta^2 = 2\sqrt{\lambda}\, S + \left(4-S+\frac{3}{2} S^2\right) + \frac{1}{\sqrt{\lambda}} \left(\frac{15}{4} S + \frac{3-24\zeta_{3}}{8} S^2 - \frac{3}{8} S^3\right) + \Op\left(\frac{1}{\lambda}\right)\, ,
\eeq
at strong coupling.

We can now express this data in terms of our quantity $\D_{J_{1}J_{2}J}$. Accounting for the difference in prefactors with ref.~\cite{Caron-Huot:2024tzr}, we find
\beq
\D_{222}^2 = \mathcal{R} \times \tilde{\lambda}_{\Delta, S}^2\, ,
\eeq
where
\beq
\mathcal{R} = \left(\frac{\sqrt{\lambda} \,S}{2}\right)^S \, \frac{\Delta (\Delta-S)}{\left(\Delta+S\right)^2 } \frac{\Gamma\left(\frac{1+\Delta+S}{2}\right)\Gamma\left(\frac{\Delta-S}{2}\right)^3}{\Gamma\left(\frac{1+\Delta-S}{2}\right)\Gamma\left(\frac{\Delta+S}{2}\right)^3}\, .
\eeq
At strong coupling, the dimension $\Delta$ becomes large, allowing us to evaluate $\mathcal{R}$ using the asymptotic expansion of the Gamma functions. Applying these expansions, along with eqs.~\eqref{eq:tilde-lambda} and~\eqref{eq:Delta2}, we extract the two-loop expression for $\log{\D_{222}}$ as
\begin{equation}
\begin{aligned}\label{eq:D222}
\log{\D_{222}} &= \frac{1}{\sqrt{\lambda}} \left[\frac{5}{8}\, S - \frac{7-4\zeta_{3}}{16}\, S^2\right] \\
&\,\,\,\,\, + \frac{1}{\lambda} \left[-\frac{13+24\zeta_{3}}{32}\, S - \frac{49-8\zeta_{3}}{64}\, S^2 + \frac{25-12\zeta_{3}-12\zeta_{5}}{64}\, S^3\right]\\
&\,\,\,\,\, + \Op\left(\frac{1}{\lambda^{3/2}}\right)\, .
\end{aligned}
\end{equation}
Note that the expansion proceeds in integer powers of $1/\sqrt{\lambda}$, or equivalently, in powers of $1/\Delta^2$, as in eq.~\eqref{eq:tilde-lambda}. This structure reflects the fact that $\log{\mathcal{R}}$ itself admits such an expansion.

A key simplification of the expression in~\eqref{eq:D222} is that the result is now manifestly regular at  $S=0$. This contrasts with $\log{\tilde{\lambda}_{\Delta, S}}$ in~\eqref{eq:tilde-lambda}, which contains a term scaling as $S/\Delta^4 \sim 1/S$, at small spin. Moreover, we verify that the $n$-loop coefficient in~\eqref{eq:D222} is a polynomial in $S$ of degree $n+1$, vanishing at $S=0$, in full agreement with the structure of the ansatz in~\eqref{eq:polynomial-ansatz}.

We can carry out similar checks for the structure constants of operators with higher lengths. These structure constants were recently constructed in~\cite{Alday:2023flc} by matching correlation functions of CPOs with the Virasoro-Shapiro amplitude in the flat-space limit. One-loop results were also derived in ref.~\cite{Fardelli:2023fyq} for various lengths.%
\footnote{Namely, $(J_{1},J_{2}, J) = (p,p,2)$ and $(J_{1},J_{2}, J) = (2,p,p)$, with $p\geqslant 2$.}
The outcome for the $\D$-coefficients is straightforward: after factoring out the Gamma functions, these structure constants become exactly the same as for $222$,
\beq\label{eq:DJJJ-one-loop}
\log{\D_{J_{1}J_{2}J}} = \log{\D_{222}} + \Op\left(\frac{1}{\lambda}\right)\, .
\eeq
Although the data in~\cite{Fardelli:2023fyq} was obtained for specific values of $J_{1}, J_{2}$ and $J$, our regularity assumptions suggest that the above result should hold for any lengths. In particular, the absence of $J_{1},J_{2}$ dependence at one loop follows from the small-spin analysis, which predicts that these quantum numbers first appear at $\Op(1/\lambda)$; see eq.~\eqref{eq:PJell}. Independence from $J$ can be established by considering the classical limit, which also serves as a consistency check for the coefficients proportional to $S^n/(\sqrt{\lambda})^{1+n}$ in the string data, as discussed in the next section. 

\section{Classical limit}\label{sec:classical}

In this section, we examine our ansatz in the classical limit, where $S, \sqrt{\lambda} \rightarrow \infty$ with $\mathcal{S} = S/\sqrt{\lambda}$ held fixed. For reasons that will become clear later, we also adopt a similar scaling for the R-charges, introducing $\J = J/\sqrt{\lambda} = \Op(1)$, and likewise for $\mathcal{J}_{1,2} = J_{1,2}/\sqrt{\lambda}$. In this regime, scaling dimensions correspond to the energies of classical spinning strings, which have been extensively studied using worldsheet techniques (see e.g.~refs.~\cite{Gubser:2002tv,Frolov:2002av}) and classical integrability~\cite{Kazakov:2004qf}. Structure constants have also been computed in this limit~\cite{Kazama:2016cfl} by solving a minimal surface problem in $AdS_{3}\times S^{3}$,
\beq
\log{C}_{123} = \sqrt{\lambda} \,\, \Area + \Op(1)\, ,
\eeq
where ``$\Area$'' refers to the area of a classical string worldsheet ending on the three operators at the AdS boundary. In what follows, we recall the expression for this area and explain how to evaluate it in the limit $\S, \J, \J_{1}, \J_{2}\rightarrow 0$.

\subsection{Area and resolvent}

In our case, for one non-BPS and two half-BPS operators, the total area consists of three contributions,
\beq\label{eq:area}
\Area = \A^{\cl} + \B^{\cl} + \N^{\cl}\, ,
\eeq
each directly linked to factors in the hexagon formula,
\beq
\log{\A} = \sqrt{\lambda}\, \A^{\cl} + \Op(1)\, ,
\eeq
with analogous expressions for $\B$ and $\N$. According to~\cite{Kazama:2016cfl} the three components can be expressed as integrals over the resolvent $R(x)$, which encodes the infinite tower of conserved charges of the excited operator. For the state of interest, this function obeys the finite-gap equation~\cite{Kazakov:2004qf}
\beq
V'(x) = \R(x+i0)+ \R(x-i0)\, , \qquad V'(x) = \textrm{sgn}(x)-\frac{2\J x}{x^2-1}
\eeq
for $x \in (a, b)\cup (-b, -a)$, with solution given by the elliptic integral~\cite{Kostov:1992pn,Casteill:2007ct,Gromov:2011bz,Gromov:2011de}
\beq\label{eq:int-res}
\R(x) = 2x \int_{a}^{b} \frac{dy\, V'(y)}{x^2-y^2} \sqrt{\frac{(x^2-b^2)(x^2-a^2)}{(b^2-y^2)(y^2-a^2)}}\, .
\eeq
By construction, the resolvent is smooth outside the cuts $(a,b)\cup (-b,-a)$. This includes $x=0$ and $x=\infty$, where it behaves as
\beq\label{eq:Res-asym}
\R(x) \sim 2\pi x\, \left(\S- \E+\J\right)\, , \qquad \R(x) \sim \frac{2\pi}{x} \left(\S+\E-\J\right)\, ,
\eeq
respectively, and $\E = \Delta/\sqrt{\lambda}$ is the classical energy of the string. Imposing~\eqref{eq:Res-asym} on the solution~\eqref{eq:int-res} fixes the relation between the global quantum numbers of the string and the gap parameters $a,b$~\cite{Gromov:2011bz,Gromov:2011de}
\beq\label{eq:EandK}
\begin{aligned}
\mathcal{J} &= \frac{\sqrt{(a^2-1)(b^2-1)}}{\pi b} \, K\left(1-\frac{a^2}{b^2}\right)\, , \\
\mathcal{S} &= \frac{ab+1}{2\pi ab} \left[b\, E\left(1-\frac{a^2}{b^2}\right) - a\, K\left(1-\frac{a^2}{b^2}\right)\right]\, , \\
\mathcal{E} &= \frac{ab-1}{2\pi ab}  \left[b\, E\left(1-\frac{a^2}{b^2}\right) + a\, K\left(1-\frac{a^2}{b^2}\right)\right]\, ,
\end{aligned}
\eeq
where $K$ and $E$ are the complete elliptic integrals of the first and second kind, respectively. Lastly, we quote the expression for the density $\rho(x)$, defined as the discontinuity of the resolvent across the cut. It reads
\beq\label{eq:int-rho}
\rho(x)  = \frac{\R(x-i0)-\R(x+i0)}{2\pi i}= -\frac{2x}{\pi} \dashint_{a}^{b} \frac{dy\, V'(y)}{x^2-y^2} \sqrt{\frac{(b^2-x^2)(x^2-a^2)}{(b^2-y^2)(y^2-a^2)}}\, ,
\eeq
for $x\in [a, b]$, with $\rho(-x) = \rho(x)$ by symmetry.

Given the resolvent and density, the structure constant components are expressed as
\beq
\begin{aligned}\label{eq:ABNcl}
&\A^{\cl} = \A^{\cl}_{\textrm{asy}} +  I_{1}[\L_{A}] +  I_{1}[\J-\L_{A}]\, , \\
&\,\,\,\,\,\,\,\,\, \B^{\cl} =    I_{-1}[\L_{B}] +  I_{1}[\J+\L_{B}]\, , \\
&\,\,\,\,\,\,\,\,\,\,\,\,\,\,\,\,\, \N^{\cl} = \N^{\cl}_{\textrm{asy}} -  I_{2}[\J] \, , \\
\end{aligned}
\eeq
where the classical bridge lengths are
\beq
\L_{A} = \frac{\J_{1}-\J_{2}+\J}{2}\, , \qquad \L_{B} = \frac{\J_{1}+\J_{2}-\J}{2}\, .
\eeq
The key ingredient is the wrapping integral
\beq\label{eq:IqL}
\begin{aligned}
I_{q}[\L] &= \int_{U^{-}} \frac{dx\left(x-1/x\right)}{8\pi^2 x}\left[\textrm{Li}_{2}\left(e^{\frac{4\pi i \L x}{x^2-1} + iq\R(x)}\right)+  \textrm{Li}_{2}\left(e^{\frac{4\pi i \L x}{x^2-1}-iq\R(1/x)}\right) \right] \\
&\qquad \,\,\,\,\,\, - (\textrm{same with}\,\, R\rightarrow 0)\, ,
\end{aligned}
\eeq
where the integration contour runs along the lower half of the unit circle,
\beq
U^{-} = \{x\in \mathbb{C}: |x| = 1,  \textrm{Im} \, x \leqslant 0\}\, ,
\eeq
Here, $\textrm{Li}_{2}(z) = \sum_{n=1}^{\infty} z^n/n^2$ is the dilogarithm function. This integral depends on the resolvent $R(x)$, the length $\L$ and an auxiliary parameter $q\in \mathbb{R}$.

The remaining terms in eq.~\eqref{eq:ABNcl} correspond to the asymptotic contributions, $\A^{\cl}_{\textrm{asy}}$ and $\N^{\cl}_{\textrm{asy}}$. Their general forms were originally derived at weak coupling in the classical spin-chain limit~\cite{Gromov:2011jh,Kostov:2012jr}, and they remain valid in the classical string regime~\cite{Basso:2022nny,Jiang:2016ulr}. Specifically, $\A^{\cl}_{\textrm{asy}}$ can be written as the integral $I_{1}[\L_{A}]$, with a contour of integration encircling the cuts of $R(x)$. The asymptotic part of the norm, by contrast, is most naturally expressed in terms of the spectral density $\rho$, as
\beq\label{eq:calNasy}
\N^{\cl}_{\textrm{asy}} =  -\int_{a}^{b}\frac{dx(x-1/x)}{4\pi^2 x} \left[\textrm{Li}_{2}\left(e^{-2\pi \rho(x)}\right)+\pi^{2} \rho^{2}(x) -\zeta_{2}\right]\, ,
\eeq
where the endpoints $a,b$ are defined implicitly in~\eqref{eq:EandK}. 

In this section, we focus on the small-spin limit, $\S\rightarrow 0$. In this regime, the resolvent simplifies, and its support contracts to single points, $a, b \rightarrow \alpha$, where $\alpha$ is the zero of $V'(x)$,
\beq
V'(\alpha) = 0 \qquad \Rightarrow \qquad \alpha = \J+\sqrt{1+\J^2}\, .
\eeq
Ultimately, we are also interested in taking $\J\rightarrow 0$. However, this step is more delicate, as $\alpha \rightarrow 1$, causing the support of the resolvent to collide with the singularity in $V'(x)$. This complication makes direct integration in eq.~\eqref{eq:ABNcl} difficult, see e.g.~ref.~\cite{Gromov:2011de}. To circumvent this issue, we adopt the approach used in the spectral problem~\cite{Gromov:2011bz}: first expanding the resolvent and integrals at small spin while keeping $\J$ finite, then taking the limit $\J\rightarrow 0$. As we will see later, the singularities that arise in this final step cancel out when considering the $\D$-coefficients~\eqref{eq:main-ansatz}.

Small-spin expressions at finite $\J$ can be directly derived from the integral representation of the resolvent~\eqref{eq:int-res}. Inverting the parametrization~\eqref{eq:EandK}, we obtain
\beq
\begin{aligned}
a &= \alpha - 2\alpha \sqrt{\frac{\alpha}{\alpha^2+1}}\, \sqrt{\S} + \frac{\alpha^2 (3 \alpha^4+6 \alpha^2-1)}{(\alpha^2-1) (\alpha^2+1)^2} \, \S + \Op\left(\S^{3/2}\right)\,, \\
b &= \alpha + 2\alpha \sqrt{\frac{\alpha}{\alpha^2+1}}\, \sqrt{\S} + \frac{\alpha^2 (3 \alpha^4+6 \alpha^2-1)}{(\alpha^2-1) (\alpha^2+1)^2} \, \S + \Op\left(\S^{3/2}\right)\,. \\
\end{aligned}
\eeq 
Substituting these into~\eqref{eq:int-res}, we find that $\R$ admits a regular expansion
\beq
\R(x) = \R^{(1)}(x) \, \S + \R^{(2)}(x)\, \S^2 + \ldots\, ,
\eeq
where the first two coefficients are
\beq\label{eq:R12}
\begin{aligned}
&\R^{(1)} = \frac{4\pi x \alpha^2}{(\alpha^2-1)(x^2-\alpha^2)}\, , \\
&\R^{(2)} =  - \frac{4 \pi x \alpha^3 ((x^4+\alpha^6) (\alpha^4 + 6 \alpha^2 +1) - x^2 \alpha^2 (5\alpha^6+3\alpha^4+3\alpha^2+5))}{(\alpha^2-1)^3 (\alpha^2+1)^2 (x^2 - \alpha^2)^3} \, .
\end{aligned}
\eeq
Higher-order terms follow from~\eqref{eq:int-res} but are more cumbersome. We provide the next two terms in Appendix~\ref{app:classical}.

Similarly, one may easily construct the expression for the density at small spin, using the integral~\eqref{eq:int-rho}. In the right variable, the latter exhibits the familiar Wigner circle law, corrected by polynomial corrections at higher orders. Explicitly,
\beq\label{eq:rhoS}
\begin{aligned}
\rho = \sqrt{1-t^2}\, &\bigg[\frac{2\sqrt{\alpha(\alpha^2+1)}}{\alpha^2-1} \, \sqrt{\S} - \frac{4 \alpha^3 (\alpha^2+3)\, t}{(\alpha^2-1)^2 (\alpha^2+1)}\, \S \\
& + \frac{\alpha^{3/2}\left(16\alpha^2 ( \alpha^4 + 
        6 \alpha^2 +1)(\alpha^2+1) \, t^2- (9 \alpha^8 + 48 \alpha^6 + 70 \alpha^4+1)\right)}{2(\alpha^2-1)^3 (\alpha^2+1)^{5/2}}\,  \S^{3/2} \\
& + \Op\left(\S^2\right) \bigg]\, ,
\end{aligned}
\eeq
with $t = \frac{2x-(a+b)}{b-a}$. Note that, unlike the resolvent, the density has an expansion in half-integer powers of the spin.
\subsection{Small spin limit}

We now consider the limit of small quantum numbers in the classical formulas~\eqref{eq:ABNcl}. As mentioned earlier, we first expand at small $\S$ while keeping the lengths $\J$ and $\L$ fixed. In this regime, the wrapping integral~\eqref{eq:IqL} is small and admits a regular expansion in $\S$,
\beq\label{eq:IqLS}
I_{q}[\L] = q\, I[\L] \, \S + \Op(\S^2)\, .
\eeq
Using the leading-order expression for the resolvent~\eqref{eq:R12}, we immediately find 
\beq\label{eq:intIL}
\begin{aligned}
I[\L]  = \int_{U^{-}} \frac{dx}{2\pi i} \frac{x\left(x-1/x\right)^2\alpha^2\left(\alpha^2+1\right)}{\left(x^2-\alpha^2\right)\left(\alpha^2 x^2-1\right)\left(\alpha^2-1\right)}  \log{\left(1-e^{\frac{4\pi i \L x}{x^2-1}}\right)}\, .
\end{aligned}
\eeq
To evaluate this integral, we perform a change of variable from $x$ to $E$,
\beq
x = \frac{-i\pm \sqrt{E^2-1}}{E} \qquad \Rightarrow \qquad E = \frac{-2ix}{x^2-1}\, .
\eeq
This transformation maps the lower half of the unit circle $U^-$ to the interval $E>1$. The `$\pm$' indicates that we should sum over the two branches of the square root to account for $x$'s with positive and negative real parts, respectively. By performing the change of variables and using $\alpha = \J+\sqrt{1+\J^2}$, we find
\beq\label{eq:IinE}
I[\L]  = \oint \frac{i\sqrt{1+\J^2}dE}{\pi \J E \sqrt{1-E^2}(1+\J^2 E^2)} \log{\left(1-e^{-2\pi \L E}\right)}\, ,
\eeq
with the contour running clockwise around the interval $(1,\infty)$. We may then integrate by parts and, after massaging the result, apply Cauchy's theorem to express $I[\L]$ in terms of the residues at $E = \pm i/\J$ and $E = i m/\L$, with $m\in \mathbb{Z}$. We find
\beq\label{eq:masterI}
I[\J] = I_0[\J] + I_{\zeta}[\J]\, ,
\eeq
where
\beq\label{eq:I0}
\begin{aligned}
&I_0[\L] = \log{\Gamma\left(\frac{\L}{\J}\right)} + \frac{\sqrt{1+\J^2}+\J}{2\J} \log{\L} +\frac{(2\L-\J)}{2\J}\log{\J}+\frac{\L}{\J}\gamma_{\textrm{E}} \\
&\,\, -\frac{\L}{\J}\log{\left(1+\sqrt{1+\J^2}\right)}-\frac{1}{2}\log{\left(\frac{\sqrt{1+\J^2}+\J}{2}\right)}+\frac{\sqrt{1+\J^2}-\J}{2\J}\log{(4\pi)}\, ,
\end{aligned}
\eeq
and
\beq\label{eq:Iz}
\begin{aligned}
&I_{\zeta}[\L] \\
&= \sum_{m\, =\, 1}^{\infty} \left[\frac{\sqrt{1 + \J^2}}{\J} \log{\left[\frac{\L +\sqrt{\L^2 + m^2}}{m}\right]} - \log{\left[\frac{\J \sqrt{\L^2 + m^2}+\L\sqrt{1 + \J^2}}{\J m+\L}\right]} -\frac{\L}{\J m}\right]\, .
\end{aligned}
\eeq
This representation is useful for studying the behavior of the integral as $\L \rightarrow 0$. In this limit, $I_0[\L]$ scales logarithmically with $\L$, while the second component, $I_{\zeta}[\L]$, is regular. This can be seen by expanding the summand in~\eqref{eq:Iz} as a power series in $\L$ and performing the sum over $m$ term by term. It shows that $I_{\zeta}[\L]$ can be re-expressed as a sum over odd $\zeta$ values,
\beq\label{eq:ItoZeta}
I_{\zeta}[\L] = \frac{1}{\J}\sum_{k\, =\, 1}^{\infty} \frac{\zeta_{2k+1}}{2k+1} c_{k}(\J)\L^{2k+1}\, ,
\eeq
with the coefficients
\beq\label{eq:ckJ}
c_{k}(\J) = \sqrt{1+\J^2}\sum_{n\, =\, k}^{\infty} (-1)^{n}\frac{\Gamma(\frac{1}{2}+n)}{\Gamma(\frac{1}{2})\Gamma(1+n)} \J^{2(n-k)}\, .
\eeq
The behavior becomes more intricate when both $\L$ and $\J$ are small. In this case, we encounter logarithms and poles in $\I_{0}[\L]$, as well as a problematic dependence on $\L/\J$ through $\log{\Gamma(\L/\J)}$. There is also a pole in $\I_{\zeta}[\L]$ at small $\J$, as shown in eq.~\eqref{eq:Iz}. However, we will soon see that all these undesirable behaviors cancel out in the final result.

A similar analysis applies to the term $\A^{\cl}_{\textrm{asy}}$ in eq.~\eqref{eq:ABNcl},
\beq
\A^{\cl}_{\textrm{asy}}= I_{\textrm{asy}}[\L_{A}]\, \S+\Op\left(\S^2\right)\, .
\eeq
Its leading coefficient can be expressed as the integral in~\eqref{eq:IinE}, up to an overall factor of $1/2$ and the choice of the integration contour, which encircles the poles at $E = \pm i/\J$. This yields
\beq\label{eq:masterIasy}
I_{\textrm{asy}}[\L_{A}] = -\log{\Gamma\left(\frac{\L_{A}}{\J}\right)\Gamma\left(1-\frac{\L_{A}}{\J}\right)} + \log{(2\pi)}\, .
\eeq

Combining the pieces, we can compute $\A^{\cl}$ and $\B^{\cl}$ at small spin. Substituting~\eqref{eq:IqLS} into eq.~\eqref{eq:ABNcl}, we find that both quantities are regular,
\beq
\A^{\cl} = \A^{\cl}_{1}\, \S + \Op\left(\S^2\right)\, , \qquad \B^{\cl} = \B_{1}^{\cl}\, \S + \Op\left(\S^2\right)\, ,
\eeq
with
\beq\label{eq:AnBclII1}
\A^{\cl}_{1} = I_{\textrm{asy}}[\L_{A}] + I[\L_{A}] + I[\J-\L_{A}]\, , \qquad \B^{\cl}_{1} =  I[\J+\L_{B}]-I[\L_{B}]\, .
\eeq
Using eqs.~\eqref{eq:masterI} and~\eqref{eq:masterIasy}, we observe that the $\log{\Gamma}$ terms cancel out, leaving us with
\beq\label{eq:AnBcl1}
\begin{aligned}
\A^{\textrm{cl}}_{1} &= \frac{1+\delta_{1}}{2}\log{\left(\L_{A}(\J-\L_{A})\right)}+ \delta_{1}\log{(4\pi)} + \gamma_{\textrm{E}} \\
&\,\,\,\,\,\, - \log{\left[\J(1+\delta_{1})(1+\J\delta_{1})\right]} + I_\zeta[\L_{A}] + I_{\zeta}[\J-\L_{A}]\, , \\
\B^{\textrm{cl}}_{1} &= \frac{1+\delta_{1}}{2}\log{\left(\L_{B}+\J\right)} + \frac{1-\delta_{1}}{2}\log{\L_{B}} + \gamma_{\textrm{E}} \\
&\,\,\,\,\,\, - \log{\left(1+\J \delta_{1}\right)} - I_\zeta[\L_{B}] + I_{\zeta}[\J+\L_{B}]\, ,
\end{aligned}
\eeq
where $\delta_{1} = \sqrt{1+\J^2}/\J$.

Finally, we have the normalization factor $\N^{\cl}$, which is also regular at small $\S$, except for a simple logarithmic correction $\sim \S\log{\S}$,
\beq\label{eq:NclS}
\N^{\cl} = -\frac{1}{2}\left(\log{(\S/2)}-1\right)\, \S + \N^{\cl}_{1} \, \S + \Op\left(\S^2\right)\, .
\eeq
The logarithm in this equation arises from the integral over the density in~\eqref{eq:calNasy}. At small spin, the density $\rho$ is small (see eq.~\eqref{eq:rhoS}) and, by expanding the dilogarithm in~\eqref{eq:calNasy} around $\rho = 0$, we find that $\N_{\textrm{asy}}$ contains the non-analytic piece 
\beq
\begin{aligned}
-\int_{a}^{b}\frac{(x-1/x) dx}{2\pi x} \rho(x)\log{\left(\frac{2\pi\rho(x)}{e}\right)} = -\frac{1}{2} \log{\left[\frac{2 \sqrt{1 + \J^2} \, \pi^2 \, \S}{e\, \J^2}\right]} \, \S + \Op\left(\S^2\right) \, .
\end{aligned}
\eeq
Higher-order corrections to $\N_{\textrm{asy}}$ are analytic in $\rho$ and produce contributions that are regular at $\S = 0$. Including the wrapping integral, the coefficient $\N^{\cl}_{1}$ in eq.~\eqref{eq:NclS} becomes
\beq\label{eq:Ncl1}
\begin{aligned}
\N^{\textrm{cl}}_{1} =& -(1+\delta_{1})\log{\J}-\delta_{1}\log{(4\pi)}-2\gamma_{\textrm{E}} \\
& -\frac{1}{4}\log{\left(1+\J^2\right)}+ \log{\left[\J\left(1+\delta_{1}\right)\left(1+\J \delta_{1}\right)^2\right]}-2I_{\zeta}[\J]\, .
\end{aligned}
\eeq

We now focus on the reduced structure constant in the classical limit, $\D^{\cl}$, see eq.~\eqref{eq:Dcl}.
To compute it, we add $\A^{\cl}, \B^{\cl}$ and $\N^{\cl}$, while removing the contribution from the Gamma functions in eq.~\eqref{eq:main-ansatz}. In this limit, the Gamma functions have large arguments and can be evaluated using Stirling's approximation, along with the formula for the classical energy, $\E-\J = \delta_{1} \S + \Op\left(\S^2\right)$.

This leads to a remarkable simplification: all problematic logarithms cancel out. For instance, the term $\S \log{\S}$ in eq.~\eqref{eq:NclS} is eliminated by $\Gamma\left(1+S/2\right)$ in eq.~\eqref{eq:main-ansatz}. Similarly, other logarithms, such as $\frac{1}{2}(1+\delta_{1})\log{\L_{A}}, ...\, $ in eqs.~\eqref{eq:AnBcl1} and~\eqref{eq:Ncl1}, are canceled by the prefactor in eq.~\eqref{eq:main-ansatz}.

In summary, $\D^{\cl}$ is regular at $\S = 0$,
\beq
\D^{\cl} = \D^{\textrm{cl}}_{1} \, \S + \Op\left(\S^2\right)\, ,
\eeq
and is free from undesirable logarithms in the lengths. Combining all elements, we find
\beq\label{eq:D1logII}
\D^{\textrm{cl}}_{1} = -\frac{1}{4}\log{\left(1+\J^2\right)} + \sum_{\L \, \in\,  L} I_{\zeta}[\L] -2 I_{\zeta}[\J]\, ,
\eeq
where the sum ranges over $L = \{-\L_{B}, \J+\L_{B}, \L_{A}, \J-\L_{A}\}$. In line with our previous discussion, we verify that $\D^{\cl}_1$ is a smooth function of the quantum numbers. Notably, the pole $\sim 1/\J$ in $I_{\zeta}[\L]$ cancels out in the sum over $\L$ in equation~\eqref{eq:D1logII}.%
\footnote{To be precise, the poles cancel out pairwise in the sum $\I_{\zeta}[\pm \L]+\I_{\zeta}[\J\mp\L]$.} This cancellation becomes evident when substituting the sum representation~\eqref{eq:ItoZeta} into the equation. It gives
\beq
\D^{\textrm{cl}}_{1} = -\frac{1}{4}\log{\left(1+\J^2\right)} + \sum_{k\, =\, 1}^{\infty} \frac{\zeta_{2k+1}}{2k+1}\, c_{k}(\J) P_{k}(\J_{1}, \J_{2}, \J)\, ,
\eeq
where $c_k$ is given in eq.~\eqref{eq:ckJ} and where $P_{k}$ is a homogeneous polynomial of degree $2k$,
\beq
P_{k} = -2 \J^{2k}+ \frac{1}{2^{2k+1}\J}\sum_{\sigma_{1}, \sigma_{2} \, =\,  \pm}\left(\J+\sigma_{1} \J_{1}+\sigma_{2}\J_{2}\right)^{2k+1}\, .
\eeq
This polynomial is symmetric under $\J_{1}\leftrightarrow \J_{2}$, as expected. It is also even in each length individually. This property stems from the symmetry of equation~\eqref{eq:D1logII} under $\L_{A} \leftrightarrow -\L_{B}$, consistent with the observations made in Section~\ref{sec:hexagon-sc}.

\subsection{Higher-order corrections}\label{sec:diff-op}

We can proceed in a similar way to analyze the power-suppressed corrections in $\S$. To this end, it is useful to observe that the higher-order contributions $R^{(k)}(x)$ to the resolvent can be derived from the leading-order solution $R^{(1)}(x)$ through the action of a differential operator. Specifically, for any function $F$ that does not depend on the quantum numbers, we can write
\beq\label{eq:F-to-Delta}
F\left[\frac{4\pi\L x}{x^2-1}+\R(x)\right]-F\left[\frac{4\pi\L x}{x^2-1}\right] = \Delta(\partial_{\J}, \partial_{\L})\, \left[\R^{(1)}(x)F'\left[\frac{4\pi\L x}{x^2-1}\right]\right]\, ,
\eeq
where $F'$ is the derivative of $F$ and where
\beq\label{eq:diff-op}
\Delta(\partial_{\J}, \partial_{\L}) = \S + \sum_{n\, =\, 2}^{\infty} \S^{n} \Delta^{(n)}(\partial_{\J}, \partial_{\L})
\eeq
is a series of differential operators in $\L$ and $\J$. In particular, by selecting $F$ as the integrand in~\eqref{eq:IqL}, we obtain a representation for the wrapping integral at small spin,
\beq
I_{q}[\L] = q\, \Delta(\partial_{\J}, q\partial_{\L}) I[\L]\, ,
\eeq
where $I[\L]$ is the integral analyzed earlier.

The operators in eq.~\eqref{eq:diff-op} exhibit several remarkable properties. First, $\Delta^{(n)}(\partial_{\J}, \partial_{\L})$ is of degree $2(n-1)$, with coefficients that depend only on $\J$,
\beq
\Delta^{(n)}(\partial_{\J}, \partial_{\L}) = \sum_{j+l \, =\, 0}^{2(n-1)} c^{(n)}_{j, l}(\J)\, \partial_{\J}^{\, j}\partial_{\L}^{\, l}\, .
\eeq
Moreover, from the definition in~\eqref{eq:F-to-Delta} and basic properties of the resolvent, it follows that the actual degree in $\partial_{\L}$ is $n-1$, and the operator contains no constant term,
\beq
c^{(n)}_{j,l\, \geqslant \, n} = c^{(n)}_{0,0} = 0\,  .
\eeq
As an illustration, substituting the small-spin resolvent from eq.~\eqref{eq:F-to-Delta} and comparing both sides yields
\beq\label{eq:Delta-diff-one}
\Delta^{(2)}(\partial_{\J}, \partial_{\L}) = \frac{\sqrt{1+\J^2}}{4}\left(\partial_{\J}+\partial_{\L}\right) \partial_{\J} + \frac{\sqrt{1+\J^2}}{\J} \left(\partial_{\J}+\frac{1}{2}\partial_{\L}\right)\, ,
\eeq
which clearly exhibits the two stated properties. The expressions at the next two orders ($n=3,4$) are more involved and are presented in Appendix~\ref{app:classical}, along with the formulae for $R^{(3)}$ and $R^{(4)}$ used in their derivation.

Another interesting property is that $\Delta$ has a simple commutation relation with the operator that shifts $\L$ by $\J$,
\beq\label{eq:FCR}
e^{\J \partial_{\L}}\, \Delta(\partial_{\J}, \partial_{\L}) = \Delta(\partial_{\J}, -\partial_{\L})\, e^{\J \partial_{\L}} \,\,\, \Rightarrow \,\,\, \Delta(\partial_{\J}, \partial_{\L}) = \Delta(\partial_{\J}+\partial_{\L}, -\partial_{\L})\, .
\eeq
This observation is useful as it allows us to apply the same operator to the two $I$-integrals contributing to $\A^{\cl}$ in eq.~\eqref{eq:AnBclII1}, and similarly for $\B^{\cl}$. Since these sums of $I$-integrals have simpler properties that their individual components, this remark significantly simplifies the calculation. It yields
\beq
\A^{\textrm{cl}} = \Delta\left(\partial_{\J}, \partial_{\L_{A}}\right)\, \A^{\textrm{cl}}_{1}\, , \qquad \B^{\textrm{cl}} = \Delta\left(\partial_{\J}, -\partial_{\L_{B}}\right)\, \B^{\textrm{cl}}_{1}\, ,
\eeq
with $\A^{\cl}_{1}$ and $\B^{\cl}_{1}$ given in eq.~\eqref{eq:AnBclII1}. We must be more careful with the normalization factor $\N^{\cl}$. The differential operator cannot be used to calculate its asymptotic part $\N^{\cl}_{\textrm{asy}}$, as it is not exactly of the type~\eqref{eq:F-to-Delta}. Nonetheless, we can use our operators to construct its wrapping integral,
\beq
\N^{\cl} = \N^{\cl}_{\textrm{asy}}-2\left[\Delta(\partial_{\J}, 2\partial_{\L}) I[\L] \right]_{\L = \J}\, .
\eeq

The last important observation is that $\Delta$ acts simply on the logarithms of $\L$'s in eq.~\eqref{eq:AnBcl1},
\beq\label{eq:Delta-logGamma}
\begin{aligned}
&\Delta (\partial_{\J}, \pm \partial_{\L}) \left[\frac{1 \pm \delta_{1}}{2}\, \log{\L} \right] \\
&\qquad \,\,\, = \frac{2\L+\S \pm (\E-\J)}{2} \log{\left[\frac{2\L+\S\pm (\E-\J)}{2}\right]} - \L \log{\L} - \frac{\S\pm (\E-\J)}{2}\, .
\end{aligned}
\eeq
The nice thing here is that this combination of logarithms, with $\L = \L_{B}, \L_{A}, ...$, is exactly what the ratio of Gamma functions in eq.~\eqref{eq:main-ansatz} produces in the classical limit. As a result, we can easily eliminate the contributions from these Gamma functions, by removing the corresponding logarithms in our previous expressions.

Putting all pieces together and simplifying further the result, we find that $\D^{\cl}$ can be cast into the form
\beq\label{eq:Drhozeta}
\D^{\cl} = \D^{\cl}_{\rho} + \D^{\cl}_{\zeta}\, ,
\eeq
where $\D_{\rho}$ depends on the density~\eqref{eq:int-rho},
\beq
\D^{\cl}_{\rho} = -\int_{a}^{b}\frac{\left(x-1/x\right)dx}{4\pi x} \rho(x)\log{\left[\frac{(x^2-1)^2\rho^2(x)}{2e\S x^2}\right]}\, ,
\eeq
and with
\beq\label{eq:Dcl-diff-final}
\begin{aligned}
\D^{\cl}_{\zeta} = \sum_{\L \, \in\,  L}\Delta(\partial_{\J}, \partial_{\L}) I_{\zeta}[\L] -2\left[\Delta(\partial_{\J}, 2\partial_{\L}) I_{\zeta}[\L]\right]_{\L\, =\, \J}\, .
\end{aligned}
\eeq
$I_{\zeta}[\L]$ is given in eq.~\eqref{eq:ItoZeta} and the sum ranges over the same set of lengths as in eq.~\eqref{eq:D1logII}. The $\rho$ contribution, $\D_{\rho}^{\cl}$, only depends on $\J$ and is free from transcendental numbers. The $\zeta$ contribution, on the other hand, depends on all the lengths in the problem and admits an expansion in odd zeta values,
\beq\label{eq:Dcl-zeta-sum}
\D^{\cl}_{\zeta} = \sum_{k=1}^{\infty} \, \zeta_{2k+1}\, \D^{\cl}_{\zeta_{2k+1}}\, .
\eeq
The regularity of $\D^{\cl}$ at small charges is evident in both $\S$ and $\L$, as all the components are regular in these variables. The regularity in $\J$ is less obvious, but it can be easily verified. For example, evaluating $\D_{\rho}^{\cl}$ through $\Op\left(\S^4\right)$ using the density~\eqref{eq:int-rho}, we find
\beq\label{eq:Dcl-diff}
\begin{aligned}
\D^{\cl}_{\rho} =  &-\frac{1}{4}\log{\left(1+\J^2\right)}\,\, \S -\frac{7+4\mathcal{J}^2}{16 \left(1+\mathcal{J}^2\right)^{3/2}} \,\, \S^2 + \frac{150+120\mathcal{J}^2+29\mathcal{J}^4}{384\left(1+\mathcal{J}^2\right)^{3}}\,\, \S^3 \\
&-\frac{1785 + 1748 \J^2 + 640 \J^4 + 86 \J^6}{3072 \left(1 + \J^2\right)^{9/2}} \,\, \S^{4} + \Op\left(\S^5\right)\, ,
\end{aligned}
\eeq
which is regular at $\J = 0$. Individual terms in $\D_{\zeta}^{\cl}$ are singular at small $\J$ due to the presence of poles $\sim 1/\J$ in $I_{\zeta}[\L]$ and in the differential operator $\Delta(\partial_{\J}, \partial_{\L})$, see eq.~\eqref{eq:Delta-diff-one}. However, these singularities cancel out when all terms are summed in~\eqref{eq:Dcl-diff-final}. For instance, for the coefficient in front of $\zeta_{3}$ in~\eqref{eq:Dcl-zeta-sum}, we obtain
\beq
\begin{aligned}
\D_{\zeta_{3}}^{\cl} &= \frac{\sqrt{1+\J^2}-1}{2\J^2} \left(\J^2-\vec{\J}^{\, 2}\right) \, \S \\
 &\,\,\,\, + \left[\frac{\sqrt{1+\J^2} \left(3\J^2 - \vec{\J}^{\, 2}\right)}{4 \J^4}  - \frac{\J^2 \left(3 + 2 \J^2\right) \left(2 + \J^2 - \vec{\J}^{\, 2}\right) - 2 \vec{\J}^{\, 2}}{8 \J^4 \left(1 + \J^2\right)} \right]\, \S^2 \\
&\,\,\,\, + \Op\left(\S^{3}\right)\, ,
\end{aligned}
\eeq
with $\vec{\J}^{\, 2} = \J_{1}^2+\J_{2}^2$. Despite the inverse powers of $\J$, this expression is regular at $\J = 0$. The terms of order $\Op\left(\S^{k>2}\right)$ in this equation and the coefficients of higher $\zeta$ values follow directly from eq.~\eqref{eq:Dcl-diff-final}. They are considerably more involved, so we refrain from presenting them here. However, we have verified that they are also regular at $\J = 0$. Thus, the final result is smooth in all its variables, as expected.

\subsection{New integral representation}

Before presenting explicit expressions for $\D^{\cl}$ at small charges, let us add a comment about our representation~\eqref{eq:Dcl-diff}. It turns out that the differential operator $\Delta(\partial_{\J},\partial_{\L})$ can be eliminated using a suitable integral representation for $I_{\zeta}[\L]$. To express this, we introduce the integral
\beq\label{eq:Z-int}
\Z\left[\R_{\mathcal{L}}\right] = \int\limits_{-i\infty}^{i\infty} \frac{\left(x+1/x\right)}{(2\pi)^2 i} \left[\frac{1}{2} \log{\frac{\Gamma\left(1-\frac{\R_{\mathcal{L}}(x)}{2\pi}\right)}{\Gamma\left(1+\frac{\R_{\mathcal{L}}(x)}{2\pi}\right)}}-\frac{\gamma_{\textrm{E}} \R_{\mathcal{L}}(x)}{2\pi}\right] d\R_{\mathcal{L}}(x) - (R_{\mathcal{L}} \rightarrow \widehat{R}_{\mathcal{L}})\, ,
\eeq
where
\beq
\R_{\mathcal{L}}(x) = \frac{4\pi \L x}{x^2-1} + \R(x)\, , \qquad \widehat{\R}_{\mathcal{L}}(x) = \frac{4\pi \L x}{x^2-1}\, ,
\eeq
and $R(x)$ is the resolvent. In contrast to the original integrals, the integration contour in~\eqref{eq:Z-int} runs along the imaginary axis. In addition, the integrand contains a ratio of Gamma functions, which is absent in the wrapping integral. Nevertheless, we assert that this integral correctly computes the $\zeta$-part of the final answer,
\beq
Z\left[\R_{\mathcal{L}}\right] = \Delta(\partial_{\J}, \partial_{\L}) \, I_{\zeta}[\L]\, .
\eeq
In other words, $\D^{\cl}_{\zeta}$ can be equivalently written as
\beq
\D^{\textrm{cl}}_{\zeta} =  \sum_{\L \, \in \, L}  \Z\left[\R_{\mathcal{L}}\right] - \Z\left[2\R_{\J/2}\right]\, ,
\eeq
with the sum range as in eq.~\eqref{eq:Dcl-diff-final}.

To prove this relation,  it suffices to consider it at leading order for small spin. The generalization to finite spin then follows from the properties of the differential operator.

The proof proceeds as follows. First, to establish a connection with the sum in eqs.~\eqref{eq:ItoZeta} and~\eqref{eq:Dcl-zeta-sum}, we expand the integrand in \eqref{eq:Z-int} in odd zeta values, using
\beq
\frac{1}{2} \log{\frac{\Gamma\left(1-\frac{\R_{\mathcal{L}}(x)}{2\pi}\right)}{\Gamma\left(1+\frac{\R_{\mathcal{L}}(x)}{2\pi}\right)}}-\frac{\gamma_{\textrm{E}} \R_{\mathcal{L}}(x)}{2\pi} = \sum_{k=1}^{\infty}\, \frac{\zeta_{2k+1}}{2k+1} \left(\frac{\R_{\mathcal{L}}(x)}{2\pi}\right)^{2k+1}\, .
\eeq
Substituting this series in the integral~\eqref{eq:Z-int} and integrating by parts, we get
\beq
Z\left[\R_{\mathcal{L}}\right] = \sum_{k=1}^{\infty} \, \frac{\zeta_{2k+1}}{2k+1} \, Z_k \left[\R_{\mathcal{L}}\right]\, ,
\eeq
with
\beq\label{eq:Zk}
Z_k \left[\R_{\mathcal{L}}\right] = -\int\limits_{-i\infty}^{i\infty} \frac{dx \, (x-1/x)}{4\pi i x (k+1)} \left[\bigg(\frac{\R_{\L}(x)}{2\pi}\bigg)^{2k+2} - \bigg(\frac{\widehat{\R}_{\L}(x)}{2\pi}\bigg)^{2k+2}\right]\, .
\eeq
Extracting the term in $\S$ at small $\S$ should then yield the coefficients $c_{k}(\J)$ in eq.~\eqref{eq:ItoZeta},
\beq
c_{k}(\J) = -\int\limits_{-i\infty}^{i\infty} \frac{dx \, (x-1/x)}{2\pi i x} \frac{R^{(1)}(x)}{2\pi} \left(\frac{2x}{x^2-1}\right)^{2k+1}\, ,
\eeq
with $R^{(1)}(x)$ given in eq.~\eqref{eq:R12}. Now, the right-hand side of this equation is simply the integral of a rational function with poles at $x=\pm 1$ and $x=\pm \alpha$. It can be evaluated by closing the integration contour at $x=\infty$ and computing the residues at $x=1$ and $x=\alpha$. This yields perfect agreement with the expression for the coefficient $c_{k}(\J)$ in~\eqref{eq:ckJ}, thereby concluding the proof.

The advantage of the $Z$ integral is that we no longer need to subtract logarithms or construct the differential operator, a task that becomes more challenging at higher orders. It only requires the expression for the resolvent, as shown in~\eqref{eq:Zk}. Moreover, the $Z$ integral is not limited to the small-spin analysis and could be useful for exploring other regimes, such as the large-spin limit.

\subsection{Final expressions}

To conclude our analysis, we present the expressions for $\D^{\cl}$ in the regime where all classical charges are small. We recall that
\begin{equation}\label{eq:Dcl123}
\D^\cl = \D_{1}^{\textrm{cl}}\, \S + \D_{2}^{\textrm{cl}}\, \S^2 + \D_{3}^{\textrm{cl}}\, \S^3 + \ldots \, ,
\end{equation}
where the coefficients are functions of the classical lengths, $\J, \J_1, \J_2$. Now, when all three lengths are small, the leading contribution at small spin reads
\begin{equation}
\D_{1}^{\textrm{cl}} = -\frac{\J^2}{4} + \frac{(\J^2-\vec{\J}^{\, 2})}{4} \zeta_{3} +\Op\left(\J^4, \J_{1,2}^{4}, \J_{1}^2\J_{2}^2, \J^{2}\J_{1,2}^2\right)\, ,
\end{equation}
with $\vec{\J}^{\, 2} = \J_{1}^2+\J_{2}^2$. After restoring the units, $\S = S/\sqrt{\lambda}, \J = J/\sqrt{\lambda}, \ldots\,$, this behavior indicates that the term linear in $S$ in $\log{\D}$ is suppressed at strong coupling, in agreement with the short-string data~\eqref{eq:D222}.

In contrast, the power-suppressed corrections~$\sim \S^{k>1}$ in~\eqref{eq:Dcl123} remain nonzero at small lengths. They are given by
\beq\label{eq:Dcl234}
\begin{aligned}
\D_{2}^{\textrm{cl}} &= -\frac{7}{16} + \frac{1}{4} \zeta_{3} +\Op\left(\J^2, \vec{\J}^{\, 2}\right)\, , \\
\D_{3}^{\textrm{cl}} &= \,\,\,\,\, \frac{25}{64}-\frac{3}{16}\zeta_{3} -\frac{3}{16} \zeta_{5} +\Op\left(\J^2, \vec{\J}^{\, 2}\right) \, , \\
\D_{4}^{\textrm{cl}} &= -\frac{595}{1024} +\frac{29}{128}\zeta_{3} +\frac{9}{32} \zeta_{5} + \frac{45}{256}\zeta_{7}+\Op\left(\J^2, \vec{\J}^{\, 2}\right) \, ,
\end{aligned}
\eeq
in perfect agreement with the two-loop data~\eqref{eq:D222} for the terms proportional to $S^{k}/(\sqrt{\lambda})^{k-1}$ with $k= 2,3$. We note that higher zeta values appear at increasing $k$, indicating that the string theory result at $(k-1)$ loops involves transcendental numbers of weight up to $2k-1$. Moreover, these numbers fall into the class of single-valued zeta values — of which odd zeta values are the simplest representatives — as expected from the worldsheet analysis~\cite{Alday:2022xwz,Alday:2023jdk}.

\section{Short-string regime}\label{sec:two-loops}

In this section, we complete the analysis of the structure constants at small spin using the hexagon formalism and obtain a prediction at two loops at strong coupling for operators with arbitrary lengths.

\subsection{Normalization factor}

To complete our calculation of the structure constants at small spin, we must determine the normalization factor $\N$ appearing in the hexagon formula. As mentioned earlier, a key property of this factor is that it only depends on the spin $S$ and length $J$ of the excited operator, not on the quantum numbers of the half-BPS operators. According to the hexagon proposal~\cite{Basso:2015zoa,Basso:2022nny}, it can be expressed in terms of the solution $\{u_k, k=1, \ldots, S\}$ to the Bethe ansatz equations for the excited operator~\cite{Beisert:2005fw},
\begin{equation}\label{eq:Bethe-Ansatz}
 1= e^{ip_{k} J} \prod_{j\, \neq\, k}^{S} S_{kj}\, ,
\end{equation}
where $p_k$ is the momentum of the root $u_k$,
\beq\label{eq:momentum}
p_k = -i \log{\left(\frac{x_{k}^{+}}{x_{k}^{-}}\right)}\, , \qquad x_k^{\pm} = x(u_k\pm i/2)\, ,
\eeq
and $S_{kj}$ is the magnon S-matrix. In this setting, the formula for the normalization factor is given by
\beq\label{eq:NequalHGW}
\N = \sqrt{\frac{H}{G}} \times e^{W}\, ,
\eeq
where $G$ is the Gaudin determinant, corresponding to the Bethe equations~\eqref{eq:Bethe-Ansatz},
\begin{equation}\label{eq:Gaudin-G}
G = \underset{1\,\leqslant\, j, k\,\leqslant\, S}{\textrm{det}}\, \left[\frac{\partial }{\partial u_j} \left(J p_k - i\sum_{l\,\neq\, k}^{S}\log{S_{kl}}\right)\right]\, ,
\end{equation}
and $H$ is a simple prefactor,
\beq\label{eq:prefactorH}
H = \prod_{k\, <\, j}^{S} \frac{(u_{k}-u_{j})^2}{(u_{k}-u_{j})^2+1} \exp{\left[\sum_{k, j}^{S} \log{\frac{\left(x_{k}^{+}x_{j}^{-}-1\right)\left(x_{k}^{-}x_{j}^{+}-1\right)}{\left(x_{k}^{+}x_{j}^{+}-1\right)\left(x_{k}^{-}x_{j}^{-}-1\right)}}\right]}\, .
\eeq
The term $W$ in~\eqref{eq:NequalHGW} represents wrapping corrections. It accounts for finite-size modifications to the norm formula when the operator is short. It also includes similar modifications to the Bethe equations~\cite{Bajnok:2008bm,Bajnok:2008qj}, which were omitted from equation~\eqref{eq:Bethe-Ansatz}  to avoid clutter but are discussed in Appendix~\ref{App:normalization}.

A well-known problem with this type of representation is that the infinite tower of wrapping corrections becomes unwieldy at strong coupling, when the length $J = \Op(1)$. Studying the small spin limit helps mitigate this issue, by suppressing the higher wrapping corrections. However, the difficulty lies in analytically continuing the above formula in the spin $S$. Notably, while it seems natural to expect the determinant $G$ to admit a regular expansion at small spin, it is not obvious how to extrapolate it beyond integer spin.

To address these issues, we first assume that $J$ is large, scaling as $\sqrt{\lambda}$ at strong coupling. (We will later extend our results to finite $J$ guided by our regularity assumptions.) Specifically, we work with
\beq
\J = J/\sqrt{\lambda} = \Op(1)\, ,
\eeq
while keeping the spin fixed. In this case, the effective string length is large, the anomalous dimension is of order $\Op(1)$, see eq.~\eqref{eq:dim-BMN} below, and the wrapping corrections remain under control. Moreover, the interactions among the roots weaken in this regime ($S_{kj} = 1 + \Op(1/\sqrt{\lambda})$), reducing the Bethe equations~\eqref{eq:Bethe-Ansatz} to matrix-model equations~\cite{Arutyunov:2004vx,Beisert:2005fw,Gromov:2012eg},
\begin{equation}\label{eq:Bethe-eqs-lin}
n_{k} = \frac{2\J x_{k}}{x_{k}^2-1} + \sum_{j\, \neq\, k}^{S}\frac{4x_k x_j (x_k x_j-1)}{\sqrt{\lambda}(x_k^2-1)(x_k-x_j)(x_j^2-1)} + \Op\left(\frac{1}{\lambda}\right)\, ,
\end{equation}
where the integer $n_k$ is the mode number of the root $x_k = x(u_k)$. This simplification allows for explicit calculations at virtually any spin.

For the states of interest, the spin is even, the roots are paired as $x_k = -x_{S-k} >0$ for $k=1, \ldots, S/2$, with mode numbers $\pm 1$ for positive and negative roots, respectively. At leading order, the interactions in~\eqref{eq:Bethe-eqs-lin} can be neglected, yielding
\begin{equation}
x^{(0)}_k = \alpha =  \J + \sqrt{1+\J^2}\, , \qquad k=1,\ldots, S/2\, ,
\end{equation}
for all positive roots. This degeneracy among roots with same mode number enhances the interactions $\propto (x_k-x_j)^{-1}$ in~\eqref{eq:Bethe-eqs-lin} at higher orders, enforcing $1/\lambda^{1/4}$ as the expansion parameter for the solution,
\begin{equation}\label{eq:exp-x}
x_k = x_k^{(0)} + \frac{x_k^{(1/2)}}{\lambda^{1/4}} + \frac{x^{(1)}_k}{\lambda^{1/2}} + \frac{x^{(3/2)}_k}{\lambda^{3/4}} + \ldots\, .
\end{equation}
The degeneracy is lifted at the order $x_k^{(1/2)}$, where the analysis reduces to the Gaussian matrix model, with the solution given by the roots of the $S/2$-th Hermite polynomial,
\begin{equation}\label{eq:Hermite}
H_{S/2} \left(\frac{x_{k}^{(1/2)}}{\sqrt{2} \, \beta}\right) = 0\, , \qquad k=1, \ldots, S/2\, ,
\end{equation}
with $\beta^2 = \alpha^2/\sqrt{1+\J^2}$.  In particular, $x_{1}^{(1/2)} = 0$ for spin $S=2$, $x_{1,2}^{(1/2)} = \pm \beta$ for spin $S=4$, and so on. Higher-order terms in~\eqref{eq:exp-x} are more intricate but can be derived recursively from the Bethe equations by tracking the loop corrections in~\eqref{eq:Bethe-eqs-lin}, as discussed in Appendix~\ref{App:normalization}.

Given the roots, one can determine the normalization factor $\N$, starting with the determinant $G$. Substituting the right-hand side of the Bethe equations~\eqref{eq:Bethe-eqs-lin} into $G$, we find that the matrix in~\eqref{eq:Gaudin-G} becomes block diagonal at strong coupling,
\beq
G = G_{+} G_{-} \left(1 + \Op\left(\frac{1}{\lambda}\right)\right)\, ,
\eeq
where the sub-determinants $G_{\pm}$ correspond to the mode numbers $\pm 1$, respectively. The off-diagonal terms, which arise from roots with different mode numbers, are delayed until two loops. In our case, the blocks are identical ($G_{+} = G_{-}$), leading to
\begin{equation}
\begin{aligned}\label{eq:G-LO}
\sqrt{G} \approx &\left[\frac{4\pi^2 \sqrt{1+\J^2}}{\J^2\sqrt{\lambda}}\right]^{S/2} \\
&\times \underset{1\,\leqslant\, j, k\,\leqslant\, S/2}{\textrm{det}} \left[ \delta_{k=j} \left(1+\sum_{l\,\neq\, k}^{S/2}\frac{2\beta^2}{\sqrt{\lambda} \left(x_{k}-x_{l}\right)^2}\right)-\delta_{k\neq j}\frac{2\beta^2}{\sqrt{\lambda} \left(x_{k}-x_j\right)^2}\right]\, ,
\end{aligned}
\end{equation}
in the Gaussian approximation. The `singular' terms proportional to $(\sqrt{\lambda}\,(x_k-x_j)^2)^{-1}$ cannot be neglected, as the difference between rapidities is $\propto 1/\lambda^{1/4}$, see eq.~\eqref{eq:exp-x}. Remarkably, when evaluated at the roots of the Hermite polynomial~\eqref{eq:Hermite}, the determinant in eq.~\eqref{eq:G-LO} can be computed exactly, yielding $\Gamma(1+S/2)$ at spin $S$.

Simplifications also occur in the loop corrections. Computing $G$ in~\eqref{eq:Gaudin-G} up to $\Op(1/\lambda)$ for various values of the spin, we find that it can be written as
\beq\label{eq:Gexp}
\sqrt{G} = \left[\frac{4\pi^2\sqrt{1+\J^2}}{\J^2 \sqrt{\lambda}}\right]^{S/2} \Gamma\left(1+\frac{S}{2}\right) \, \exp{\left(\sum_{n\, \geqslant\, 1} \frac{P_{n}^{G}(S)}{(\sqrt{\lambda})^{n}}\right)}\, ,
\eeq
where $P_{n}^{G}(S)$ is a polynomial of degree $n+1$ in $S$, with no constant term, $P^{G}_{n}(0) = 0$, and coefficients depending on $\J$. A similar structure is found for the factor $H$ in eq.~\eqref{eq:NequalHGW}, which takes the form
\beq\label{eq:Hexp}
\sqrt{H} = \exp{\left(\sum_{n\, \geqslant\, 1} \frac{P^{H}_{n}(S)}{(\sqrt{\lambda})^{n}}\right)}\, ,
\eeq
where $P_{n}^{H}(S)$ follows the same polynomial pattern as $P_{n}^{G}(S)$. Explicit expressions for $P_{n}^{G}$ and $P_{n}^{H}$ are provided in Appendix~\ref{App:normalization} for $n=1,2$, see equations~\eqref{eq:PG} and~\eqref{eq:PH}. In this form, analytic continuation in the spin becomes straightforward. In particular, we verify that the loop corrections to $G$ and $H$ are consistent with the general structure described earlier.

Finally, we consider the wrapping corrections, $W$. These corrections are typically expressed as (mutiple) integrals over the rapidities of mirror magnons, similar to those studied earlier. Explicit forms for the relevant integrals are provided in Appendix~\ref{App:normalization}. Here, we simply state the results at strong coupling for the terms linear in $S$, which correspond to the one-mirror-magnon approximation,
\beq\label{eq:W012}
W = \left(W^{(0)} + \frac{W^{(1)}}{\sqrt{\lambda}} + \frac{W^{(2)}}{\lambda} + \ldots \right) \, S + \Op\left(S^2\right)\, .
\eeq
The leading term is given by the same integral as in the classical limit,
\beq
W^{(0)} = -2I[\J]\, ,
\eeq
see equation~\eqref{eq:IinE}, while the subsequent terms can be derived by applying suitable linear operators to this integral,
\beq\label{eq:hatDeltaW}
W^{(i)} = -2 \left[\widehat{\Delta}^{(i)} I[\L]\right]_{\L \, =\, \J}\, .
\eeq
The first operator is given by
\beq\label{eq:hatDelta1}
\begin{aligned}
&\widehat{\Delta}^{(1)} =\\
& -\frac{\sqrt{1+\J^2}}{2} (\partial_{\J}+\partial_{\L})\partial_{\J} - \frac{\sqrt{1+\J^2}}{\J}\left(\partial_{\J}+\frac{1}{2}\partial_{\L}\right) + \frac{1}{2\J} \left(\J\sqrt{1+\J^2} -\mathcal{R}\right) \partial_{\L}^{2}\, ,
\end{aligned}
\eeq
where $\mathcal{R}$ extracts the residue at $\J = 0$ of the function it acts on,
\beq\label{eq:Res-op}
\mathcal{R} [f] = \underset{\J\, =\,0}{\textrm{res}} f(\J)\, .
\eeq
The second operator $\widehat{\Delta}^{(2)}$ is more involved and includes the effect of wrapping corrections to the Bethe equations, emerging at two loops. Its form is detailed in Appendix~\ref{App:normalization}.

The subleading $\Op(S^{k\geqslant 2})$ contributions in $W$ are significantly more difficult to analyze and will not be discussed here. However, we expect them to follow the same pattern as the other corrections, becoming increasingly suppressed at strong coupling.

\subsection{Short-string limit}

We now extract the expression for the term linear in $S$ at finite $J$. To simplify access to the reduced structure constant $\D$ in eq.~\eqref{eq:main-ansatz}, we first strip off a few factors from $\N$ and define the quantity $\D_{\N}$ as follows,
\begin{equation}\label{eq:DN}
\N = \frac{2^{2S}\lambda^{3S/2+\gamma/2}}{\left(4\pi\right)^{S+\gamma} e^{2\gamma_{\textrm{E}}S}} \sqrt{\frac{\Gamma(J)\Gamma(J-1)}{\Gamma(\Delta+S)\Gamma(\Delta+S-1)}} \times \frac{\D_{\N}}{\lambda^{S/4}\Gamma\left(1+\frac{S}{2}\right)}\, ,
\end{equation}
where $\gamma = \Delta-J-S$. The prefactor above is chosen to ensure that
\beq
\D = \D_{\N} \D_{\A} \D_{\B}\, ,
\eeq
where $\D_{\A}$ and $\D_{\B}$ are defined in eq.~\eqref{eq:DABfirst}. It can be expanded at small spin using the scaling dimension formula~\eqref{eq:slope}, which reads
\begin{equation}\label{eq:dim-BMN}
\Delta = \sqrt{\lambda}\, \J + S\left[\frac{\sqrt{1+\J^2}}{\J} -\frac{1}{2\J (1+\J^2)\sqrt{\lambda}} + \frac{4\J^2-1}{8\J (1+\J^2)^{5/2} \lambda}\right] + \Op \left(S^2, \frac{S}{\lambda^{3/2}}\right)\, ,
\end{equation}
at strong coupling, in the regime where $\J = \Op(1)$.

Taking the logarithm of the equation~\eqref{eq:DN} and incorporating our findings for $\log{\N}$, we obtain
\begin{equation}\label{eq:logDN}
\log{\D_{\N}} = S \left[c^{(0)} + \frac{c^{(1)}}{\sqrt{\lambda}} + \frac{c^{(2)}}{\lambda}\right] + \Op\left(S^2, \frac{S}{\lambda^{3/2}}\right)\, , 
\end{equation}
where the coefficients $c^{(i)}$ are functions of the charge $\J$. These coefficients receive contributions from the polynomials in eqs.~\eqref{eq:Gexp} and~\eqref{eq:Hexp}, the wrapping integrals~\eqref{eq:W012}, and the prefactors in eqs.~\eqref{eq:Gexp} and~\eqref{eq:DN}. After combining all these contributions, we obtain

\begin{equation}
\begin{aligned}\label{eq:a-coefficients}
c^{(0)} &= -\frac{1}{4}\log{\left(1+\J^2\right)} + K[\J] \, ,\\
c^{(1)} &= \frac{5 + 2 \J^2}{8 (1 + \J^2)^{3/2}} +\left[\widehat{\Delta}^{(1)} K[\L] \right]_{\L\, =\, \J}\, , \\
c^{(2)} &= \frac{57 - 12 \J^2 + 4 \J^4}{96 (1 + \J^2)^3} + \left[\widehat{\Delta}^{(2)} K[\L] \right]_{\L\, =\, \J}\, ,
\end{aligned}
\end{equation}
where, to simplify the expressions, we introduced
\beq
K[\L] = \log{\left[\frac{1}{4}(1+\sqrt{1+\J^2})^2(\J+\sqrt{1+\J^2})\right]} - 2I_{\zeta}[\L]\, .
\eeq
Here, $I_{\zeta}[\L]$ is the sum over odd $\zeta$ values in eq.~\eqref{eq:Iz} and the operators $\widehat{\Delta}^{(i)}$ are given in eqs.~\eqref{eq:hatDelta1} and~\eqref{eq:hatDelta-final}.

Finally, to get the formula for $\log\,{\D_{\N}}$ at finite $J$, we expand~\eqref{eq:a-coefficients} for small $\J$ and substitute $\J \rightarrow J/\sqrt{\lambda}$. It yields

\begin{equation}\label{eq:logDNfinal}
\log\,{\D_{\N}} = S \left[-\frac{1}{J}-\frac{\frac{7}{8}-J}{\sqrt{\lambda}}-\frac{\left(\frac{15}{32}-\frac{J}{2}-\frac{J^2}{4}\right)+\left(\frac{5}{24}-\frac{J^2}{3}\right)\zeta_{3}}{\lambda} \right] + \Op\left(S^2, \frac{S}{\lambda^{3/2}}\right) \, .
\end{equation}
We note that $\log{\D_{\N}}$ is polynomial in $J$, except for a pole in $1/J$. This pole cancels out a similar term from $\log{\D_{\A}}$, see eqs.~\eqref{eq:DABsecond} and~\eqref{eq:PJell}. More generally, both $\log{\D_{\N}}$ and $\log{\D_{\A}}$ contain odd powers of $J$, which cancel in their sum. Bringing all the pieces together, using eqs.~\eqref{eq:logDNfinal},~\eqref{eq:DABsecond} and~\eqref{eq:PJell}, we obtain
\beq\label{eq:D1S2loops}
\begin{aligned}
\log{\D} &= \log{\D_{\N}} + \log{\D_{\A}} + \log{\D_{\B}} \\
&= S \left[\frac{5}{8\sqrt{\lambda}} + \frac{\left(19-8J^2\right)+8\left(1+J^2-J_{1}^2-J_{2}^2\right)\zeta_{3}}{32 \lambda}\right] + \Op\left(S^2, \frac{S}{\lambda^{3/2}}\right)\, .
\end{aligned}
\eeq
This gives our final expression for the leading small-spin behavior of the $\D$-coefficient at strong coupling. It agrees perfectly with the stringy results in eqs.~\eqref{eq:D222} and~\eqref{eq:DJJJ-one-loop}.

\subsection{Two-loop prediction}

By combining our results, we can formulate a two-loop prediction for the structure constants of three operators with arbitrary lengths at strong coupling. This extends to any length the short-string data obtained for $J_{1} = J_{2} = J = 2$ in ref.~\cite{Alday:2024ksp} and expressed in our notation in eq.~\eqref{eq:D222}. It reads
\begin{equation}
\begin{aligned}\label{eq:two-loop-pred}
\log&\,\D_{J_{1}J_{2}J}\\
& = \frac{1}{\sqrt{\lambda}} \left[\frac{5}{8}\, S - \frac{7-4\zeta_{3}}{16}\, S^2\right] \\
&\,\,\,\,\, + \frac{1}{\lambda} \left[\frac{\left(19-8J^2\right)+8\big(1+J^2-\vec{J}^{\, 2}\big)\zeta_{3}}{32}\, S - \frac{49-8\zeta_{3}}{64}\, S^2 + \frac{25-12\zeta_{3}-12\zeta_{5}}{64}\, S^3\right] \\
&\,\,\,\,\, + \Op\left(\frac{1}{\lambda^{3/2}}\right)\, ,
\end{aligned}
\end{equation}
with $\vec{J}^{\, 2} = J_{1}^2+J_{2}^{2}$. The terms linear in $S$ follow directly from eq.~\eqref{eq:D1S2loops}, while the terms proportional to $S^{n+1}/(\sqrt{\lambda})^{n}$ originate from the classical string analysis (see eq.~\eqref{eq:Dcl234}).

The only term in~\eqref{eq:two-loop-pred} that remains undetermined in our combined analysis is the one proportional to $S^2/\lambda$. Based on our general assumptions, its coefficient can depend at most linearly on $J_{1},J_{2},$ and $J$. However, since the classical limit showed no such linear dependence, we conclude that this coefficient is independent of the lengths and must match the short-string data~\eqref{eq:D222} for $J_{1} = J_{2} = J = 2$.

Finally, we should note that our results provide some insight into the answer at three loops. For example, the (maximal) contribution proportional to $S^{4}/\lambda^{3/2}$ can be extracted from the classical result in~\eqref{eq:Dcl234}. In addition, the small-spin results from Section~\ref{sec:hexagon-sc} determine the dependence of the term proportional to $S/\lambda^{3/2}$ on $J_{1}$ and $J_{2}$. It would be interesting to see if these integrability-based predictions can be tested by extending the powerful worldsheet method from refs.~\cite{Alday:2024ksp,Wang:2025pjo} to higher loops.

\section{Conclusion}\label{sec:conclusion}

In this paper, we examined structure constants of single-trace operators as functions of the coupling constant and spin. Our starting point was the hexagon representation, whose study at small spin proved remarkably simple thanks to the exact solution of the QSC equations. This approach led to concise all-loop expressions for structure constants of operators of any length, up to a normalization factor that remains challenging to study and is currently accessible only in the large-length limit.

Moreover, much like scaling dimensions, the small-spin limit offers valuable insight into structure constants at strong coupling. In particular, it indicates that, after factoring out a specific ratio of Gamma functions, the structure constants simplify and take a polynomial form in spin and R-charges at each order in the strong coupling expansion. We showed that this structure is fully consistent with existing data for short strings and smoothly interpolates between the small-spin expansion and the classical limit. Building on this observation, we extended recent two-loop results for the shortest operators to those of arbitrary length.

However, this approach, based on small-spin data and classical string results, did not fully determine all coefficients in the string-length expansion. While it proved sufficient at one loop, one coefficient remained undetermined at two loops, with additional missing terms expected at higher loops.

A more complete analysis would require extending certain strong-coupling calculations. For example, the method used in Section~\ref{sec:two-loops} to extract the normalization factor at finite spin could, in principle, be applied to the full structure constants at higher loops, avoiding the need for an exact finite-coupling analysis of the $\A$ and $\B$ factors. The main obstacle lies in developing a systematic expansion of the hexagon formula in this regime, incorporating all necessary wrapping corrections to the normalization factor and transfer matrices.

Another promising direction is the construction of higher-loop corrections to the structure constants in the semiclassical limit. As we have seen, the classical expressions contain a wealth of information and are more tractable than the full hexagon formulation. While little is currently known about semiclassical loop corrections on the worldsheet side, exact expressions might be attainable through integrability-based methods, building on the approaches used in refs.~\cite{Gromov:2011jh,Kostov:2012jr,Jiang:2016ulr}.

Recent advances have also enabled the extraction of structure constants for states on subleading trajectories~\cite{Julius:2023hre,Julius:2024ewf}, using precision QSC data for the low-lying spectrum~\cite{Gromov:2023hzc,Ekhammar:2024rfj} to disentangle the string-theory sum rules. These studies have uncovered intriguing polynomial structures and selection rules for structure constants that remain to be fully understood, whether from the worldsheet perspective or directly through integrability. Applying the methods developed in this paper to these higher-string states could offer new insights into these phenomena.

Finally, it would be interesting to understand the simplifications observed in this paper directly from a field-theoretic perspective. In recent years, significant progress has been made in studying CFT correlators through analytic continuation in the spin, not only at the level of spectral data using the QSC formalism~\cite{Gromov:2014bva,Alfimov:2014bwa,Klabbers:2023zdz,Ekhammar:2024neh,Brizio:2024nso}, but also in terms of non-local light-ray operators, see for example \cite{Homrich:2024nwc} and references therein. However, this description can sometimes be challenging to apply. In this context, the similarities between small-spin results and expressions found for correlation functions in the presence of a Wilson line (see e.g.~\cite{Pufu:2023vwo,Billo:2023ncz,Billo:2024kri,Dempsey:2024vkf}) are particularly intriguing. They suggest that an interpretation in terms of nonlocal operators may be more tractable at small spin.

\acknowledgments

We thank Simon Ekhammar, Giulia Fardelli, Vasco Gon\c calves, Nikolay Gromov, Shota Komatsu, Gregory Korchemsky, Michelangelo Preti, Istv\'an Sz\'ecs\'enyi, Pedro Vieira, Congkao Wen and Alexander Zhiboedov for interesting discussions. 
A.~G.~is supported by a Royal Society funding, URF\textbackslash{R}\textbackslash221015.

\appendix

\section{Structure constants at weak coupling}\label{App:PerturbTest}

In this appendix, we compare our results at weak coupling with the direct analytic continuation to small spin of known perturbative expressions for $\A$ and $\B$. To generate this data at finite spin, we evaluate the hexagon formulae~\cite{Basso:2015eqa,Basso:2017khq} order by order in the weak coupling expansion, across a broad range of spins, and fit the $L$-loop results to ans\"atze constructed from nested harmonic sums of weight up to $2L$.%
\footnote{To reduce the number of data points needed to obtain general-spin results at higher loops, we follow the strategy outlined in ref.~\cite{Velizhanin:2013vla}}
These sums are defined recursively as
\beq \label{eq:harmonicSumsDef}
\s{a}(x)= \sum_{i\, =\, 1}^{x} \frac{(\textrm{sgn}\, {a})^i}{i^a}\, , \qquad \s{a,b,c,\ldots}(x)= \sum_{i\, =\, 1}^{x} \frac{(\textrm{sgn}\, {a})^i}{i^a}\, \s{b,c,\ldots} (i)\, ,
\eeq
where $a,b,c, \ldots$ are integers defining the sum's weight, $w = |a|+|b|+|c|+\ldots\,$. In our case, the argument $x$ will the spin of the operator, which we denote as $s$ throughout this appendix to distinguish it from our notation for nested sums.

Since we are interested in the small-spin limit, we need to expand the harmonic sums~\eqref{eq:harmonicSumsDef} as $x\rightarrow0$. To do so, it is useful to express these sums as iterated integrals.%
\footnote{See ref.~\cite{Zerbini:2015rss} for further applications and properties.}
This method can be illustrated with a simple example,
\beq
\s{2,3}(s)= \sum_{i_1 \leq i_2 \leq s } \frac{1}{i_2^2 i_1^3}= \sum_{i_1 \leq i_2\leq s } \int_0^\infty \prod_{n=1}^2 dt_n~\frac{t_1^{2} t_2}{2}  e^{-t_1 i_1- t_2 i_2}\, ,
\eeq
where we used Schwinger's trick to rewrite the harmonic sum as a simple geometric sum. The sum can be evaluated directly, and by applying the change of variables $t_i = - \log{x_i}$, we obtain
\beq
\s{2,3}(s)= \int_0^1 \prod_{n=1}^2 \frac{dx_n}{x_n}~ \frac{x_1 x_2 (1 - x_2 - (-1 + x_1) x_1^s x_2^{1 + s} + x_1^s (-1 + x_1 x_2)) \log{x_1} \log^{2}{x_2}}{2 (-1 + x_1) (-1 + x_2) (-1 + x_1 x_2)}\, .
\eeq
In this form, expanding around $s=0$ becomes straightforward. Performing the integration, we obtain the leading-order term as
\beq
\s{2,3}(s)= \left( \frac{8}{7} \z{2}^3 - 2 \z{3}^2 \right)s+\mathcal{O}\left(s^2\right)\, .
\eeq
The result can be checked against available code that performs the same manipulations (see for example ref.~\cite{Ablinger:2009ovq}).

Also, we recall that $s$ is even for operators on the leading Regge trajectory. This property is essential for handling alternating nested sums with negative indices, as they introduce factors of $(-1)^s$ when extended to non-integer spins.

Lastly, our analysis below is restricted to operators of length $J=2$, as only in this case do the results admit a parametrization in terms of harmonic sums; see ref.~\cite{Eden:2012rr} and references therein.

\subsection{Bottom bridge}

We begin with the bottom-channel contributions for $\ell_{B} = 1,2$, which are obtained by evaluating the mirror integrals in refs.~\cite{Basso:2015eqa,Basso:2015zoa} for a large set of spin values.

\paragraph{Bridge length 1.} In this case, we obtain
\begin{equation}\label{eq:lb1g6}
    \B (\ell_{B}=1) = 1+ g^4 \left(c_{1|4}+c_{2|4}\z{3} \right)+g^6\left(c_{1|6}+c_{2|6}\z{3}+c_{3|6}\z{5} \right) +\Op\left(g^{8}\right)\, ,
\end{equation}
where we defined
\begin{align}
c_{1|4}=&\, 4\left(\s{-2}^2 - 2\s{-3}\s{1} - 2\s{-2}\s{1}^2 -2\s{1}\s{3} - \s{4} + 2\s{-3, 1} + 4\s{1}\s{-2, 1} + 2\s{-2, 2} +2\s{3, 1} - 4\s{-2, 1, 1}\right)\, , \nonumber \\
c_{2|4}=&\, 24 \s{1}\, , \nonumber \\
c_{1|6}=&\, \frac{32}{3}\left(-6\s{-6} + 3\s{-3}^2 - 30\s{-5}\s{1} - 6\s{-4}(\s{-2} -3\s{1}^2) + 5\s{3}^2 + 6\s{6} + 30\s{-5, 1} - 12\s{-4, 2} - 24\s{-3, 3} +\right. \nonumber \\  & \s{1}^3(\s{3} - 6\s{-2, 1}) - 48\s{3}\s{-2, 1} + 12\s{-2, 1}^2 - 3\s{-3}(4\s{-2}\s{1} - 3\s{1}^3 + 11\s{1}\s{2} + 4(-4\s{3} + \s{-2, 1})) + \nonumber \\ &54\s{4, -2} - 6\s{4, 2} + 6\s{5, 1} - 48\s{-4, 1, 1} + 6\s{-2}(\s{1}^2\s{2} - 9\s{4} - 10\s{1}\s{2, 1} + 2(\s{-3, 1} + \s{-2, 2} + 4\s{3, 1} - \nonumber \\  &2\s{-2, 1, 1})) + 12\s{2}(3\s{-3, 1} + \s{3, 1} - 6\s{-2, 1, 1}) + 6\s{1}^2(2\s{4} - 5\s{-3, 1} - 4\s{-2, 2} - \s{3, 1} + 6\s{-2, 1, 1}) + \nonumber \\ &36\s{-2, 2, 2} - 48\s{-2, 3, 1} - 36\s{2, -3, 1} - 12\s{2, 3, 1} - 48\s{3, 1, -2} + 12\s{4, 1, 1} - 72\s{-3, 1, 1, 1} - 3\s{1}(3\s{2}(\s{3}- \nonumber \\ &6\s{-2, 1}) + 2(\s{5} + 5\s{-4, 1} - 6\s{-2, 3} + 4\s{2, -3} - 2\s{2, 3} + 3\s{4, 1} - 11\s{-3, 1, 1} - 2\s{-2, 1, -2} - 10\s{-2, 2, 1} - \nonumber \\ &10\s{2, 1, -2} - 3\s{3, 1, 1} + 18\s{-2, 1, 1, 1})) + 72\s{2, -2, 1, 1} - 24\s{3, 1, 1, 1} + 144\s{-2, 1, 1, 1, 1})\, , \nonumber \\
c_{2|6}=&\,  -\frac{32}{3}\left(6\s{-3} + 15\s{-2}\s{1} - 4\s{1}^3 + 9\s{1}\s{2} + 4\s{3} - 12\s{-2, 1}\right) \, ,\nonumber \\
c_{3|6}=&\,-240 \s{1} \, .
\end{align}
By expanding the nested sums at small argument with the method described earlier, we obtain
\beq
\B(\ell_{B}=1) = 1 + s \left[ 3g^4( 4 \z{2} \z{3} + 5 \z{5})-48g^6(\z{3}\z{4} + \z{2} \z{5} + 7 \z{7})\right]+\Op\left(sg^8, s^2\right)\, ,
\eeq
in agreement with the formula in Table~\ref{tab:C123BJ2}.
\paragraph{Bridge length 2.} When the bottom bridge length is equal to 2, we have 
\begin{equation}\label{eq:lb2g6}
    \B(\ell_{B}=2) = 1 + g^6\left(c_{1|6}+c_{2|6}\z{3}+c_{3|6}\z{5} \right) +\Op\left(g^{8}\right)\, ,
\end{equation}
where 
\begin{align}
   c_{1|6}=\, & \frac{8}{3}(-6\s{-6} + 9\s{-3}^2 - 12\s{-5}\s{1} - 7\s{3}^2 + 24\s{-5, 1} -
6\s{-4, 2} - 12\s{-3, 3} + 2\s{-3}(\s{1}^3 - 9\s{1}\s{2} + \nonumber \\ &11\s{3} -6\s{-2, 1}) - 20\s{3}\s{-2, 1} + 12\s{-2, 1}^2 - 2\s{1}^3(\s{3} +2\s{-2, 1}) + 6\s{-2}(-5\s{4} + 2\s{-3, 1} + \nonumber \\ &\s{-2, 2} + 4\s{3, 1}) +30\s{4, -2} + 12\s{4, 2} - 12\s{5, 1} - 36\s{-4, 1, 1} - 12\s{-3, -2,1} - 12\s{-3, 1, -2} +\nonumber \\ & 6\s{1}^2(\s{-4} - \s{4} - 2(\s{-3, 1} + \s{-2, 2} - \s{3, 1} - 2\s{-2, 1, 1})) + 24\s{2}(\s{-3, 1} - \s{3, 1} -2\s{-2, 1, 1}) - \nonumber \\ & 6\s{-2, 2, -2} + 24\s{-2, 2, 2} - 36\s{-2, 3, 1} -24\s{2, -3, 1} + 24\s{2, 3, 1} - 24\s{3, 1, -2} - 12\s{4, 1, 1} - \nonumber \\ &48\s{-3, 1, 1, 1} - 24\s{-2, -2, 1, 1} + 6\s{1}(2\s{5} - 2\s{-4, 1} - \s{-2}\s{-2, 1} + 3\s{2}(\s{3} + 2\s{-2, 1}) + \nonumber \\ & 3\s{-2, 3} - 3\s{2,-3} - 6\s{-2}\s{2, 1} - 3\s{2, 3} + 3\s{4, 1} + 6\s{-3, 1, 1} +\s{-2, 1, -2} + 6\s{-2, 2, 1} + \nonumber \\ & 6\s{2, 1, -2} - 6\s{3, 1, 1} -12\s{-2, 1, 1, 1}) + 48\s{2, -2, 1, 1} + 48\s{3, 1, 1, 1} + 96\s{-2,1, 1, 1, 1})\, , \nonumber \\
   c_{2|6}=\, & \frac{16}{3}(3\s{-3} + 3\s{-2}\s{1} + \s{1}^3 - \s{3} - 6\s{-2, 1})\, , \nonumber \\
   c_{3|6}=\, & 80 \s{1}\, .
\end{align}
Taking the small spin limit of eq.~\eqref{eq:lb2g6} we obtain
\beq
\B(\ell_{B}=2) = 1 + 4s g^6(-3 \z{3}\z{4} + 9 \z{2} \z{5} + 14 \z{7})+\mathcal{O}\left(sg^{8}, s^2\right)\, ,
\eeq
which also matches with the result reported in Table~\ref{tab:C123BJ2}.

\subsection{Adjacent bridge}

Let us now consider the contribution from the adjacent channel. According to the hexagon proposal~\cite{Basso:2015zoa}, it admits the decomposition
\beq\label{eq:app-A}
\mathcal{A}= \mathcal{A}_{\text{asy}} + \delta \A\, ,
\eeq
where $\delta \A$ represents the mirror corrections. The leading contribution at weak coupling comes from the asymptotic term,
\beq\label{eq:partitions}
\mathcal{A}_{\text{asy}}= \sum_{\alpha\, \cup\, \bar{\alpha} \, =\, \textbf{u}}(-1)^{|\bar{\alpha}|} \prod_{j \in \bar{\alpha}} e^{i p_j \ell_A} \prod_{i \in \alpha,j \in \bar{\alpha}} \frac{1}{h_{ij}}\, ,
\eeq
where the sum runs over all partitions of the set of Bethe roots, $\alpha\cup \bar{\alpha} = \textbf{u} =\{u_k, k=1, \ldots , S\}$. Here, $p_{j}$ is the momentum of the root $u_j$, and $h_{ij} = h(u_i, u_j)$ is the hexagon form factor; see ref.~\cite{Basso:2015zoa} for explicit expressions.

Again, we can interpolate between different spin values using nested sums for $J=2$, which implies $\ell_A=1$. Through $3$ loops, we obtain%
\footnote{A.G. thanks V. Gon\c calves for sharing unpublished results.}
\beq\label{eq:Aasy}
\mathcal{A}_{\text{asy}} = \frac{(2s)!}{(s!)^2} \left(1+g^2 c_{1|2}+g^4 c_{1|4}+g^6(c_{1|6}+c_{3|6}\z{3}) + \Op\left(g^{8}\right)\right)\, ,
\eeq
where
\begin{align}
c_{1|2}=&-4( \s{2} + 2 \s{1}( \s{1} -  \sC{1}))\, , \nonumber \\ 
c_{1|4}=&\,2(4 \s{-4} + 4 \s{-2}^2 + 4 \s{-3} \s{1} + 16 \s{1}^4 + 40 \s{1}^2 \s{2} + 4 \s{2}^2 + 15 \s{1} \s{3} -  \s{4} + 4 \s{-3, 1} +\nonumber \\ & 8 \s{1}\s{-2, 1}+ 8 \s{-2, 2} + 5 \s{1, 3} + 9 \s{3, 1} - 24 \s{-2, 1, 1} - \nonumber \\ & 8 \s{-3} \sC{1} - 32 \s{1}^3 \sC{1} - 32 \s{1} \s{2} \sC{1} - 8 \s{3} \sC{1} + 16 \s{-2, 1} \sC{1} + 16 \s{1}^2 \sC{1}^2 + 4 \s{-2}(\s{1}^2 +  \s{2} -\nonumber \\ & 4 \s{1} \sC{1}) - 16 \s{1}^2 \sC{2})\, , \nonumber \\
c_{1|6}=&-\frac{32}{3}(12 \s{-6} - 6 \s{-3}^2 + 12 \s{-4} \s{-2} + 51 \s{-5} \s{1} + 24 \s{-4} \s{1}^2 + 6 \s{-2}^2 \s{1}^2 + 6 \s{-2} \s{1}^4 + 8 \s{1}^6 +\nonumber \\ &9 \s{-4} \s{2} + 3 \s{-2}^2 \s{2} + 27 \s{-2} \s{1}^2 \s{2} + 48\s{1}^4 \s{2} + 3 \s{-2} \s{2}^2 + 48 \s{1}^2 \s{2}^2 +  \s{2}^3 + 18 \s{-2} \s{1} \s{3} +\nonumber \\ & 45 \s{1}^3 \s{3} + 48 \s{1} \s{2} \s{3} + \s{3}^2 + 114 \s{-2} \s{4} + 30 \s{1}^2 \s{4} + 3 \s{2} \s{4} + 27\s{1} \s{5} -  \s{6} - 54 \s{-5, 1} +\nonumber \\ & 6 \s{1} \s{-4, 1} + 6 \s{-4, 2} - 24 \s{-2} \s{-3, 1} + 12 \s{1}^2 \s{-3, 1} - 57 \s{2} \s{-3, 1} + 66 \s{-3, 3} - 6 \s{-2} \s{1} \s{-2, 1} + \nonumber \\ &2 \s{1}^3 \s{-2, 1} - 72 \s{1} \s{2} \s{-2, 1} + 100 \s{3} \s{-2, 1} - 24 \s{-2, 1}^2 - 24 \s{-2} \s{-2, 2} + 12 \s{1}^2 \s{-2, 2} -\nonumber \\ & 6 \s{2} \s{-2, 2} - 42 \s{1} \s{-2, 3} + 36 \s{1} \s{2, -3} + 60 \s{-2} \s{1} \s{2, 1} - 12 \s{1} \s{2, 3} - 108 \s{-2} \s{3, 1} +\nonumber \\ & 12 \s{1}^2 \s{3, 1} - 9 \s{2} \s{3, 1} - 108 \s{4, -2} + 18 \s{1} \s{4, 1} + 6 \s{4, 2} - 6 \s{5, 1} + 120 \s{-4, 1, 1} - 66 \s{1} \s{-3, 1, 1} -\nonumber \\ & 6 \s{1} \s{-2, 1, -2} + 48 \s{-2} \s{-2, 1, 1} - 24 \s{1}^2 \s{-2, 1, 1} + 150 \s{2} \s{-2, 1, 1} - 60 \s{1} \s{-2, 2, 1} - 60 \s{-2, 2, 2} +\nonumber \\ & 108 \s{-2, 3, 1} + 48 \s{2, -3, 1} - 60 \s{1} \s{2, 1, -2} + 12 \s{2, 3, 1} + 108 \s{3,1, -2} - 18 \s{1} \s{3, 1, 1} - 12 \s{4, 1, 1} +\nonumber \\ & 120 \s{-3, 1, 1, 1}+ 108 \s{1} \s{-2, 1, 1, 1} - 144 \s{2, -2, 1, 1} + 24 \s{3, 1, 1, 1} - 336 \s{-2, 1, 1, 1, 1} + 18 \s{-5} \sC{1} - \nonumber \\ &54 \s{-4} \s{1} \sC{1} - 12 \s{-2}^2 \s{1} \sC{1} - 30 \s{-2} \s{1}^3 \sC{1} - 24 \s{1}^5 \sC{1} - 42 \s{-2} \s{1} \s{2} \sC{1} - 84 \s{1}^3 \s{2} \sC{1} - 30 \s{1} \s{2}^2 \sC{1} -\nonumber \\ & 12 \s{-2} \s{3} \sC{1} - 54 \s{1}^2 \s{3} \sC{1} - 12 \s{2} \s{3} \sC{1} - 24 \s{1} \s{4} \sC{1} - 6 \s{5} \sC{1} + 36 \s{-4, 1} \sC{1} + 66 \s{1} \s{-3, 1} \sC{1} +\nonumber \\ & 36 \s{1}^2 \s{-2, 1} \sC{1} - 48 \s{2} \s{-2, 1} \sC{1} + 48 \s{1} \s{-2, 2} \sC{1} - 36 \s{-2, 3} \sC{1} + 36 \s{2, -3} \sC{1} + 72 \s{-2} \s{2, 1} \sC{1} - \nonumber \\ & 6 \s{1} \s{3, 1} \sC{1} - 72 \s{-3, 1, 1} \sC{1} - 60 \s{1} \s{-2, 1, 1} \sC{1} - 72 \s{-2, 2, 1} \sC{1} - 72 \s{2, 1, -2} \sC{1} + 144 \s{-2, 1, 1, 1} \sC{1} +\nonumber \\ & 24 \s{-2} \s{1}^2 \sC{1}^2 + 24 \s{1}^4 \sC{1}^2 + 36 \s{1}^2 \s{2} \sC{1}^2 + 12 \s{1} \s{3} \sC{1}^2 - 24 \s{1} \s{-2, 1} \sC{1}^2 - 8 \s{1}^3 \sC{1}^3 +  \s{-3}(24 \s{-2} \s{1} +\nonumber \\ & 13 \s{1}^3 - 54 \s{1}^2 \sC{1} + 4(-25 \s{3} + 6 \s{-2, 1} + 6 \s{2} \sC{1}) + 12 \s{1}(4 \s{2} +  \sC{1}^2 -  \sC{2})) - 24 \s{-2} \s{1}^2 \sC{2} - \nonumber \\ &24 \s{1}^4 \sC{2} - 36 \s{1}^2 \s{2} \sC{2} - 12 \s{1} \s{3} \sC{2} + 24 \s{1} \s{-2, 1} \sC{2} + 24 \s{1}^3 \sC{1} \sC{2} - 16 \s{1}^3 \sC{3})\, , \nonumber \\
c_{3|6}=&-32 \s{1}( \s{-2} +  \s{2}) \, ,
\end{align}
with $\sC{a}=\s{a}(2s)$. By expanding these coefficients at small spin and using the facts that the tree-level result is $1+\Op(s^2)$ and that $c_{3|6}=\Op(s^2)$, we obtain
\beq
\begin{aligned}\label{eq:spAsy}
\mathcal{A}_{\text{asy}}= 1&+s\left[-8\z{3} g^2 +g^4(-32 \z{2} \z{3} + 90 \z{5})+g^6(160 \z{3} \z{4} + 448 \z{2} \z{5} - 1120 \z{7})\right]\\
&+\Op\left(s g^8, s^2\right)\, .
\end{aligned}
\eeq

The mirror corrections $\delta\A$ in eq.~\eqref{eq:app-A} appear at order $\Op\left(g^6\right)$ for $\ell_{A} = 1$. The relevant integral was studied in refs.~\cite{Basso:2015eqa,Eden:2015ija}, where it was shown to be proportional to the L\"uscher correction $\delta \gamma$ to the anomalous dimension~$\gamma$,
\begin{equation}
\frac{\delta \A}{\A_\text{asy}} = \frac{2\delta \gamma}{\gamma} + \mathcal{O}\left(g^8\right)\, .
\end{equation}
To leading order at weak coupling, $\gamma = 8g^2 S_{1}(s) + \Op\left(g^4\right)$, and the finite-spin expression for the L\"uscher correction can be found in ref.~\cite{Bajnok:2008qj}, see eq.~(18) therein. It yields
\begin{equation}
\frac{\delta \A}{\A_\text{asy}} = \left(-160\, \s{1} \z{5}+\ldots\right) g^{6} + \Op\left(g^{8}\right)\, ,
\end{equation}
where the dots represent terms that are multilinear in the nested sums.
The harmonic sums on the right-hand side can be analytically continued to the complex spin plane and then expanded at small spin, as described earlier. This gives
\begin{equation}
\frac{\delta \A}{\A_\text{asy}} = -160\zeta_{2} \z{5} \, g^6 \, s + \mathcal{O}\left(g^8 s, g^{6}s^2\right)\, .
\end{equation}
Substituting this, along with~\eqref{eq:spAsy}, into eq.~\eqref{eq:app-A}, gives the full perturbative contribution up to $\Op\left(g^6\right)$, in perfect agreement with eq.~\eqref{eq:Adjacent-weak-coupling}.

\section{Test at large $J$}\label{app:largeJ}

In this appendix, we test our formula in the large-$J$ regime at weak coupling, $g^2\rightarrow 0$. In this limit, the hexagon representation is dominated by the asymptotic sum in eq.~\eqref{eq:partitions}. To evaluate it, we must determine the Bethe roots $\{u_k, k=1, \ldots, S\}$. When $g^2 = 0$, these roots satisfy the Bethe equations for the $sl(2)$ spin chain,
\beq
2\pi n_{k} = -i J \log{\left(\frac{u_k+\frac{i}{2}}{u_{k}-\frac{i}{2}}\right)} - i \sum_{j\neq k}^{S} \log{\left(\frac{u_k-u_j+i}{u_k-u_j-i}\right)}\, .
\eeq
Here, $S$ is even, and the roots are symmetric, with mode numbers $n_{k} = \pm 1$ for the positive and negative roots, respectively. In the large-$J$ limit, the solution admits an expansion in $1/\sqrt{J}$,
\beq
u_k = \pm \frac{J}{2\pi} \left(1 + \frac{u^{(1/2)}_k}{\sqrt{J}} + \frac{u^{(1)}_{k}}{J} + \Op\left(\frac{1}{J^{3/2}}\right) \right)\, .
\eeq
This structure closely parallels the one encountered in Section~\ref{sec:two-loops} at strong coupling, with $1/\sqrt{J}$ here playing the role of $1/\lambda^{1/4}$ there. In particular, the leading corrections are given by the roots of the $S/2$-th Hermite polynomial, $H_{S/2}(u^{(1/2)}_k /\sqrt{2}) = 0$. Higher-order terms can be computed iteratively for any fixed spin.

By substituting the roots into the sum over partitions in eq.~\eqref{eq:partitions}, and using the weak-coupling expressions
\beq
e^{ip_k} = \frac{u_k+\frac{i}{2}}{u_k-\frac{i}{2}} + \Op\left(g^2\right)\, , \qquad h(u_k, u_j) = \frac{u_k-u_j}{u_k-u_j-i}+\Op\left(g^2\right)\, ,
\eeq
we can compute the asymptotic contribution $\A_{\textrm{asy}}(\ell_{A})$ at large $J$ for various values of the spin $S$ and bridge length $\ell_A$. The results are well captured by the ansatz
\beq\label{eq:A-largeJ}
\A_{\textrm{asy}}(\ell_{A}) = \left(\frac{2\pi}{J}\right)^{S}\frac{\Gamma(\ell_{A}+S)}{\Gamma(\ell_{A})} \, \exp{\left(\sum_{k\, =\, 1}^{\infty} \frac{P_{k}(S, \ell_{A})}{J^{k}}\right)}\, ,
\eeq
where $P_{k}(S, \ell_{A})$ is a polynomial of degree $k+1$ in $S$ and $\ell_{A}$, vanishing at $S=0$. The first two polynomials are
\beq\label{eq:PellS}
P_{1}(S, \ell_{A}) = \frac{S(2-3S)}{4}\, , \qquad P_{2}(S, \ell_{A}) = \frac{S(4  - 24 S  + 17 S^2+4\pi^2 (S- 2\ell_{A}(\ell_{A}+1)))}{48}\, .
\eeq
This result allows for a straightforward expansion at small $S$. For convenience, we express it in terms of the generating function
\beq\label{eq:genA}
\widehat{\A}_{\textrm{asy}}(z) = \sum_{\ell_{A}\, = \, 1}^{\infty} z^{\ell_{A}} \A_{\textrm{asy}}(\ell_{A})\, ,
\eeq
where each coefficient corresponds to a specific bridge length $\ell_{A}$. We find
\beq\label{eq:data-largeJ}
\widehat{\A}_{\textrm{asy}}(z)  = \frac{z}{1-z}\left(1 + S \, a_{J}(z) +\Op\left(S^2\right)\right)\, ,
\eeq
where
\beq\label{eq:aJz}
a_{J}(z)= \log{\left(\frac{2\pi}{J(1-z)}\right)} + \psi(1) + \frac{1}{2J} + \frac{1}{J^2}\left(\frac{1}{12} - \frac{\pi^2}{3(1-z)^{2}}\right) + \Op\left(\frac{1}{J^3}\right)\, ,
\eeq
after substituting eqs.~\eqref{eq:A-largeJ} and~\eqref{eq:PellS} into~\eqref{eq:genA} and performing the sum over $\ell_{A}$ in the large-$J$, small-$S$ limit.

Turning now to our general formula~\eqref{eq:fofell}, we find in the limit $g^2\rightarrow 0$,
\beq\label{eq:fLO}
f_{J}(\ell) = \sum_{k\, =\, 1}^{J/2-1} \frac{(-1)^{k+\ell+1}\Gamma(J)\Gamma(2k)\zeta_{2k+1}}{(2\pi)^{2k}\Gamma(-\ell)\Gamma(J-2k)\Gamma(1+\ell+2k)}\, .
\eeq
With its help we can evaluate $\A$ to leading order at small $S$. Using~\eqref{eq:A-to-F} and~\eqref{eq:F-to-f}, we obtain 
\beq\label{eq:generating-F}
\sum_{\ell_{A} \, =\, 1}^{\infty} z^{\ell_{A}} F_{J}(-\ell_{A}) = \widehat{f}_{J}(z) + z^{J} \widehat{f}_{J}(1/z)\, , 
\eeq
for the generating function, with
\beq
\widehat{f}_{J}(z) = \sum_{\ell_{A}\, =\, 1}^{\infty} z^{\ell_{A}} f_{J}(-\ell_{A})\, .
\eeq
By substituting the integral representation~\eqref{eq:ttozeta} for $\zeta_{2k+1}$ in eq.~\eqref{eq:fLO} and summing over $k$ and $\ell_{A}$, we can further write
\beq\label{eq:int-Fofz}
\widehat{f}_{J}(z) = \frac{z}{2(1-z)} \int_{0}^{\infty}\frac{dt}{e^{t}-1} \left[\left(1+(1-z) \frac{it}{2\pi}\right)^{J-1} + \left(1-(1-z)\frac{it}{2\pi}\right)^{J-1}-2\right] \, ,
\eeq
which is valid for any $J$ at weak coupling.

To evaluate~\eqref{eq:generating-F} at large $J$, we observe that only the first term, $\widehat{f}_{J}(z)$, on the right-hand side needs to be considered, as the second term, proportional to $z^{J}$, is exponentially suppressed when $z<1$. To proceed with $\widehat{f}_{J}(z)$, we rewrite the factors inside the integral~\eqref{eq:int-Fofz} as
\beq
\left(1\pm (1-z) \frac{it}{2\pi}\right)^{J-1} = D_{\tau}\cdot e^{\pm i \tau t}\, ,
\eeq
where
\beq\label{eq:tau-J}
\tau = \frac{(J-1)(1-z)}{2\pi}\, ,
\eeq
and $D_{\tau}$ is the differential operator
\beq
D_{\tau} = \exp{\left[- \sum_{n\, = \, 2}^{\infty} \frac{(J-1)}{n} \left(-\frac{\tau}{J-1}\right)^{n} \partial_{\tau}^{n}\right]}\, .
\eeq
The resulting integral over $t$ can be computed exactly and expressed in terms of $\psi$-functions,
\beq
\widehat{f}_{J}(z) = \frac{z}{2(1-z)} \, D_{\tau}\cdot (2\psi(1)-\psi(1+i\tau)-\psi(1-i\tau))\, .
\eeq
Finally, we can expand this expression at large $J$ to any desired order $\Op(1/J^{k})$ by truncating the sum in $D_{\tau}$ at $n=k+1$ and applying the asymptotic expansion for the $\psi$-functions at large $\tau\rightarrow \infty$, together with~\eqref{eq:tau-J}. In this way, it is straightforward to verify that
\beq
\widehat{f}_{J}(z) = \frac{z}{1-z}\, a_{J}(z)\, ,
\eeq
in perfect agreement with eq.~\eqref{eq:data-largeJ}.

\section{Regge limit}\label{app:Regge}

In this appendix, we test the ansatz~\eqref{eq:main-ansatz} in the Regge limit at strong coupling. This limit explores the behavior of structure constants when the spin $S$ is small, of order $\Op(1/\sqrt{\lambda})$, while the dimension $\Delta$ remains of order $\Op(1)$, and arbitrary.%
\footnote{Note that at strong coupling, the small-spin limit analyzed in Section~\ref{sec:hexagons} arises as a special case of this regime, characterized by $\Delta \rightarrow J$. This contrasts with weak coupling, where the Regge limit is instead dominated by spins given by $S = 1-J+\Op(\lambda)$, see~refs.~\cite{Ekhammar:2024neh,Klabbers:2023zdz,Gromov:2015vua} for recent studies.}
It connects to the physical regime of integer spins through an analytic continuation in the spin.

A detailed investigation of the Regge behavior of structure constants at strong coupling was recently carried out in ref.~\cite{Alday:2024xpq} through higher-loop calculations using a worldsheet-like representation for the four-point function. Our analysis will be more limited in scope, focusing on the one-loop correction. In addition, we will follow the conformal Regge theory developed in refs.~\cite{Costa:2012cb,Costa:2013zra,Costa:2017twz} to relate our results to the four-point function of chiral primary operators.

For simplicity, we focus on cases where two operators have length 2 and two have length $p\geqslant 2$. In this setup, after factoring out overall weight factors and color factor $1/N^2$, the connected part of the planar four-point function is described by a single function of the spacetime cross ratios, see e.g.~ref.~\cite{Binder:2019jwn},%
\footnote{To be precise, superconformal Ward identities require that $\G_{22pp} = \G^{\textrm{free}}_{22pp} +\I\,\mathcal{H}_{22pp}$, where $\G^{\textrm{free}}_{22pp}$ and $\I$ are known functions of the cross ratios and $SU(4)$ polarizations of the chiral primary operators, while $\mathcal{H}_{22pp}$ is a function of the cross ratios alone. However, in the Regge limit, $\G^{\textrm{free}}_{22pp}\rightarrow 0$ and $\I\rightarrow 1$, and therefore $\G_{22pp} \sim \mathcal{H}_{22pp}(U,V)$.}
\beq
\langle \textrm{Tr}\,  Z_{1}^{2}(x_1)\, \textrm{Tr}\, Z^{2}_{2}(x_2) \, \textrm{Tr}\, Z_{3}^{p}(x_{3}) \, \textrm{Tr}\, Z_{4}^{p}(x_{4}) \rangle_{\textrm{conn}} \propto \G_{22pp} (U, V)\, ,
\eeq
where
\beq
U = \frac{x_{12}^{2}x_{34}^2}{x_{13}^{2}x_{24}^{2}} = z\bar{z}\, , \qquad V = \frac{x_{14}^{2}x_{23}^{2}}{x_{13}^{2}x_{24}^{2}} = (1-z)(1-\bar{z})\, ,
\eeq
with $z$ and $\bar{z}$ being the usual cross-ratio variables. The Regge limit corresponds to taking $z, \bar{z} \rightarrow 0$ while keeping their ratio fixed. Using the variables introduced in ref.~\cite{Costa:2013zra}
\beq
z = \sigma e^{\rho}\, , \qquad \bar{z} = \sigma e^{-\rho}\, ,
\eeq
the Regge limit is realized as $\sigma \rightarrow 0$ with $\rho$ held fixed.

In Euclidean kinematics, where $z$ and $\bar{z}$ are complex conjugates, this limit is controlled by the standard Operator Product Expansion, with the lightest operators providing the dominant contribution,
\beq
\G_{22pp} \propto \sigma^2 \rightarrow 0\, .
\eeq
The situation becomes more interesting in Minkowskian kinematics. To access this domain, following~\cite{Costa:2013zra}, one must lift the reality condition on $z$ and $\bar{z}$ and analytically continue $z$ counterclockwise around the branch point at $z=1$, while keeping $\bar{z}$ fixed. In this process, the correlation function develops a singular behavior as $\sigma\rightarrow 0$, which is governed by the leading Regge trajectory of length-2 operators.

To be precise, according to conformal Regge theory~\cite{Costa:2012cb}, the correlation function in this limit is expressed as an integral over the imaginary scaling dimension, $\Delta = i\nu$, of the exchanged operator,
\beq
\G_{22pp} = 2\pi i \int_{-\infty}^{\infty} d\nu\, \sigma^{-1-S(\nu)} \alpha(\nu) \Omega_{i\nu}(\rho) + \ldots\, ,
\eeq
where the dots represent subleading contributions when $\sigma\rightarrow 0$. Here, $S(\nu)$ is the spin of the operator, treated as a function of its scaling dimension $\Delta=i\nu$, see eq.~\eqref{eq:Delta2},
\beq
S =-\frac{\nu^2+4}{2\sqrt{\lambda}}\left(1+\frac{1}{2\sqrt{\lambda}}\right) + \Op\left(\frac{1}{\lambda^{3/2}}\right)\, ,
\eeq
and
\beq\label{eq:Omega}
\Omega_{i\nu}(\rho) = \frac{\nu\sin{(\nu\rho)}}{4\pi^2\sinh{\rho}}\, .
\eeq
The information about the structure constants, $b_{2+S}$, is contained in the Regge residue~\cite{Costa:2013zra,Costa:2012cb}
\beq
\alpha(\nu) = -\frac{2^{S-1}\pi^{2} S' e^{\frac{i\pi S}{2}}}{\nu \sin{\left(\frac{\pi S}{2}\right)}} \gamma_{S}(\nu)\gamma_{S}(-\nu)K_{2+\Delta, 2+S}\, b_{2+S}\, ,
\eeq
where $S' = dS/d\nu$,
\beq\label{eq:gamma-Regge}
\gamma_{S}(\nu) = \Gamma\left(2 + \frac{S+i\nu}{2}\right) \Gamma\left(p+ \frac{S+i\nu}{2}\right)\, ,
\eeq
and $K$ is a ratio of Gamma functions that arise from the normalization of conformal blocks in $\nu$-space,
\beq\label{eq:K-factor}
K_{2+\Delta, 2+S} = \frac{\Gamma(\Delta+S+4)\Gamma(\Delta+S+3)}{4^{S+1}\Gamma\left(2+\frac{\Delta+S}{2}\right)^5 \Gamma\left(p+\frac{\Delta+S}{2}\right)\Gamma\left(2+\frac{S-\Delta}{2}\right)\Gamma\left(p+\frac{S-\Delta}{2}\right)}\, .
\eeq
Here, we have taken into account that the formula in~\cite{Costa:2012cb,Costa:2013zra} is written in terms of the quantum numbers and structure constants of the primary operator with maximal spin in a given supermultiplet. This operator has dimension $2+\Delta$ and spin $2+S$, where $\Delta$ and $S$ label the $sl(2)$ primary $\textrm{Tr}\, D^{S}Z^{2}$. The relation for the structure constants follows from the decomposition of superconformal blocks into conformal blocks~\cite{Dolan:2001tt}, leading to
\beq
b_{2+S} = \frac{2^{S-6}(\Delta+S)^2 (\Delta+S+2)^2}{(\Delta+S-1)(\Delta+S+1)^2 (\Delta+S+3)} \times C_{222} (S)C_{pp2} (S)\, ,
\eeq
where $C_{J_1 J_2 J}(S)$ denotes the structure constant of the $sl(2)$ primary in our normalization.

We can now re-express this data in terms of our $\D$-coefficients in~\eqref{eq:main-ansatz}. Remarkably, after combining all pieces, the Gamma functions in the $K$-factor~\eqref{eq:K-factor} cancel against those in the prefactor $\Gamma(AdS)$ in~\eqref{eq:main-ansatz}, leaving a simpler expression for the Regge residue,
\beq
K_{2+\Delta, 2+S}\, b_{2+S} = \frac{\D_{222}\D_{pp2}}{Z(p)(2\sqrt{\lambda})^{S}\Gamma\left(1+\frac{S}{2}\right)^2}\, ,
\eeq
where $Z(p)$ arises from the normalization of the structure constants and the sphere factor in~\eqref{eq:main-ansatz},
\beq
Z(p) = \frac{(p-1)!(p-2)!}{2}\, .
\eeq
Taking this into account, the Regge formula simplifies to
\beq\label{eq:Regge-formula}
\mathcal{G}_{22pp} = \frac{\pi^2 i\sqrt{\lambda}}{Z(p)}\int_{-\infty}^{\infty}\frac{S' d\nu}{\nu} \frac{\gamma_{S}(\nu) \gamma_{S}(-\nu)\Omega_{i\nu}(\rho)}{(\sqrt{\lambda\,}\sigma)^{1+S}} \frac{e^{\frac{i\pi S}{2}}\Gamma\left(-\frac{S}{2}\right)}{\Gamma\left(1+\frac{S}{2}\right)} \D_{222}\D_{pp2}\, ,
\eeq
where $S=S(\nu)$ throughout the equation.

We now expand this expression to one loop at strong coupling. Since the spin is small, the $\D$-coefficients simplify as $\D = 1+\Op(1/\lambda)$. Expanding the remaining terms, we get
\beq
\mathcal{G}_{22pp} = \mathcal{G}^{\textrm{LO}}_{22pp} + \frac{1}{\sqrt{\lambda}} \mathcal{G}_{22pp}^{\textrm{NLO}} + \Op\left(\frac{1}{\lambda}\right)\, ,
\eeq
where
\beq
\mathcal{G}^{\textrm{LO/NLO}}_{22pp} = -\frac{(2\pi)^2 i}{Z(p) \sigma}\int d\nu \, \Omega_{i\nu} (\rho) \gamma_{0}(\nu) \gamma_{0}(-\nu) f^{\textrm{LO/NLO}}\, ,
\eeq
with
\beq
f^{\textrm{LO}} = \frac{1}{\nu^2+4}\, ,
\eeq
and 
\beq\label{eq:f-NLO}
f^{\textrm{NLO}} = \frac{1}{4}\left[\log{\big(\sqrt{\lambda}\, \sigma\big)} - \frac{i\pi}{2}-\gamma_{\textrm{E}} - \psi\left(2+\frac{i\nu}{2}\right) - \psi\left(p+\frac{i\nu}{2}\right) + (\nu\rightarrow -\nu)\right]\, .
\eeq
These expressions can be compared with known results for correlation functions at strong coupling. In particular, the leading-order term, $\G^{\textrm{LO}}$, correctly reproduces the Regge limit of the SUGRA correlator studied in ref.~\cite{Costa:2012cb}.

The situation becomes more intriguing for the next term, $\G^{\textrm{NLO}}$, which corresponds to stringy corrections. To establish a match in this case, we must resum an infinite series of corrections to the SUGRA correlator. This requirement is indicated by the presence of the logarithmic term $\sim \log{(\sqrt{\lambda}\, \sigma)}$ in $f^{\textrm{NLO}}$. More generally, the appearance of the factor $(\sqrt{\lambda}\, \sigma)^{-S}$ in eq.~\eqref{eq:Regge-formula} suggests that the Regge limit at strong coupling demands not only $\sigma \ll 1$ but the stricter condition $\sigma \ll 1/\sqrt{\lambda} \ll 1$.

To perform the required summation of the stringy corrections, we can use the Mellin amplitude $\mathcal{M}_p(s, t)$ given in ref.~\cite{Binder:2019jwn}. The relevant regime corresponds then to the double-scaling limit where $t, \sqrt{\lambda} \rightarrow \infty$ with $t/\sqrt{\lambda}$ and $s$ held fixed. Explicit expressions in this limit follow from the mapping to the Virasoro-Shapiro amplitude and can be written in terms of differential operators acting on the SUGRA correlator $\G^{\textrm{LO}}$. The final result is expressed as a sum over $\nu$-integrals,
\beq\label{eq:G-string}
\G^{\textrm{DS}}_{22pp} \approx -\frac{2\pi^2 i}{Z(p)}\sum_{k\, =\, 1}^{\infty} \left(\frac{1}{\xi}\right)^{1+2k} \zeta(1+2k) \int_{-\infty}^{\infty} d\nu \, \Omega_{i\nu}(\rho)\gamma_{2k}(\nu)\gamma_{2k}(-\nu)\, ,
\eeq
where $\xi = \sqrt{\lambda}\, \sigma$ is kept fixed as $\sqrt{\lambda}, \sigma \rightarrow \infty$, and $\Omega$ and $\gamma$ are defined in eqs.~\eqref{eq:Omega} and~\eqref{eq:gamma-Regge}.

The $k$-th term in this sum originates from the stringy correction to $\mathcal{M}_p(s,t)$, which scales as $t^{2k}/(\sqrt{\lambda})^{2k+1}$ at large $t$. It represents contributions from operators of spin $2k$. However, unlike the actual Regge limit, which is dominated by the leading Regge trajectory, the $k$-th coefficient in eq.~\eqref{eq:G-string} incorporates contributions from an infinite number of trajectories.

Agreement with the conformal Regge formula~\eqref{eq:Regge-formula} is found in the stricter limit $\xi = \sqrt{\lambda}\, \sigma \rightarrow 0$. To see that, we apply the Sommerfeld-Watson transformation to convert the sum into an integral in eq.~\eqref{eq:G-string},
\beq
\sum_{k\, =\, 1}^{\infty} f(k) = \int\limits_{\epsilon-i\infty}^{\epsilon+i\infty} \frac{ie^{i\pi k}dk}{2\sin{\pi k}} f(k)\, ,
\eeq
where the contour runs parallel to the imaginary axis with a small real part, $\epsilon>0$. The leading behavior at $\xi = 0$ originates from the (double) pole at $k = 0$. By evaluating the residue at this point, we obtain an exact agreement with the NLO Regge formula~\eqref{eq:f-NLO}.

\section{Classical limit}\label{app:classical}

In this appendix, we present the expressions that we used to perform calculations through fourth order at small spin in the classical limit.

\subsection{Resolvent at small spin}

As explained earlier, the resolvent can be expressed as a series in integer powers of $\S$ at small spin by expanding the integral~\eqref{eq:int-res} around $a, b =  \alpha$,
\beq
\R(x) = \sum_{k=1}^{\infty}\, \S^{k}\, \R^{(k)}(x)\, .
\eeq
The first two terms in this expansion are given in eq.~\eqref{eq:R12}. The next two terms are more bulky. They read
\beq
\begin{aligned}
&\R^{(3)} = \frac{2\pi x\alpha^4}{(\alpha^2-1)^5(\alpha^2+1)^5 (x^2-\alpha^2)^5}\\
&\,\,\, \times \big[\,\big(\alpha^{10}+x^{8}\big) \big(3 + 34 \alpha^2 + 157 \alpha^4 + 124 \alpha^6 + 
   157 \alpha^8 + 34 \alpha^{10} + 3 \alpha^{12}\big) \\
& \qquad -x^2  \alpha^6  \big(-13 - 5 \alpha^2 + 275 \alpha^4 + 835 \alpha^6 + 417 \alpha^8 + 297 \alpha^{10} + 217 \alpha^{12} + 25\alpha^{14}\big) \\
& \qquad +x^4  \alpha^4  \big(73 + 111 \alpha^2 + 189 \alpha^4 + 1163 \alpha^6 + 1163 \alpha^8 + 189 \alpha^{10} + 111 \alpha^{12} + 73 \alpha^{14}\big) \\
& \qquad -x^6  \alpha^2  \big(25 + 217 \alpha^2 + 297 \alpha^4 + 417 \alpha^6 + 835 \alpha^8 + 275 \alpha^{10} - 5 \alpha^{12} - 13  \alpha^{14}\big)\, \big]\, ,\\
&R^{(4)} = -\frac{\pi x  \alpha^5}{2(\alpha^2-1)^7 (\alpha^2+1)^{8}(x^2 - \alpha^2)^7}\\
&\,\,\, \times \big[\,\big(\alpha^{14}+x^{12}\big) \big(21 + 342 \alpha^2 + 2689 \alpha^4 + 10536 \alpha^6 + 13674 \alpha^8 + 27396 \alpha^{10} \\
&\,\,\,\,\,\, \qquad \qquad + 13674 \alpha^{12} + 10536 \alpha^{14} + 2689 \alpha^{16} + 342 \alpha^{18} + 21 \alpha^{20}\big)\\
&\qquad - x^2\alpha^{10} \big(1 + 107 \alpha^2 + 2147 \alpha^4 + 20353 \alpha^6 + 
 73514 \alpha^8 + 94222 \alpha^{10} \\
&\,\,\,\,\,\, \qquad \qquad + 161318 \alpha^{12} + 
 77986 \alpha^{14} + 38053 \alpha^{16} + 20087 \alpha^{18} + 
 3495 \alpha^{20} + 237 \alpha^{22}\big) \\
 &\qquad + x^4\alpha^{8} \big(-1167 - 3288 \alpha^2 + 13145 \alpha^4 + 88042 \alpha^6 + 
 208498 \alpha^8 + 204872 \alpha^{10} \\
&\,\,\,\,\,\, \qquad \qquad + 363178 \alpha^{12} + 
 257948 \alpha^{14} + 59581 \alpha^{16} + 24704 \alpha^{18} + 12125 \alpha^{20} + 1162 \alpha^{22}\big) \\
 &\qquad - 2x^6\alpha^{6} (1+\alpha^2) \big(1169 + 350 \alpha^2 + 2573 \alpha^4 + 55048 \alpha^6 + 
 127330 \alpha^8 \\
&\,\,\,\,\,\, \qquad \qquad + 36660 \alpha^{10} + 127330 \alpha^{12} + 
 55048 \alpha^{14} + 2573 \alpha^{16} + 350 \alpha^{18} + 
 1169 \alpha^{20}\big) \\
 &\qquad + x^8\alpha^{4} \big(1162 + 12125 \alpha^2 + 24704 \alpha^4 + 59581 \alpha^6 + 
  257948 \alpha^8 + 363178 \alpha^{10} \\
&\,\,\,\,\,\, \qquad \qquad + 204872 \alpha^{12} + 208498 \alpha^{14} + 88042 \alpha^{16} + 13145 \alpha^{18} - 3288 \alpha^{20} - 1167 \alpha^{22}\big) \\
&\qquad - x^{10}\alpha^{2} \big(237 + 3495 \alpha^2 + 20087 \alpha^4 + 38053 \alpha^6 + 
 77986 \alpha^8 + 161318 \alpha^{10} \\
&\,\,\,\,\,\, \qquad \qquad + 94222 \alpha^{12} + 73514 \alpha^{14} + 20353 \alpha^{16} + 2147 \alpha^{18} + 107 \alpha^{20} + \alpha^{22}\big)\, \big] \, ,
\end{aligned}
\eeq
with $\alpha = \J+\sqrt{1+\J^2}$.

\subsection{Differential operators}

The spin-suppressed corrections to the classical structure constant can be constructed using differential operators, as shown in eq.~\eqref{eq:diff-op}. The first non-trivial operator is given in eq.~\eqref{eq:Delta-diff-one}. The following two operators are given by
\beq
\begin{aligned}
&\Delta^{(3)}(\partial_{\J}, \partial_{\L}) \\
&\,\, = \left[\frac{1+\J^2}{48} (\partial_{\J}+\partial_{\L})\partial_{\J} + \frac{6 + 7 \J^2}{24 \J} \left(\partial_{\J}+\frac{1}{2}\partial_{\L}\right) + \frac{8 + 22 \J^2 + 13 \J^4}{16 \J^2 (1 + \J^2)}\right] (\partial_{\J}+\partial_{\L})\partial_{\J} \\
&\,\,\,\,\,\,\,\, + \frac{3 + 4 \J^2}{24 \J^2} \partial_{\L}^2 - \frac{2+\J^2}{4\J^3(1+\J^2)} \left(\partial_{\J}+\frac{1}{2}\partial_{\L}\right)\, , \\
&\Delta^{(4)}(\partial_{\J}, \partial_{\L})\\
&= \frac{(1+\J^2)^{3/2}}{1152} (\partial_{\J}+\partial_{\L})^{3}\partial_{\J}^{3} + \frac{\sqrt{1 + \J^2} (4 + 5 \J^2)}{192 \J} (\partial_{\J}+\partial_{\L})^{2}\partial_{\J}^{2}\left(\partial_{\J}+\frac{1}{2}\partial_{\L}\right) \\
& + \frac{48 + 137 \J^2 + 86 \J^4}{384 \J^2 \sqrt{1 + \J^2}}(\partial_{\J}+\partial_{\L})^{2}\partial_{\J}^{2} + \frac{12 + 31 \J^2 + 19 \J^4}{384 \J^2 \sqrt{1 + \J^2}}(\partial_{\J}+\partial_{\L})\partial_{\J}\partial_{\L}^{2} \\
& + \frac{8 + 132 \J^2 + 103 \J^4}{192 \J^3 \sqrt{1 + \J^2}}(\partial_{\J}+\partial_{\L})\partial_{\J}\left(\partial_{\J}+\frac{1}{2}\partial_{\L}\right) + \frac{\sqrt{1 + \J^2} (1 + 2 \J^2)}{24 \J^3} \left(\partial_{\J}+\frac{1}{2}\partial_{\L}\right)\partial_{\L}^2 \\
& - \frac{64 + 160 \J^2 + 133 \J^4 + 32 \J^6 - \J^8}{128 \J^4 (1 + \J^2)^{5/2}} (\partial_{\J}+\partial_{\L})\partial_{\J} - \frac{2 + \J^2}{16 \J^4 \sqrt{1 + \J^2}} \partial_{\L}^2  \\
& + \frac{8 + 20 \J^2 + 13 \J^4 + 3 \J^6}{16 \J^5 (1 + \J^2)^{5/2}} \left(\partial_{\J}+\frac{1}{2}\partial_{\L}\right)\, .
\end{aligned}
\eeq
They are easily seen to obey the commutation relation~\eqref{eq:FCR}. One can also check equation~\eqref{eq:Delta-logGamma}, using the known expression for the energy of a short classical string~\cite{Frolov:2002av}
\beq
\E = \J + \delta_{1} \, \S + \delta_{2} \, \S^{2} + \delta_{3}\, \S^{3} + \delta_{4}\, \S^{4} + \ldots\, ,
\eeq
with
\beq
\begin{aligned}
&\delta_{1} = \frac{\sqrt{1+\J^2}}{\J}\, , \qquad \delta_{2} = -\frac{2+\J^2}{4\J^{3}(1+\J^{2})}\, ,\\
&\delta_{3} = \frac{8 + 20 \J^2 + 13 \J^4 + 3 \J^6}{16 \J^5 (1 + \J^2)^{5/2}}\, ,\\
&\delta_{4} = -\frac{80 + 336 \J^2 + 540 \J^4 + 385 \J^6 + 138 \J^8 + 21 \J^{10}}{128 \J^7 (1 + \J^2)^4}\, .
\end{aligned}
\eeq

\section{Normalization factor at strong coupling}\label{App:normalization}

In this appendix, we provide the details of the computation of the normalization factor through two loops at strong coupling.

\subsection{Gaudin norm}

To get the roots and the Gaudin determinant at strong coupling, we need to expand the Bethe equations~\eqref{eq:Bethe-Ansatz}. The momentum $p_k$ in~\eqref{eq:momentum} is easily expanded,
\beq
p_{k} = \frac{4\pi x_k}{\sqrt{\lambda}(x_k^2-1)} -\frac{16\pi^3 x_k^3 (x_k^{4} + 4 x_k^2 + 1)}{3\lambda^{3/2} (x_k^2-1)^5} + \Op\left(\frac{1}{\lambda^{5/2}}\right)\, ,
\eeq
using
\beq\label{eq:xpm-sc}
x_k^{\pm} = x_k \pm \frac{2\pi i x_k^2}{\sqrt{\lambda}(x_k^2-1)} + \frac{4\pi^2 x_k^3}{\lambda (x_k^2-1)^{3}} \mp \frac{8\pi^3i x_k^4 (x_k^2+1)}{\lambda^{3/2} (x_k^2-1)^5} + \Op\left(\frac{1}{\lambda^{2}}\right)\, .
\eeq
For the S-matrix, we have the general expression~\cite{Beisert:2005fw} 
\beq\label{eq:Smatrix}
S_{kj} = \frac{u_k-u_j+i}{u_k-u_j-i} \left(\frac{1-1/x_{k}^{-}x_{j}^{+}}{1-1/x_{k}^{+}x_{j}^{-}}\right)^{2} \frac{1}{\sigma_{kj}^{2}}\, ,
\eeq
where $\sigma_{kj}$ is the dressing phase~\cite{Beisert:2006ez}. The factor that depends on the difference between rapidities $(u_k-u_j)$ requires special treatment, as it produces poles at $x_k = x_j$,
\beq\label{eq:singular-sum}
-i \log{\left[\frac{u_k-u_j+i}{u_k-u_j-i}\right]} = 2\sum_{n\, =\, 1}^{\infty} \frac{(-1)^{n+1}}{2n-1}\left[\frac{4\pi}{\sqrt{\lambda}\, (x_{k}-x_{j})(1-1/x_{k}x_{j})}\right]^{2n-1}\, .
\eeq
These poles lead to an enhancement of the loop corrections for roots with the same mode numbers and to the emergence of the parameter $1/\lambda^{1/4}$ in the solution,
\beq\label{eq:enhacement}
\frac{1}{\sqrt{\lambda}\, (x_{k}-x_{j})} = \Op\left(\frac{1}{\lambda^{1/4}}\right)\, ,
\eeq
Therefore, to find the roots up to $x_{k}^{(L+1/2)}$ in eq.~\eqref{eq:exp-x} and construct the Gaudin determinant~\eqref{eq:Gaudin-G} at $L$ loops, we must retain all terms up to $n = L+1$ in the sum~\eqref{eq:singular-sum}. In particular, at two loops ($L=2$), we need the terms with $n=1,2,3$. No such precaution is needed for the remaining factors in~\eqref{eq:Smatrix}, as they are smooth at $x_k = x_j$. We obtain
\beq\label{eq:easy-part-S}
-2i \log{\left(\frac{1-1/x_{k}^{-}x_{j}^{+}}{1-1/x_{k}^{+}x_{j}^{-}}\right)} = \frac{8\pi (x_k - x_j) (x_k x_j+1)}{\sqrt{\lambda}(x_k^2 -1) (x_k x_j-1)(x_j^2 - 1)} + \Op\left(\frac{1}{\lambda^{3/2}}\right)\, ,
\eeq
and
\beq
2i \log{\sigma_{kj}} = \frac{\delta^{\textrm{AFS}}_{kj}}{\sqrt{\lambda}} + \frac{\delta^{\textrm{HL}}_{kj}}{\lambda} + \Op\left(\frac{1}{\lambda^{3/2}}\right)\, ,
\eeq
where $\delta_{kj}^{\textrm{AFS}}$ and $\delta_{kj}^{\textrm{HL}}$ originate from the Arutyunov-Frolov-Staudacher (AFS) and the Hernández-López (HL) dressing phases, respectively. They are given by~\cite{Arutyunov:2004vx,Hernandez:2006tk}
\beq\label{eq:HLphase}
\begin{aligned}
\delta^{\textrm{AFS}}_{kj} &= -\frac{8\pi(x_{k}-x_{j})}{(x_k^{2}-1)(x_{k}x_{j}-1)(x_{j}^2-1)}\, , \\
\delta^{\textrm{HL}}_{kj} &= -\frac{16\pi x_{k}^2 x_{j}^2}{(x_{k}^2-1)(x_{j}^2-1)}\bigg[\frac{2}{(x_{k}-x_{j})(x_{k}x_{j}-1)} \\
&\qquad\qquad\qquad\qquad\,\,\,\,\,\,\,\,\, +\left(\frac{1}{(x_{k}-x_{j})^2}+\frac{1}{(x_{k}x_{j}-1)^{2}}\right)\log{\frac{(x_{k}+1)(x_{j}-1)}{(x_{k}-1)(x_{j}+1)}}\bigg]\, .
\end{aligned}
\eeq
Lastly, for the prefactor $H$ in eq.~\eqref{eq:prefactorH}, we should use
\beq
\frac{1}{2}\log{\frac{(u_k-u_j)^2}{(u_k-u_j)^2 + 1}} \approx \sum_{n\, =\, 1}^{L} \frac{(-1)^n}{2n} \left[\frac{4\pi}{\sqrt{\lambda}\, (x_{k}-x_{j})(1-1/x_{k}x_{j})}\right]^{2n}\, ,
\eeq
at $L$ loops. The remaining factor in~\eqref{eq:prefactorH} is smooth at $x_k = x_j$ and is therefore unaffected by the enhancement~\eqref{eq:enhacement}. For it, we find
\beq
\log{\frac{\left(x_{k}^{+}x_{j}^{-}-1\right)\left(x_{k}^{-}x_{j}^{+}-1\right)}{\left(x_{k}^{+}x_{j}^{+}-1\right)\left(x_{k}^{-}x_{j}^{-}-1\right)}} = -\frac{16\pi^2 x_k^2 x_j^2}{\lambda (x_k^2-1)(x_k x_j-1)^2(x_j^2-1)}+ \Op\left(\frac{1}{\lambda^{2}}\right)\, ,
\eeq
using~\eqref{eq:xpm-sc}.

After solving the Bethe equations recursively to the desired order for low spin values and substituting the solution into the Gaudin determinant~\eqref{eq:Gaudin-G}, we find that the result admits the form given in eq.~\eqref{eq:Gexp}, with the first two polynomials given by
\beq\label{eq:PG}
\begin{aligned}
&P^{G}_{1}(S) = \\
&\left[\frac{4 + 3 \J^2 + 2 \J^4}{8  \J^2  (1 + \J^2)^{3/2}} + \frac{\sqrt{1 + \J^2} \pi^2}{12  \J^2}\right] S - \left[\frac{12 + 17 \J^2 + 8 \J^4}{16 \J^2 (1 + \J^2)^{3/2}} + \frac{\sqrt{1 + \J^2} \pi^2}{24 \J^2}\right] S^2\, , \\
&P^{G}_{2}(S) =\\
&-\left[\frac{24 + 88 \J^2 + 153 \J^4 + 12 \J^6 - 4 \J^8}{96  \J^4  (1 + \J^2)^3} + \frac{(16 + 23 \J^2 + 6  \J^4) \pi^2}{24 \J^4 (1 + \J^2)} - \frac{(1 + \J^2) \pi^4}{360  \J^4}  \right] S \\
& - \left[\frac{32 + 104 \J^2 + 79 \J^4 + 76 \J^6 + 25 \J^8}{64 \J^4 (1 + \J^2)^3} + \frac{\pi^2}{48  \J^2} + \frac{7 (1 + \J^2) \pi^4}{720  \J^4}\right] S^2\\
&+ \left[\frac{240 + 784 \J^2 + 834 \J^4 + 456 \J^6 + 107 \J^8}{384 \J^4 (1 + \J^2)^3} + \frac{(4 + 6 \J^2 + 3 \J^4) \pi^2}{96 \J^4 (1 + \J^2)} + \frac{(1 + \J^2) \pi^4}{240\J^4}\right] S^3 \\
& + P_{2}^{\textrm{HL}}(S)\, .
\end{aligned}
\eeq
The polynomial $P_{2}^{\textrm{HL}}(S)$ is the contribution coming from the HL dressing phase~\eqref{eq:HLphase}, which kicks in at two loops,
\beq
\begin{aligned}
P_{2}^{\textrm{HL}}(S) &= \frac{S}{3 \J^4 \sqrt{1 + \J^2}} \\
&\,\,\,\, + \left[-\frac{2 + 11 \J^2 + 3 \J^4}{12 \J^4 (1 + \J^2)^{3/2}} + \frac{4 + 3 \J^2 + \J^4}{4  \J^2 (1 + \J^2)^2} \log{\left(\frac{1 + \sqrt{1 + \J^2}}{\J}\right)}\right] S^2\, .
\end{aligned}
\eeq
Similarly, the prefactor $H$ in~\eqref{eq:prefactorH} takes the form~\eqref{eq:Gexp}, with the polynomials
\beq\label{eq:PH}
\begin{aligned}
&P^{H}_{1} = \frac{\sqrt{1 + \J^2} \pi^2}{4\J^2} S - \frac{\sqrt{1 + \J^2} \pi^2}{8\J^2} S^2\, , \\
&P^{H}_{2} = -\left[\frac{(6 + 9 \J^2 + 2 \J^4) \pi^2}{8 \J^4 (1 + \J^2)} - \frac{(1 + \J^2) \pi^4}{72 \J^4}\right] S -\left[\frac{(2 + 3 \J^2) \pi^2}{16 \J^4} + \frac{(1 + \J^2)\pi^4}{48  \J^4}\right] S^2 \\
&\qquad \,\,\,\,\,+\left[\frac{(4 + 6 \J^2 + 3 \J^4) \pi^2}{32 \J^4 (1 + \J^2)} + \frac{(1 + \J^2) \pi^4}{144\J^4}\right] S^3\, .
\end{aligned}
\eeq

\subsection{Wrapping corrections}

According to the proposal in~\cite{Basso:2022nny}, when wrapping corrections are included, the Gaudin determinant is replaced by a Fredholm determinant corresponding to the infinite system of Thermodynamic Bethe Ansatz (TBA) equations for the excited operator.%
\footnote{The formula for the normalization factor in ref.~\cite{Basso:2022nny} also includes corrections to the prefactor $H$. However, these corrections are explicitly suppressed by the Y functions, which decay as $S^2$ at small spin and can therefore be disregarded.} The evaluation of this determinant is generally challenging due to the complex structure of the TBA equations. However, at small spin, the determinant is expected to simplify, and its expansion truncates to the first wrapping corrections.

This expectation is justified by the fact that higher-wrapping contributions involve the exchange of more mirror magnons, all of which are supertraced over their respective flavors. Since these traces are power-suppressed at small spin, processes involving a single mirror magnon should dominate in this regime, as in Section~\ref{sec:hexagons}. Therefore, to leading order at small spin, we can rely on the leading Lüscher formula from ref.~\cite{Basso:2017muf}, which is significantly simpler to study than the full Fredholm determinant.

This formula predicts two type of wrapping corrections,
\beq\label{eq:WFphi}
W = W^{F} + W^{\Phi}\, .
\eeq
The first one, $W^{F}$, directly arises from the expansion of the Fredholm determinant. It takes the form of an integral over the rapidity of a mirror magnon, with the integrand expressed in terms of the (derivative of the) S-matrix,
\beq
W^{F} = \frac{1}{2}\sum_{a\, = \, 1}^{\infty} \int \frac{du}{2\pi} e^{-J\Em_{a}(u)}\, \textrm{STr}\, \mathbb{K}_{aa}(u, u; \textbf{u})\, .
\eeq
Here,
\beq
\mathbb{K}_{ab}(u, v; \textbf{u}) = -i\, \mathbb{S}_{ba}(v, u)\partial_{u} \mathbb{S}_{ab}(u, v) \prod_{k=1}^{S} \mathbb{S}_{a1}(u, u_{k})\mathbb{S}_{b1}(v, u_{k})\, ,
\eeq
where $\mathbb{S}_{ab}(u, v)$ is the S-matrix for two mirror magnons, with rapidities $u, v$ and spins $a, b$, and $\mathbb{S}_{a1}(u, u_k)$ is the S-matrix between a mirror magnon and a Bethe root.
The integrand involves a supertrace `$\textrm{STr}$' over the flavors of the mirror magnon, similar to the one entering the definition of transfer matrices. Evaluating this trace, using the algorithm in~\cite{Basso:2017muf}, and taking the small spin limit, we find that the integral is linear in $S$,
\begin{equation}
W^F = W_{1}^{F} S + \Op\left(S^2\right)\, ,
\end{equation}
with
\begin{equation}\label{eq:int-deltaG}
\begin{aligned}
W_{1}^{F} = \sum_{a\, =\, 1}^{\infty} \int \frac{du}{2\pi i} \frac{1}{(x^{[+a]}x^{[-a]})^{J}}
&\bigg[\frac{\Sigma_{+}}{a} \left(\frac{x^{[+a]}-1/x^{[+a]}}{x^{[-a]}-1/x^{[-a]}} +\frac{x^{[-a]}-1/x^{[-a]}}{x^{[+a]}-1/x^{[+a]}}\right) \\
& - \frac{iJ\Sigma_{-}}{g} \left(\frac{1}{x^{[+a]}-1/x^{[+a]}} + \frac{1}{x^{[-a]}-1/x^{[-a]}}\right)\bigg]\, .
\end{aligned}
\end{equation}
Here,
\begin{equation}\label{eq:Sigma-pm}
\Sigma_{\pm} = \sum_{n\, =\, 1}^{\infty} \frac{2\pi I_{J+2n-1}(\sqrt{\lambda})}{J I_{J}(\sqrt{\lambda})} \left[\left(x^{[+a]}\right)^{1-2n}\mp \left(x^{[-a]}\right)^{1-2n}\right]\, ,
\end{equation}
with $I$ the modified Bessel function. As written, formula~\eqref{eq:int-deltaG} is valid at any coupling $g = \sqrt{\lambda}/(4\pi)$. However, here we focus solely on the strong coupling regime, $g\rightarrow \infty$.

When $\sqrt{\lambda} \rightarrow \infty$ with $J \sim \sqrt{\lambda}$, the integral~\eqref{eq:int-deltaG} is dominated by rapidities $u \in [-2g, 2g]$. In terms of the Zhukovsky variable~\eqref{eq:x-u}, we have $x\in U^{-}$, where $U^{-}$ is the lower half of the unit circle. To express the integrand in this variable, we first apply the transformation $x^{[+a]} \rightarrow 1/x^{[+a]}$ throughout and then expand at strong coupling, using
\beq
x^{[\pm a]} = x \pm \frac{2\pi ia x^2}{\sqrt{\lambda}(x^2-1)} + \frac{4\pi^2 a^2 x^3}{\lambda (x^2-1)^{3}} \mp \frac{8\pi^3ia^3 x^4 (x^2+1)}{\lambda^{3/2} (x^2-1)^5} + \Op\left(\frac{1}{\lambda^{2}}\right) \, .
\eeq
For the sum in~\eqref{eq:Sigma-pm}, we use the expression for the generating function of the ratios of Bessel $I$ functions, in the limit $J, \lambda \rightarrow \infty$, with $\J = J/\sqrt{\lambda}$ fixed. It reads
\begin{equation}
\sum_{n\, =\, 1}^{\infty} \frac{I_{J+2n-1}(\sqrt{\lambda})}{I_{J}(\sqrt{\lambda})} x^{1-2n} = \sum_{k\, =\, 0}^{\infty} \frac{r^{(k)}(x)}{(\sqrt{\lambda})^{k}}\, ,
\end{equation}
where the first three coefficients are given by
\begin{equation}
\begin{aligned}
&r^{(0)}(x) = \frac{\alpha x}{\alpha^2 x^2-1}\, , \\
&r^{(1)}(x) = - \frac{2\alpha^2 x}{(1+\alpha^2)^2 (\alpha^2 x^2-1)^3} [1+ \alpha^6 x^4 + 3\alpha^2 x^2(1+\alpha^2)]\, , \\
&r^{(2)}(x) = \frac{4\alpha^3 x}{(1+\alpha^2)^{5}(\alpha^2 x^2-1)^{5}} \\
& \qquad\,\,\,\,\,\,\,\, \times [ (1+\alpha^{10}x^{8})(1-3\alpha^2+\alpha^4) +\alpha^2 x^2(20+37\alpha^2+9\alpha^4+2\alpha^6) \\
& \qquad\qquad\,\, + 25\alpha^4 x^{4}(1+\alpha^2(2+\alpha^2)^2)+\alpha^{6}x^6 (2+9\alpha^2+37\alpha^4+20\alpha^6)]\, ,
\end{aligned}
\end{equation}
with $\alpha = \J + \sqrt{1+\J^2}$.

At strong coupling, the term involving $\Sigma_{+}$ dominates in~\eqref{eq:int-deltaG}, leading to
\beq
W^{F}_{1} = -\frac{1}{\J} \int_{U^{-}} \frac{dx (x^2-1)}{2\pi i x^2} \sum_{a\, = \, 1}^{\infty} \frac{e^{-2\pi a \J E(x)}}{a} \left(r^{(0)}(1/x) - r^{(0)}(x)\right) + \Op\left(\frac{1}{\sqrt{\lambda}}\right)\, ,
\eeq
with $E = -2ix/(x^2-1)$. Substituting the expression for $r^{(0)}(x)$ in this integrand and comparing the result with the integral $I[\L]$ in eq.~\eqref{eq:intIL}, we obtain
\beq
W^{F}_{1} = -2 I[\J] + \Op\left(\frac{1}{\sqrt{\lambda}}\right)\, .
\eeq
The loop corrections are obtained by expanding the various ingredients to higher orders. Ultimately, we find that these corrections can be expressed in terms of linear operators acting on $I[\L]$,
\beq
W^{F}_{1} = -2\sum_{k\, = \, 0}^{\infty} \frac{1}{(\sqrt{\lambda})^{k}} \left[\widehat{\Delta}_{F}^{(k)} I[\L] \right]_{\L\, =\, \J}\, ,
\eeq
with $\widehat{\Delta}_{F}^{(0)} = 1$. The first two operators are given by
\beq\label{eq:hatDeltaF}
\begin{aligned}
&\widehat{\Delta}_{F}^{(1)}\\
&= -\frac{\sqrt{1+\J^2}}{2} (\partial_{\J}+\partial_{\L})\partial_{\J} - \frac{\sqrt{1+\J^2}}{\J}\left(\partial_{\J}+\frac{1}{2}\partial_{\L}\right) + \frac{1}{2\J} \left(\J\sqrt{1+\J^2} -\mathcal{R}\right) \partial_{\L}^{2}\, , \\
&\widehat{\Delta}_{F}^{(2)} \\
&= \frac{1+\J^2}{8}\left(\partial_{\J}^{3}+2\partial_{\J}^{2}\partial_{\L}-\partial_{\J}\partial_{\L}^2-2\partial_{\L}^3\right)\partial_{\J}+\frac{3+5\J^2}{12\J}(2\partial_{\J}+3\partial_{\L})\partial_{\J}^2 +\frac{3}{8\J}\partial_{\J}\partial_{\L}^{2}\\
&\,\,\,\,\,+\frac{3\L+2\J^2\L-6\J^3}{24\J^2}\partial_{\L}^{3} + \frac{5+4\J^2}{4(1+\J^2)}(\partial_{\J}+\partial_{\L})\partial_{\J}+ \frac{2+4\J^2+3\J^4}{4\J^2(1+\J^2)}\partial_{\L}^2\\
&\,\,\,\,\, +\frac{1}{2\J(1+\J^2)}\left(\partial_{\J}+\frac{1}{2}\partial_{\L}\right)-\frac{\sqrt{1+\J^2}}{8\J^3} \left((2\J+\L)\partial_{\L}-1\right)\partial_{\L}^2\mathcal{R}\, ,
\end{aligned}
\eeq
where the operator $\mathcal{R}$ extracts the residue at $\J = 0$, see eq.~\eqref{eq:Res-op}.

The second term, $W^{\Phi}$, in~\eqref{eq:WFphi} arises from the fact that the wrapping corrections also modify the Bethe equations and, consequently, the Bethe roots. This effect can be incorporated as a shift in the right-hand side of the equation~\eqref{eq:Bethe-eqs-lin} by a phase $\Phi_{k}/(2\pi)$. It is described at leading order by the Bajnok-Janik formula~\cite{Bajnok:2008bm,Bajnok:2008qj}
\beq\label{eq:Phi-BJ}
\Phi_k = -\sum_{a\, =\, 1}^{\infty} \int \frac{du}{2\pi} \,  e^{-J \Em_{a}(u)}\, \textrm{STr}\left[ \mathbb{S}_{a1}(u, u_{1}) \ldots \partial_{u}\mathbb{S}_{a1}(u, u_{k}) \ldots \mathbb{S}_{a1}(u, u_{S}) \right] + \ldots\, .
\eeq
Dots represent higher-order wrapping corrections, which become increasingly suppressed at strong coupling. Morover, since these corrections involve additional mirror magnons and more graded traces, we also expect them to be further suppressed at small spin.

The phase $\Phi_{k}$ should be incorporated into the Gaudin determinant $G$ by adding $\partial_{u_j}\Phi_{k}/(2\pi)$ in eq.~\eqref{eq:Gaudin-G}. In principle, one should also correct the roots, both in $G$ and in $H$. However, since $\Phi_{k} = \Op(S)$, these corrections are subleading at small spin. The derivatives $\partial_{u_j}\Phi_{k}=\Op(1)$ and therefore their effect cannot be neglected.

Taking this into account and evaluating the result at strong coupling, we find that $\Phi$ introduces a correction of order $\Op(S)$ to the normalization factor,
\beq
W^{\Phi} = W^{\Phi}_{1} \, S + \Op\left(S^2\right)\, ,
\eeq
with
\begin{align}
&W^{\Phi}_{1} \\
&= \int_{U^{-}} dx\, \frac{4\alpha^4 (x^2-1)  ((1 + x^2) (1 + \alpha^2) - 4  x  \alpha)  (\alpha (1 + \alpha^2) (1 + x^2) - 2  x  (1 + \alpha^4))}{\lambda\,(x - \alpha)^4 (x \alpha-1)^4 (\alpha^2-1)^2 (\alpha^2+1)(e^{2\pi \J E(x)}-1)} \\
&\,\,\,\,\,+ \Op\left(\frac{1}{\lambda^{3/2}}\right)\, .
\end{align}
Like the other integrals, $W^{\Phi}_{1}$ can be rewritten as a linear operator acting on $I[\L]$,
\beq
W^{\Phi}_{1} = -\frac{2}{\lambda} \left[\widehat{\Delta}^{(2)}_{\Phi}I[\L] \right]_{\L \, =\,  \J} + \Op\left(\frac{1}{\lambda^{3/2}}\right)\, ,
\eeq
with
\beq\label{eq:hatDeltaPhi}
\widehat{\Delta}^{(2)}_{\Phi} = \left(-\frac{1 + \J^2}{12}  \partial_{\J}^{3} -\frac{\J}{4} \partial_{\J}^{2}  + \frac{3 + 2  \J^2}{4  \J^2}  \partial_{\J}+ \frac{1 + 2  \J^2}{4  \J^3} \right)\partial_{\L}\, .
\eeq

In summary, combining our findings, we conclude that the wrapping corrections to the normalization factor can be expressed in the form~\eqref{eq:hatDeltaW} with the operators given by
\beq\label{eq:hatDelta-final}
\widehat{\Delta}^{(1)} = \widehat{\Delta}^{(1)}_{F}\, , \qquad \widehat{\Delta}^{(2)} = \widehat{\Delta}^{(2)}_{F} + \widehat{\Delta}^{(2)}_{\Phi}\, , 
\eeq
where the various components are defined in eqs.~\eqref{eq:hatDeltaF} and~\eqref{eq:hatDeltaPhi}.

\bibliography{biblio}
\bibliographystyle{JHEP}

\end{document}